\documentclass[10pt]{article}
\usepackage{mystyle}

\begin{document}
\maketitle 

\begin{abstract}
    This study presents a comprehensive framework for modelling earthquake-induced landslides (EQILs) through a channel-based analysis of landslide centroid distributions. A key innovation is the incorporation of the normalised channel steepness index (\ksn) as a physically meaningful and novel covariate, inferring hillslope erosion and fluvial incision processes. Used within spatial point process models, \ksn{} supports the generation of landslide susceptibility maps with quantified uncertainty. To address spatial data misalignment between covariates and landslide observations, we leverage the \texttt{inlabru} framework \citep{lindgren2024inlabru, suen2025cohering}, which enables coherent integration through mesh-based disaggregation, thereby overcoming challenges associated with spatially misaligned data integration. Our modelling strategy explicitly prioritises prospective transferability to unseen geographical regions, provided that explanatory variable data are available. By modelling both landslide locations and sizes, we find that elevated \ksn{} is strongly associated with increased landslide susceptibility but not with landslide magnitude. The best-fitting Bayesian model, validated through cross-validation, offers a scalable and interpretable solution for predicting earthquake-induced landslides in complex terrain.
    
    
\end{abstract}
\section*{Keypoint}
\begin{enumerate}
    \item Channel-based analysis of Earthquake-induced landslide (EQIL) centroid distributions.
    \item Integration of the channel steepness index (\ksn{}) into statistical models for landslide occurrence.
    \item Generation of EQIL susceptibility maps via spatial point process modelling, addressing spatial data misalignment through the \texttt{inlabru} framework \citep{lindgren2024inlabru, suen2025cohering}.
    \item Development of transferable EQIL susceptibility models applicable to new geographical regions, conditional on covariate availability.
    \item Rigorous model evaluation using (strictly) proper scoring rules for EQIL susceptibility assessment.
\end{enumerate}

\section{Introduction}


Large earthquakes in mountainous regions are often accompanied by widespread landsliding \citep{robinson2017rapid}. One step in understanding the drivers of these landslides is to create landslide inventories: the advent of high-resolution satellite imagery has facilitated the development of both manual \citep{valagussa2021role, rosser2021changing} and automated \citep{fayne2019automated, meena2019comparison} landslide inventories. These inventories are typically characterised by high spatial precision, completeness, scalability and traceability. In these inventories, landslides are delineated as polygons with associated areas, providing a quantitative measure of landslide size. Inventory efforts are generally designed to ensure spatially unbiased mapping of landslide events and undergo rigorous checks to minimise detection errors.

Thanks to the rapid availability of seismic data from the United States Geological Survey (USGS) for all major events in formats useful for disaster management, ground shaking models have been developed, despite challenges arising from their complex spatial distributions, irregular geometries, variable sizes, and data censoring. In particular, data censoring in landslide inventories refers to the omission of small landslides, typically due to limitations in detection methods or predefined mapping thresholds. Consequently, landslides below a certain size are under-reported or not registered, leading to a truncated dataset that may bias statistical models and frequency–area distributions. The integration of spatial information on ground shaking (as a trigger), observed landslide occurrences, and a rich set of covariates offers a well-defined modelling framework for analysing both the location and sizes of landslides. As landslide inventories are typically derived from satellite imagery acquired some time after the mainshock, it is possible that some recorded landslides were in fact triggered by subsequent aftershocks. Nonetheless, in principle, such events could also be incorporated within modelling frameworks.

\subsection{Landslide Modelling}
 

There has been a growing body of research focusing on spatial and spatio-temporal modelling of landslides using the Integrated Nested Laplace Approximation (\inla) framework \citep{lombardo2018point, lombardo2020space, loche2022landslide, opitz2022high, yadav2023joint, bryce2022unified, bryce2025updated}, employing slope-unit aggregation and trigger data (see Table~\ref{tab:cov}) aggregated within these slope-unit polygons. \inla{} provides a computationally efficient approach for approximate Bayesian inference, especially in spatial and spatio-temporal models, offering a fast and accurate alternative to traditional Markov Chain Monte Carlo (MCMC) methods \citep{rue2009approximate}. Slope units, commonly used for aggregating landslide data, are delineated by geomorphological boundaries derived from Digital Elevation Models (DEMs) and typically assumed to represent homogeneous triggering conditions. The slope-unit approach aggregates data into polygons to mitigate spatial misalignment in landslide susceptibility modelling. The aforementioned studies employed both the parameterised and parameter-free versions of the \textit{r.slopeunit} algorithm developed by \citet{alvioli2016automatic, alvioli2020parameter}. However, the slope-unit method entails several inherent limitations:

\begin{enumerate}
    \item \textbf{Parameterisation Issues:} The calibration of parameters (e.g.\ slope unit shape and size) introduces subjectivity, as these choices are user-defined and lack a principled, objective foundation. Conversely, parameter-free approaches obscure the generation process, as they rely on implicit model fitting in an initial stage, with unclear spatial adaptability of the underlying parameters. The influence of these hidden modelling choices, as well as the propagation of their associated uncertainties into subsequent landslide predictions, remains poorly understood.

    \item \textbf{Modelling issues:} Landslide occurrences are often modelled as spatial point processes using crown or centroid locations. Aggregating data over slope units introduces bias by discarding point-level information and conflating covariate uncertainty, making uncertainty propagation intractable. Imposing fixed slope-unit boundaries restricts the spatial domain, conflicting with the theoretical assumptions of point pattern models. While point locations may suffer from registration errors, representing landslides as points preserves spatial resolution and enables more robust inference, particularly when jointly modelling landslide locations and sizes using a Bernoulli-log-Gaussian hurdle model \citep{bryce2022unified}. The hurdle model is a two-step approach: (1) the study domain is partitioned into slope unit polygons, with presence and absence modelled using a Bernoulli distribution based on intersection with aggregated landslide areas; and (2) landslide log-sizes are modelled using a Gaussian distribution. The transition from the Bernoulli to Gaussian components is especially sensitive to aggregation, with unclear implications for uncertainty propagation.

    To capture intra-unit variability and geometric characteristics within each slope unit, these models often include additional covariates as offsets, such as:
    \begin{enumerate*}[label=\roman*)]
        \item standard deviation of slope within the unit,
        \item slope unit area and perimeter, and
        \item maximum internal distance.
    \end{enumerate*}
    These terms aim to account for heterogeneity that would otherwise be obscured by polygonal aggregation, but further complicate model specification and interpretation.

    \item \textbf{Poor Reproducibility:} Model transferability is limited, as each new location requires reparameterisation to generate suitable slope units, which do not guarantee the original spatial model would work the same. 
    
    \item \textbf{Scalability Constraints:} The computational burden associated with iterative slope-unit generation limits the feasibility of applying this approach to large-scale regional analyses.
\end{enumerate}

These studies rely on the aggregation of landslide information over slope units, which are intended to be geomorphologically meaningful terrain partitions representing slope stability. Here, we propose an alternative approach that avoids slope-unit aggregation while still accounting for morphological covariates and addressing spatial misalignment. This enables both retrospective analysis and prospective prediction in spatial modelling. Specifically, we suggest combining:
\begin{enumerate*}[label=\roman*), itemjoin={\;}]
    \item pixel-based normalised channel steepness index (\ksn), a widely used metric for understanding erosion in mountainous terrain,
    \item pixel-based distance metrics to channels, and
    \item a mesh-based disaggregation method for point pattern models \citep{lindgren2024inlabru, suen2025cohering}.
\end{enumerate*}
The pixel-based approach ensures spatial homogeneity by tracing parent nodes both upstream and downstream, producing unique channel distance metrics per pixel. These erosion-related covariates become usable in spatial modelling by aligning responses and predictors on a shared mesh, resolving data misalignment via the \inlabru{} package, which extends \inla{} to support more flexible nonlinear models \citep{lindgren2024inlabru}.

\subsubsection{DEM-derived Landslide Susceptibility Covariates}
Traditional triggers for earthquake-induced landslides (EQILs) are summarised in Table~\ref{tab:cov}. Covariates 1 to 10 are derived from digital elevation models (DEMs) and are widely used in landslide modelling for their ability to capture key aspects of terrain morphology. However, their direct inclusion in linear statistical models is often constrained by strong nonlinear relationships, particularly for slope in degree, which has a bounded upper limit. Furthermore, as these covariates originate from the same underlying DEM, they tend to be highly correlated, both linearly and nonlinearly, introducing potential confounding in spatial models. 

\begin{table}[htbp]
\centering
\caption{Traditional covariates (triggers) used in modelling earthquake-induced landslides (EQILs)}
\renewcommand{\arraystretch}{1.3}
\begin{tabular}{lll}
\toprule
1. Elevation & 6. Profile Curvature & 11. Peak Ground Acceleration (PGA) \\
2. Slope & 7. Relative Slope Position (RSP) & 12. Peak Ground Velocity (PGV) \\
3. Sine of Aspect & 8. Topographic Wetness Index (TWI) & 13. Outcropping Lithology \\
4. Cosine of Aspect & 9. Distance to Faults & 14. Landform Classes \\
5. Planar Curvature & 10. Distance to Streams & 15. Land Cover \\
\bottomrule
\end{tabular}
\label{tab:cov}
\end{table}

Despite these limitations, DEM-based covariates offer rich geomorphic insights that can be further exploited through more advanced topographic metrics. The normalised channel steepness index (\ksn), along with a pixel-based distance-to-channel metrics, are derived from DEMs but are designed to replace traditional DEM-based covariates by capturing geomorphic processes more directly and mitigating redundancy. These covariates allow us to examine whether the active downcutting of the river channel increases susceptibility of the connected slopes (see Section \ref{sec:ksn} for technical details).

Topographic relief and erosion rates provide essential insights into the geomorphic processes that shape the Earth's surface. \citet{roering2007functional} proposed a dimensionless relationship between relief ($R$) and erosion rate ($E$), which enables comparison across different landscapes. Building on this framework, \citet{hurst2013influence} investigated uplift and erosional decay patterns along California's Dragon's Back Pressure Ridge. \citet{grieve2016long} demonstrated how high-resolution topographic data can be used to estimate relief and erosion at multiple spatial scales by applying the hilltop flow routing algorithm . In line with this approach, we quantify the spatial correspondence between channel steepness and hillslope characteristics in the context of earthquake-induced landslides (EQILs).

\subsubsection{Peak Ground Acceleration (PGA)}
Peak Ground Acceleration (PGA) is a commonly used covariate in modelling earthquake-induced landslides (EQILs), serving as a proxy for seismic shaking intensity \citep{meunier2008topographic, umar2014earthquake, nowicki2014development}. Although PGA reflects the maximum amplitude of ground motion, it does not capture shaking duration or frequency content, known to influence slope stability. For instance, \citet{baker2021seismic} emphasise the role of shaking duration (i.e.\ the number of ground motion cycles) in landslide triggering and slope displacement. Similarly, \citet{stafford2009new} advocate for metrics such as Arias Intensity and Cumulative Absolute Velocity as more comprehensive indicators of cumulative seismic energy input.

In this study, duration-related metrics are excluded due to limitations in the landslide inventory, which aggregates events from multiple shocks. As a result, we adopt a purely spatial modelling approach. While this does not incorporate temporal variation, the framework can be extended to a spatio-temporal setting when time-stamped data are available. Event aggregation hinders the consistent attribution of duration-based measures to individual landslides and risks introducing spurious associations or confounded effects if such metrics were naively included.

This simplification is a known limitation of our model, potentially under-representing the influence of sustained or repeated ground shaking in landslide triggering. Additionally, other non-earthquake mechanisms may also accumulate damage over time, further complicating the landslide process. By focusing exclusively on PGA, we risk omitting mechanisms essential for explaining failures in certain geophysical contexts, particularly where cumulative damage 
rather than peak motion governs slope instability. Nevertheless, this exclusion supports a conservative and interpretable modelling framework. Future research leveraging event-specific landslide inventories may enable the rigorous inclusion of duration-based covariates in a principled spatio-temporal setting.

\subsection{Proper Scoring Rules}

\citet{opitz2022high, yadav2023joint, dahal2024junction} assess landslide models using a single summary score, which obscures spatial variation in predictive performance. Such aggregation can be misleading in spatial settings, where prediction accuracy often varies substantially across the landscape. For instance, a low average score may mask poor performance in specific regions. \citet{dahal2024junction} implement both random (termed thinning cross-validation in Section~\ref{sec:cv}) and temporal cross-validation, yet these approaches do not adequately capture spatial structure. In contrast, \citet{opitz2022high} compute scores on pixel grids and slope-unit polygons, whose resolutions fine enough to reflect spatial clustering, yet still report only aggregated scores.

Common model selection criteria such as the Watanabe-Akaike Information Criterion (WAIC) and the Deviance Information Criterion (DIC) are similarly limited in spatially structured contexts. These metrics can favour models that are overly smooth or insufficiently regularised, and often misidentify relevant covariate effects in the presence of strong spatial dependence. Relying solely on such aggregated diagnostics may therefore lead to suboptimal or misleading conclusions.

In this work, we emphasise the rigorous application of strictly proper scoring rules for spatial predictive evaluation, assessing how well a predicted probability distribution aligns with unseen landslide occurrences. Proper and strictly proper scoring rules, especially Logarithmic Score (LS) and Continuous Ranked Probability Score (CRPS), are crucial for model diagnostics \citep{gneiting2007strictly}, particularly for evaluating the predictive performance of probabilistic models. 

\section{Data and Methods}
 In this section, we will go through the processing details of the response and explanatory variables (see summary in Table \ref{tab:data}). 

\begin{table}[htbp] 
\small
\centering
\caption{Summary of spatial covariate and response variable datasets used in the analysis. }
\begin{tabular}{lllll}
\toprule
\textbf{Response Variable} & \textbf{Acronym} & \textbf{Type} & \textbf{Units} & \textbf{Source} \\
\midrule
Landslide centroids & -- &  Point pattern & -- & \citet{valagussa_landslide_2022} \\
Log landslide size & -- &  Mark (continuous) & \(m^2\) & \citet{valagussa_landslide_2022} \\
\toprule
\textbf{Explanatory Variable} & \textbf{Acronym} & \textbf{Type} & \textbf{Units} & \textbf{Source} \\
\midrule
PGA Mean & PGA &  $1.73$km Raster (continuous) & g\tablefootnote{Gravitational acceleration, in log scale in original data} & \citet{marano2024shakemap} 
\\
Land Cover & LC &  Polygon (categorical) & -- & \citet{fao_2009}, \citet{uddin2015development} \\
Geology & Geo &  Polygon (categorical) & -- & \citet{dahal2014regional}\tablefootnote{For lithological details for geology, see Table \citet{dhital2015geology}} \\
Channel Steepness Index & \ksn & $30$m Raster (continuous) & m & \cite{copernicus_dem_2021}, \cite{mudd2022lsdtopotools}\tablefootnote{Computed by \lsd \  using DEM \label{lsdtopo}} \\
Relief to Channel & Rf2Ch & $30$m Raster (continuous) & km & \cite{copernicus_dem_2021}, \cite{mudd2022lsdtopotools}\tablefootnotemark{lsdtopo} \\
Flow Distance to Channel & Fd2Ch & $30$m Raster (continuous) & km & \cite{copernicus_dem_2021}, \cite{mudd2022lsdtopotools}\tablefootnotemark{lsdtopo} \\
Digital Elevation Model & DEM & $30$m Raster (continuous) & km & \citet{copernicus_dem_2021} \\
\bottomrule
\end{tabular}
\label{tab:data}
\end{table}

\subsection{Response Variable (Landslide Catalogue)}


The landslide inventory used in this study is from \citet{valagussa_landslide_2022}, integrating two sources: (i) \citet{roback2018size}, with 24,915 landslides, and (ii) \citet{valagussa2018pre}, with 4,300 landslides. During integration, \citet{valagussa2021role} excluded some \citet{roback2018size} landslides due to cloud cover or severe image distortion, cross-checked using Google Crisis imagery from 2 May to 6 June 2015. The \citet{roback2018size} inventory covers landslides triggered by the 25 April 2015 Mw7.8 Gorkha earthquake and its aftershock sequence, including the significant Mw7.3 aftershock on 12 May 2015, with events mapped from 26 April to 15 June 2015. 

\begin{figure}[htbp]
    \centering
    \begin{subfigure}[t]{\linewidth}
        \centering
        \includegraphics[width=\linewidth]{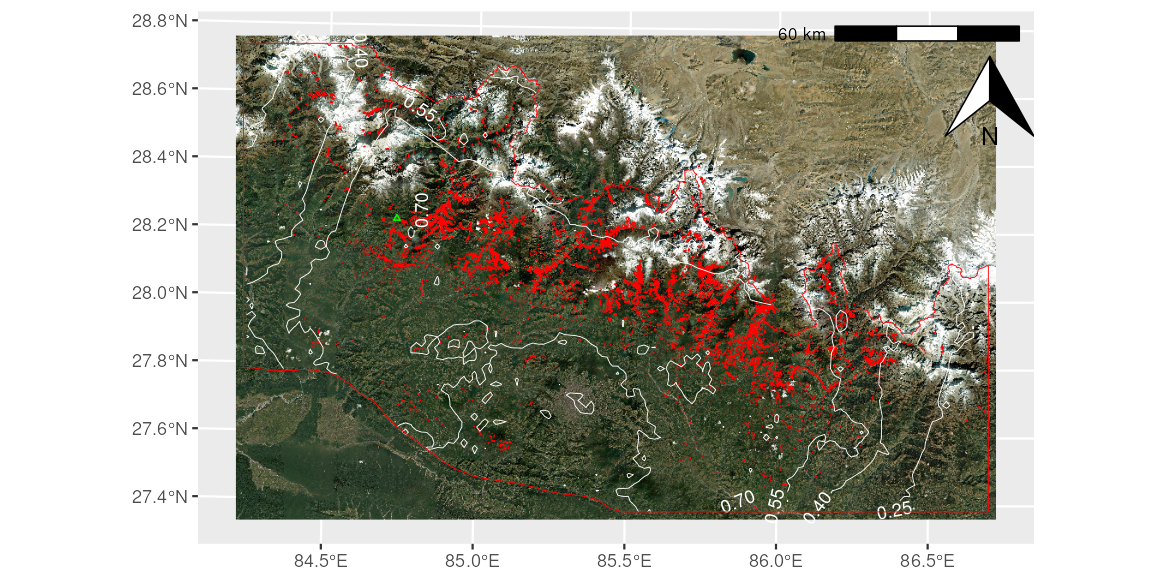}
        \caption{$M_w$ 7.8 mainshock (25 Apr 2015).}
        \label{fig:ldsize_mw78}
    \end{subfigure}
    
    \vspace{0.5em} 

    \begin{subfigure}[t]{\linewidth}
        \centering
        \includegraphics[width=\linewidth]{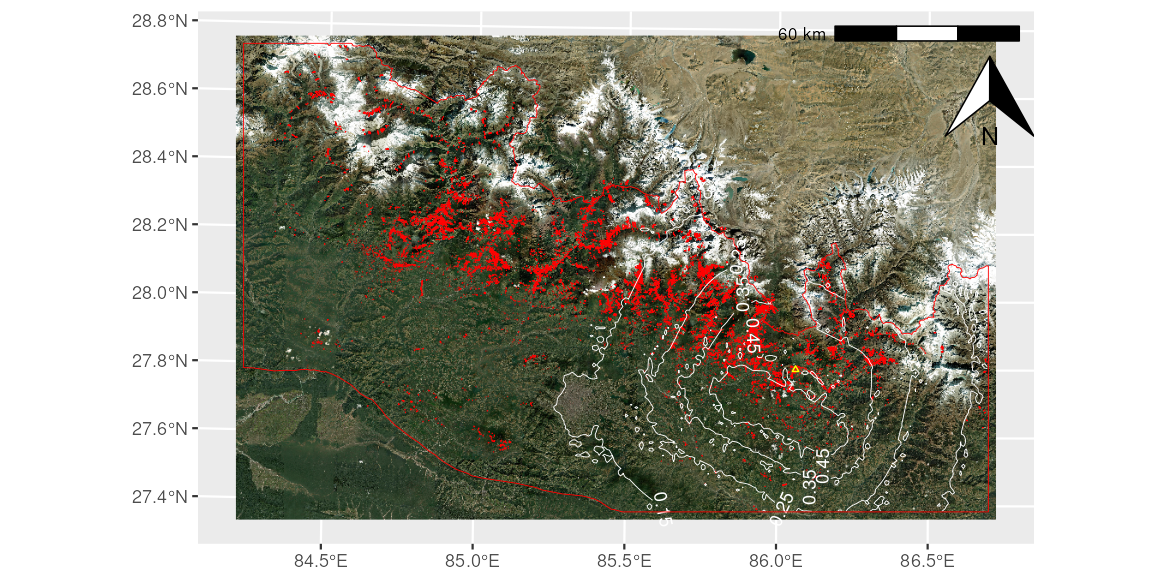}
        \caption{$M_w$ 7.3 largest aftershock (12 May 2015).}
        \label{fig:ldsize_mw73}
    \end{subfigure}

    \caption{Landslides (in red) induced by the 2015 Gorkha Earthquake \citep{valagussa_landslide_2022} within the study region, showing (a) the $M_w$ 7.8 mainshock epicentre (green triangle) and (b) the $M_w$ 7.3 largest aftershock epicentre (yellow triangle). White contours indicate Peak Ground Acceleration (unit: $g$).}
    \label{fig:ldsize}
\end{figure}

Overall, the EQIL catalogue captures landslides triggered by multiple seismic events, reflecting the cumulative geomorphic response (see Figure \ref{fig:ldsize}). The study area spans \ang{84.24}E–\ang{86.70}E longitude and \ang{27.38}N–\ang{28.74}N latitude, covering the central-eastern Himalayas most affected by the earthquake sequence.

Given the complexity of cascading and delayed failures, we focus here on the spatial characteristics of landslide occurrence. Although this study adopts a purely spatial framework, the proposed methodology is readily extensible to a spatio-temporal setting that can account for aftershock-driven remobilization and progressive failures over time. As the landslide inventory is recorded as polygons, a methodological choice is required to represent them as point patterns. We model centroids rather than crowns, as centroids are more spatially consistent with the associated covariates and avoid the subjectivity involved in delineating landslide crowns. This trade-off is of limited concern in our study, given that we are more interested in identifying susceptible areas impacted by landslides than in pinpointing their exact initiation locations.

\subsection{Exploratory Variables}

Since one of our aims is to evaluate the informativeness of \ksn{} in landslide susceptibility modelling, we include only explanatory variables that have been identified as significant in the literature. While not strictly necessary, we strive to maintain consistency in the units of continuous variables while, for interpretability, ensure that their value ranges are comparable. 


\begin{figure}[ht]
    \centering
\includegraphics[width=\linewidth]{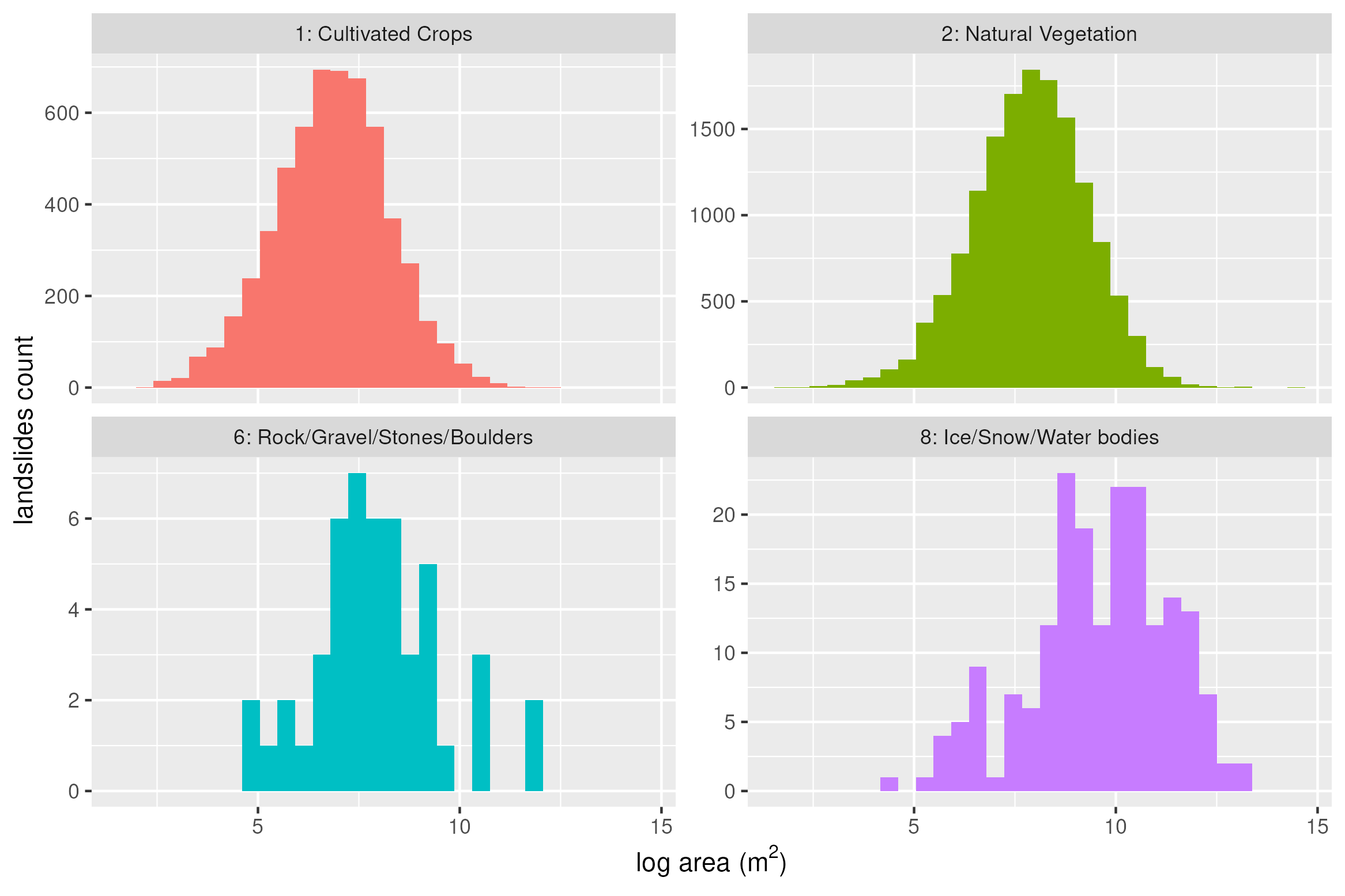}
    \caption{Histogram of landslide centroid count against $\log(\text{landslide area})$ in $m^2$, stratified by land cover. See Table \ref{tab:lcc_labels_filtered} for label description.}
    \label{fig:ldslcc}
\end{figure}

\subsubsection{Channel Steepness Index (\ksn) and Distance Metrics to Channel}\label{sec:ksn}
We download the Copernicus DEM \citep{copernicus_dem_2021} for our regions, which has a resolution of 1 arcsecond (approximately 30 metres). The source data is in a geographic coordinate system, but we measure distances so project the DEM into universal transverse Mercator (UTM) coordinates, at 30 metre grid spacing, using Lanczos resampling. We then compute topographic derivatives using LSDTopotools \citep{mudd2022lsdtopotools}. These derivatives include:

\begin{enumerate}[label=\roman*)]
    \item \textbf{Channel Steepness Index (\ksn):} This index quantifies the steepness of channels, which is an important metric for understanding erosion and sediment transport in rivers because rivers actively incise into the landscape, they strongly influence and are closely coupled with hillslope processes.
    \item \textbf{Relief to Channel (Rf2Ch):} This is defined as the change in elevation between a hilltop pixel and the nearest channel node, computed using the hilltop flow routing algorithm \citep{grieve2016long}. Hilltops are identified by extracting basin edges that share a stream order, following the techniques outlined in \citep{hurst2012using}. It provides insight into the vertical distance of the mass movement and its potential impact on landslide processes.
    \item \textbf{Flow Distance to Channel (Fd2Ch):} This is the distance following a flow path from a pixel to the nearest channel node, which is crucial for modelling debris flow and drainage patterns.
\end{enumerate}

\subsubsection*{Algorithm Overview}

The channel steepness index, \ksn, is derived from the slope-area relationship and reflects the steepness of a river channel relative to its drainage area. As a topographic metric, it serves as a local proxy for the rate of elevation loss along a channel segment, enabling comparisons of channel gradients across different catchments, independent of drainage basin size. Erosion rates and \ksn{} are observed to be positively correlated \citep{gailleton2021impact}. 







Assuming a steady-state landscape where uplift is balanced by incision, this equation can be rearranged to express channel slope in terms of drainage area:

\begin{equation}
    S = \text{\ksn} A^{-m/n},
\end{equation}

where $S = dz/dx$ is the gradient of elevation, $z$, along the channel with flow distance, $x$; 
$\theta = m/n$ represents the channel profile concavity, $\text{\ksn}$ quantifies channel steepness, and $A$ is the drainage area, based on channel profile data, offering a direct link to erosional dynamics.

A substantial body of literature supports the use of SPIM and its derivatives for understanding fluvial processes and long-term erosion (e.g.\ \citealp{howard1983channel, howard1994detachment, whipple1999dynamics, gasparini2011generalized}). Within this context, \ksn{} is a powerful diagnostic tool for identifying spatial patterns of landscape adjustment to tectonic and climatic forces. Its relevance to hazard assessment in mountainous terrain, where steep channels are often associated with landslide initiation, makes it particularly valuable for modelling EQIL susceptibility and improving process-based understanding of landscape response. For further technicality, see Appendix \ref{sec:ksn_tech}.

\subsubsection*{Erosion and Stream Power}

The stream power law suggests that erosion, normalised by drainage area, is proportional to channel steepness. The unit of the channel steepness index depends on the concavity index \( \theta = m/n \), where in the case of concavity \( \theta = 0.5 \), the unit of measurement is in metres (m). Additionally, erosion modelling incorporates a law with a concavity index of $0.5$, which introduces potential bias if not aligned with the landscape’s properties. This bias should be carefully considered when comparing erosion models to the actual landscape data. The methods described here leverage topographic metrics derived from the DEM to analyse drainage networks, basins, and erosion patterns.
\subsubsection*{Basin and Channel Identification}

Drainage areas (or basins) are delineated based on junctions, where each junction may have two or more contributing channels. The basin selected by a junction includes the channel flowing downstream, extending one pixel before the next junction. This approach allows us to delineate basins that are representative of a given Strahler order. While the model offers the flexibility to manually select outlets to identify valid basins, in this study, basins are identified using the default algorithm due to the large extent of the study area.

Channels are identified using the Strahler order system, which classifies drainage networks based on the hierarchy of tributaries. The smallest unbranched drainage is designated as a first-order stream, and the order increases as streams merge. The outlet specification can be challenging, especially if the outlet falls within a missing pixel. In such cases, a larger DEM may be needed to ensure the river is fully represented.

The algorithm identifies the river network and excludes these areas from the channel extraction, resulting in missing pixels. In the post-processing step, we fill the missing pixels with the nearest \ksn \ values. Finally, all basins are reviewed for consistency.

\subsubsection{Land cover and lithology}\label{sec:lcl}

Figure~\ref{fig:ldslcc} shows that landslides occurring within glacial landscapes constitute only a small proportion of the overall landslide inventory. Visually, their distribution appears skewed to the right, suggesting that landslides in glacial areas tend to be larger in size compared to those occurring in other land cover categories, assuming the logarithmic transformed sizes in the inventory is approximately Gaussian-distributed. However, due to the limited number of observed landslides in glacial regions, it is difficult to draw robust conclusions. This motivates the need to explicitly quantify the associated uncertainty within our modelling framework.

Glacial processes tend to flatten channel profiles, which complicates the application of standard fluvial geomorphic models. A first-order interpretation suggests that landslides situated in the upstream (left-hand) portions of catchments typically conform to the stream power law. However, this relationship often breaks down in glacial terrain, particularly in the presence of hanging valleys. In such cases, the channel profile deviates from the expected concave channel geometry.

The local steepness index, \ksn, as computed (refer to \citet{mudd2014statistical} for details), does not automatically account for the presence of hanging valleys. Because hanging valleys do not adhere to the stream power law, their inclusion leads to systematic misestimation of local \ksn\ values. To address this, we manually identify regions with clear signs of misestimation (most often corresponding to hanging valleys) and substitute the affected \ksn\ values with those from nearby geomorphologically similar regions (see Figure~\ref{fig:glacier} in Appendix \ref{sec:ksn_fix} for illustration).   

The low number of observations in this glacial region limits the statistical power of our models, making it challenging to detect meaningful spatial patterns or covariate effects. Consequently, our modelling strategy is guided by the data density, requiring methods capable of handling limited observations in this area while still providing reliable inference alongside rigorous uncertainty quantification. This limitation underscores the challenges of applying stream power-based indices in glacial landscapes and highlights the need for methodological refinement in such contexts.

\subsubsection{Peak Ground Acceleration (PGA)}
The PGA described here is extracted from the $M_w$ 7.8 mainshock. The PGA standard deviation raster (ranging from 0.394 to 0.567 in $\log$ scale), compared to the PGA mean raster (ranging from -6.071 to -3.068 in $\log$ scale) is excluded from the modelling process, as it displays limited spatial variability and appears to be systematically distributed. Its inclusion would necessitate additional assumptions to appropriately propagate the associated uncertainty. We have tried incorporating the PGA from the main aftershock but it did not improve the results. 


\subsection{Study Region}
For this study, we restrict our analysis to areas where landslide inventory data are available and only within the geographical boundaries of Nepal . Additionally, we exclude regions corresponding to the Siwalik Group, Gangetic Plain, and Sub-Himalayas from the land cover data, as these areas exhibit minimal or no landslide activity (see Figure \ref{fig:ldsize}). Including them could introduce numerical instability into the modelling process due to the lack of informative events.

\section{Model Construction and Assessment}
This section describes the modelling details. Since we model a continuous spatial domain using finite element methods (FEM), we first present the mesh construction, followed by the statistical model specification and the description of covariates used in the candidate models.

\subsection{Mesh Specification}
We construct a hexagonal mesh composed of triangular elements, each with an area of approximately 0.1~km\textsuperscript{2}. Based on numerical simulation results, this resolution is sufficiently fine to ensure that the edge length is roughly three times that of a 30~m raster \citep{suen2025cohering}. The hexagonal structure is particularly well-suited for representing a stationary underlying field due to its uniform connectivity and reduced directional bias. Using a finer mesh also enables the model to better capture potential anisotropic features that may not be explicitly represented in the available covariate data. This mesh design reflects a deliberate balance between computational efficiency and the need to accommodate the spatial complexity of the candidate models.


\subsection{Models Specification}

Let \(\mathcal{Y} = \{y_i\}_{i = 1}^n\) denote the set of landslide centroids, and \(\mathcal{Z} = \{z_i\}_{i = 1}^n\) represent the corresponding log-transformed landslide sizes observed at locations \(y_i\). Let \(\vartheta\) be a collection of parameters characterising the distribution of log-transformed landslide sizes. While \(\vartheta\) may be evaluated at the centroid locations \(y_i\), it is not necessarily restricted to these points. For simplicity, we assume that the covariate process \(\vartheta\) is evaluated at the landslide centroids \(y_i\), and this \(\vartheta\) and the intensity function \(\lambda\) are a priori independent. The landslide centroids \(\mathcal{Y}\) are modelled as a realisation of a Poisson point process with intensity function \(\lambda\), i.e.\ \(y_i \sim \textsf{Poisson}(\lambda_i)\). The log-transformed landslide sizes are modelled as \(\log z_i \sim \textsf{N}(\mu_i, \sigma^2_i)\).

More generally, the parameters \(\vartheta\) can be defined over broader spatial domains, such as the full landslide polygon, the crown, or overlapping regions between landslides. This flexibility allows the model to capture spatial correlation and dependencies among neighboring landslides. Thus, \(\vartheta\) may incorporate various spatial features and reflect influences from surrounding landslide activity. These parameters are further governed by a prior distribution that encodes structural or environmental heterogeneity.

Here we justify mathematically how to model landslide centroids and corresponding size separately as a marked point process. Letting \(\bm{y} = (y_1, \dots, y_n)^\intercal\) and \(\bm{z} = (z_1, \dots, z_n)^\intercal\), we specify the conditional model for the log-transformed landslide sizes as 
\[
\bm{z} \mid \bm{y}, \vartheta, \lambda \sim p\left(\bm{z} \mid \vartheta(\bm{y})\right).
\]

The joint density of the landslide centroids and log-sizes is then given by
\begin{align*}
p(\mathcal{Y}, \mathcal{Z} \mid \lambda, \vartheta) 
    &= p(\mathcal{Z} \mid \mathcal{Y}, \lambda, \vartheta) \, p(\mathcal{Y} \mid \lambda, \vartheta) \\
    &= p(\mathcal{Z} \mid \mathcal{Y}, \vartheta) \, p(\mathcal{Y} \mid \lambda),
\end{align*}
where the second equality follows from the assumed conditional independence of \(\mathcal{Z}\) and \(\lambda\) given \(\mathcal{Y}\) and \(\vartheta\).

The posterior distribution of \(\lambda\) and \(\vartheta\) given the observed data is
\begin{align*}
p(\lambda, \vartheta \mid \mathcal{Y}, \mathcal{Z}) 
    &= \frac{p(\mathcal{Z} \mid \mathcal{Y}, \vartheta) \, p(\vartheta) \, p(\mathcal{Y} \mid \lambda) \, p(\lambda)}{C_{\mathcal{Y}, \mathcal{Z}}} \\
    &\propto p(\mathcal{Z} \mid \mathcal{Y}, \vartheta) \, p(\vartheta) \, p(\mathcal{Y} \mid \lambda) \, p(\lambda) \\
    &= p(\lambda \mid \mathcal{Y}) \, p(\vartheta \mid \mathcal{Z}, \mathcal{Y}),
\end{align*}
where \(C_{\mathcal{Y}, \mathcal{Z}}\) is a normalising constant. Thus, the parameters \(\lambda\) and \(\vartheta\) can be estimated separately.

Finally, we define \(z(s)\) as the potential log-size of a landslide at any spatial location \(s\), conditional on the occurrence of a landslide at that location. This formulation enables spatial prediction or simulation of landslide sizes at unobserved sites.

For modelling purposes, the centroids of mapped landslide polygons are extracted to form a spatial point pattern. Each point is marked with the log-transformed landslide area (in \(m^2\)), which serves as a proxy for landslide magnitude. The distribution of log-area sizes is approximately Gaussian (see Figure~\ref{fig:ldslcc}).

In this work, we model landslide centroids and sizes separately to enhance interpretability, facilitate model comparison, and manage computational costs. While a joint modelling approach is technically feasible within the \inlabru\ framework and our proposed methodology, it introduces additional complexities. Specifically, jointly modelling locations and sizes would require careful specification of the dependence structure between spatial clustering and landslide magnitude, which remains challenging both conceptually and computationally. Furthermore, accurately propagating this dependence through the spatial domain with proper uncertainty quantification would substantially increase model complexity, potentially obscuring interpretability. By maintaining separability, we provide clearer insights into the effects of covariates on landslide occurrence and size, while ensuring that model diagnostics and uncertainty assessments remain transparent and tractable.

\subsection{Candidate Models}
Our primary objective is to construct a predictive model that generalises robustly to previously unobserved geographical regions. To this end, we explicitly exclude spatially structured random effects, such as those induced by Mat\'ern SPDE fields, from the model formulation. This modelling decision is motivated by the well-documented issue of spatial confounding, wherein spatial random effects may exhibit collinearity with observed covariates (or covariates encountered in future model updates), thereby inflating posterior uncertainty and attenuating or biasing fixed effect estimates. In contrast to models incorporating only independent noise terms, the inclusion of spatially structured components introduces additional latent uncertainty that is challenging to quantify and propagate, particularly in the context of out-of-sample predictive tasks.

While \citet{dupont2022spatial+} developed a method in the classical (frequentist) statistical framework to separate the influence of spatial random effects from covariates---an approach known as orthogonalisation, which helps avoid spatial confounding---there is currently no established equivalent in Bayesian models. Developing such methods remains an open area for future research. Even if such orthogonalisation were achievable within a Bayesian hierarchy, a key inferential challenge would persist: how to correctly propagate the uncertainty induced by latent spatial components through the joint posterior predictive distribution, without compromising the interpretability and identifiability of fixed covariate effects.

In light of these unresolved issues, we adopt a more parsimonious modelling strategy that relies exclusively on covariate effects. This enables us to maintain interpretability and predictive transferability while avoiding the complications associated with spatial confounding and the non-trivial posterior integration over latent spatial fields.

The covariates included in the candidate models used for model comparison are described below. The linear predictor is defined as
\begin{equation}
    \eta_i := 
    \begin{cases}
    \log \lambda_i, & \text{for the landslide centroid model} \\
    \mu_i, & \text{for the log-transformed landslide size model}
    \end{cases}
\end{equation}

Accordingly, the linear predictor for the \(i\)-th landslide observation is specified as
\begin{equation}
    \eta_i = \beta_0 + \beta_{\text{lc}} X_{\text{lc}} + \beta_{\text{geo}} X_{\text{geo}} + \sum_j f_j(\beta_j, X_j),
\end{equation}
where \(\beta_\cdot\) denotes the model coefficients, \(X_\cdot\) represents the explanatory variables, and \(f_j(\beta_j, X_j)\) denotes either a linear or non-linear transformation of the $j$-th covariates. The subscripts \(\text{lc}\) and \(\text{geo}\) refer to land cover and geological classes, respectively. Categorical covariates such as land cover and geology are incorporated via independent and identically distributed (i.i.d.) random effects. Table~\ref{tab:model_formulas_cov} provides a summary of the covariate-specific mappings \(f_j(\cdot)\) used across models.

We fitted six pairs of models, as outlined in Table~\ref{tab:model_formulas_cov}. Each pair consists of one model for landslide occurrence (centroid locations) and one for log-transformed landslide size, both incorporating analogous sets of covariates. However, the centroid and size models need not be strictly matched within a given pair; for instance, one may model centroids using \texttt{fit5a} and sizes using \texttt{fit6b}.

Erosion rate and \ksn{} are known to exhibit a power-law relationship \citep{howard1983channel, whipple1999dynamics}. To address the nonlinearity of this relationship, we explore two approaches: (i) using the logarithm of \ksn{} as an explanatory variable, and (ii) applying a square-root transformation of \ksn{} combined with a second-order random walk (RW2) smoothing prior. For comparison, we also consider models \texttt{fit3a} and \texttt{fit3b}, which include the logarithm of \ksn{} without smoothing.

Models \texttt{fit4a} and \texttt{fit4b} exclude \ksn{} altogether and instead include a digital elevation model (DEM) covariate, enabling us to assess the relative contributions of \ksn{} and DEM to model performance.

Finally, models \texttt{fit5a}, \texttt{fit5b}, \texttt{fit6a}, and \texttt{fit6b} incorporate relief and flow distance to the channel as covariates. In \texttt{fit5a} and \texttt{fit6a}, preliminary analyses guided modifications of the relief and flow distance variables to tame nonlinearity and improve model fit. For the log-size models, \ksn{} was removed from \texttt{fit5b} and \texttt{fit6b}, as initial findings indicated that its inclusion did not enhance predictive performance for landslide size.

We also experimented with including both relief relative to channel (Rf2Ch) and flow distance to channel (Fd2Ch) in the same model. However, these two covariates are highly correlated, leading to confounding effects that degraded model performance and interpretability.

\begin{table}[htbp]
\small
\centering
\caption{Formulas used in each model. Each cell lists the specific covariate included in that model apart from the categorical land cover and geology which are included in each model.}
\begin{tabular}{llccccc}
\toprule
\textbf{Model} & \textbf{Response} & \textbf{PGA} & $\bm{k_{sn}}$ & \textbf{Rf2Ch} & \textbf{Fd2Ch} & \textbf{DEM} \\
\midrule
\texttt{fit1a} & \multirow{6}{*}{centroids} & \multirow{6}{*}{rw2(PGA)} & $\log{(\text{\ksn})}$ & -- & -- & -- \\
\texttt{fit2a} &  &  & rw2($\sqrt{\text{\ksn}}$) & -- & -- & -- \\
\texttt{fit3a} &  &  & rw2($\log{(\text{\ksn})}$) & -- & -- & -- \\
\texttt{fit4a} &  &  & -- & -- & -- & DEM \\
\texttt{fit5a} &  &  & rw2($\log{(\text{\ksn})}$) & $\exp(-\text{Rf2Ch})$ & -- & -- \\
\texttt{fit6a} &  &  & rw2($\log{(\text{\ksn})}$) & -- & $\exp(-\text{Fd2ch})$ & -- \\
\midrule
\texttt{fit1b} & \multirow{6}{*}{log sizes} & \multirow{6}{*}{$\log(\text{PGA})$} & $\log{(\text{\ksn})}$ & -- & -- & -- \\
\texttt{fit2b} &  &  & rw2($\sqrt{\text{\ksn}}$) & -- & -- & -- \\
\texttt{fit3b} &  &  & rw2($\log{(\text{\ksn})}$) & -- & -- & -- \\
\texttt{fit4b} &  &  & -- & -- & -- & DEM \\
\texttt{fit5b} &  &  & -- & Rf2Ch & -- & -- \\
\texttt{fit6b} &  &  & -- & -- & Fd2Ch & -- \\
\bottomrule
\end{tabular}
\\
\label{tab:model_formulas_cov}
\footnotesize
\vspace{0.5em}
\raggedright
\textbf{Note:} rw2() denotes a second-order random walk (RW2) prior used for smooth effects. See Table~\ref{tab:data} for acronym definitions.
\end{table}

\subsection{Model Assessment Using Cross-Validation Scores}\label{sec:cv}
For point pattern observation models, a more appropriate assessment method is automated Grouped Cross-Validation, where correlated groups of points are left out \citep{adin2024automatic}. However, in the context of earthquake-induced landslides (EQILs), it is not straightforward to define correlation across landslides. Therefore, we consider two strategies for partitioning the data into training and test sets for both point patterns and marks (i.e.\ \(\log\) landslide sizes in \(m^2\)):

\begin{enumerate}

    \item \textbf{Thinning Cross-Validation (Thinning CV):} The data are split at the level of individual landslide events. Specifically, we randomly sample 50\% of the landslide centroids, regardless of their associated sizes, for training;  the remaining 50\% are reserved for testing. This approach is based on the assumption that landslide locations (centroids) are conditionally independent of their sizes. Consequently, the training set reflects half the total spatial intensity, enabling us to assess model performance on held-out landslide locations. While an alternative would be to flip a coin for each landslide event (i.e.\ independent Bernoulli sampling), this could result in an unbalanced split. We verified that the variability introduced by this difference is minor (only around 3\% of the data). Therefore, for simplicity and reproducibility, we adopt the fixed 50\% sampling strategy.

    \item \textbf{Grid-based Cross-Validation (Grid CV):} The study region is partitioned into square grid cells, and alternating grids are selected (i.e.\ in a chequerboard pattern, as illustrated in Figure~\ref{fig:cv_chess} in Appendix \ref{sec:cv_chess}) to form the training set, with the remaining grids constituting the test set. This approach facilitates evaluation of the model’s spatial generalizability by computing prediction scores on geographically distinct, held-out regions, effectively implementing a spatial leave-one-set-out cross-validation scheme.

\end{enumerate}

To assess the robustness of the cross-validation results, we additionally interchanged the roles of the training and test sets. We considered grid sizes of \(3~\text{km} \times 3~\text{km}\), \(5~\text{km} \times 5~\text{km}\), and \(10~\text{km} \times 10~\text{km}\), and found consistent model performance across all resolutions. For clarity and conciseness, we report results based on the \(3~\text{km} \times 3~\text{km}\) grid, as it provides the best balance between the following criteria:

\begin{enumerate}
    \item \textbf{Capturing spatial clustering:} The grid must be fine enough to distinguish spatial differences between areas with and without landslide presence.
    \item \textbf{Preserving categorical detail:} The training set must adequately cover all categories of the explanatory variables; otherwise, the model cannot predict category-specific effects during cross-validation.
    \item \textbf{Maintaining event balance:} The numbers of events in the training and test sets can be sufficiently comparable to ensure stable evaluation under two-fold cross-validation.
    \item \textbf{Computation:} The number of posterior samples to be integrated for individual grid score increases with the number of grids. 
\end{enumerate}

A summary of event counts and a visualisation of the train–test split are provided in Table~\ref{tab:cv_event_counts} and Figure~\ref{fig:cv_chess}, respectively.

\begin{table}[H]
\centering
\caption{Number of events for each cross-validation (CV) setting.}
\begin{tabular}{lcc}
\toprule
\textbf{CV Type(Training Set)} & \textbf{\# Events in Train Set} & \textbf{\# Events in Test Set} \\
\midrule
Thinning (Set A)& 10,236 & 10,235 \\
Thinning (Set B) & 10,235 & 10,236 \\
Grid (White) & 10,200 & 10,356 \\
Grid (Black) & 10,356 & 10,200 \\
\bottomrule
\end{tabular}
\label{tab:cv_event_counts}
\end{table}

\subsubsection{Prediction Scores}

We use four distinct negatively oriented (i.e.\ lower values indicate better performance) prediction scores to evaluate model performance on the $i'$-th test grid, based on $1,000$ posterior predictive samples:

\begin{align*}
S_\text{SE}(F_{i'},y_{i'})&=[y_{i'}-\E(Y_{i'}|\text{data})]^2, \\
S_\text{DS}(F_{i'},y_{i'})&=\frac{[y_{i'}-\E(Y_{i'}|\text{data})]^2}{\V(Y_{i'}|\text{data})} +
  \log[\V(Y_{i'}|\text{data})], \\
S_\text{LS}(F_{i'},y_{i'})&=-\log[\pP(Y_{i'} = y_{i'}|\text{data})], \\
S_\text{CRPS}(F_{i'},y_{i'}) &= \sum_{k=0}^\infty [\pP(Y_{i'} \leq k |\text{data}) - I(y_{i'} \leq k)]^2,
\end{align*}
where $y_{i'}$ denotes the observed event count in the $i'$-th test grid, $\E(\cdot)$ represents the expectation, $\V(\cdot)$ the variance and $I(\cdot)$ the indicator function.

The \textit{Squared Error} (SE) is a proper scoring rule that evaluates the accuracy of the predictive mean, while the \textit{Dawid--Sebastiani} (DS) score extends this by incorporating both the predictive mean and variance. The \textit{Logarithmic Score} (LS) and the \textit{Continuous Ranked Probability Score} (CRPS) are both \textit{strictly proper} scoring rules \citep{gneiting2007strictly}, meaning they are uniquely minimised when the predictive distribution coincides with the true data-generating distribution.

In principle, the CRPS involves an integration over the entire support of the predictive distribution, which often requires evaluating expectations up to infinity. In practical implementation, this is typically approximated using numerical integration up to a sufficiently large bound to ensure computational tractability.

For log-transformed size, the evaluation of the CRPS remains technically involved due to the fact that the full posterior predictive distribution is represented as a mixture of normal distributions. While closed-form expressions exist for certain scoring rules under Gaussian assumptions, deriving an exact analytical form for the CRPS in the case of a normal mixture is non-trivial. As such, CRPS values for log size are not reported here.

While the LS directly assesses the probability assigned to the observed event, it does not encourage the assignment of high probability mass to values near the observation \citep{gneiting2007strictly}. Moreover, it is sensitive to outliers, and forecasts optimised under LS may tend to be overdispersed \citep{gneiting2005calibrated}. By contrast, CRPS compares the full predictive cumulative distribution function against the empirical distribution, offering robustness against extreme values and making it particularly suited for evaluating sample-based forecasts. Notably, CRPS remains proper (and strictly proper) when the first moment of the predictive distribution is finite.

In practice, strictly proper scoring rules, such as LS and CRPS, are considered more powerful than proper scoring rules because they assess the entire predictive distribution rather than focusing only on specific moments or summary statistics \citep{pic2025proper}. This makes them particularly valuable when the goal is to identify and select the ideal probabilistic forecast.

In this study, we fit a series of spatial models incorporating different covariates and prior specifications to better understand the factors influencing landslide occurrence. Covariates were selected based on physical relevance and their statistical properties. For model evaluation, we rely primarily on the CRPS and LS, with particular attention to their behaviour across space and under different cross-validation scenarios. To identify the best-performing model, we examine multiple aspects: (i)  mean proper scores, (ii) spatial strictly proper score difference plots, and (iii)  empirical cumulative distribution functions (ECDFs) of the strictly proper scores. 
The ECDF formulates as 
\begin{equation}
    F(S) = \frac{\#\{S_{i'} < S; i= 1, \dots, n\}}{n},
\end{equation}
where $\#\{\cdot\}$ is the count and $S$ is the specified score metric.

\subsection{Coefficient of Variation (\cv)}

The Coefficient of Variation (\cv) plays a crucial role in quantifying uncertainty within the model, as it provides a normalised measure of the relative variability in the posterior sample distribution. By assessing the \cv, we can gain valuable insights into the consistency and stability of the model’s predictions, highlighting areas where uncertainty is more pronounced. This is essential for interpreting the results, especially when considering the inherent uncertainty in the model’s outputs. Specifically, the \cv{} expresses the standard deviation relative to the mean in the model's estimates. In this context, it serves as an effective diagnostic tool for evaluating the relative variability or stability of the predicted landslide intensities and for each covariate across space. Furthermore, the literature has sparked ongoing discussions about how to quantify joint uncertainty contributions using variability-based metrics. For instance, \citet{yuan2017point} propose a framework for assessing joint correlation of the fixed effects and spatial random field through space and time, providing a deeper understanding of how different sources of variability interact and propagate throughout the model. We have an approximation valid for Normal distribution for the coefficient of variation obtained from 100 posterior samples,
\begin{align*}
    \text{CV} &:= \sqrt{\frac{\V\left[\exp\left(z(\s)\beta\right)\right]}{\left\{\E\left[\exp\left(z(\s)\beta\right)\right]\right\}^2}}. 
\end{align*}


\section{Results}
We present an exploratory analysis of channel profiles and concavity in Section~\ref{sec:explore}, evaluate the prediction scores of the spatial models in Section~\ref{sec:p_score}, and conduct a case study of the best-performing model based on these scores in Section~\ref{sec:case_study}.

\subsection{Exploratory Analysis Case Study}\label{sec:explore}
We selected a subset of basins, specifically those numbered 2228, 14982, and 15329, representing both fluvial and glacial landscapes, to examine how the channel steepness index (\ksn) corresponds to landslide locations of varying sizes. These basins were chosen to include a sufficient number of landslides to enable meaningful visual interpretation, while avoiding overcrowding, thereby supporting subsequent model development. For further exploratory analysis of landslide sizes, see Appendix~\ref{sec:lds_explore}.

\subsubsection{Channel Profile Analysis}\label{sec:channel_profile}
\begin{figure}[htbp]
    \centering
    
    \begin{subfigure}[t]{0.75\linewidth}
        \includegraphics[width=\linewidth]{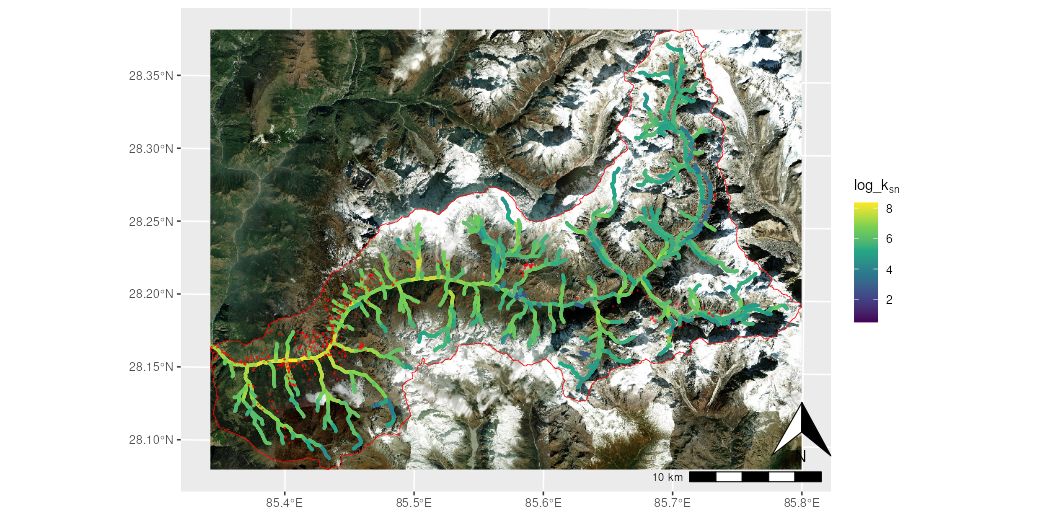}
        \caption{Channel steepness index ($\log$\ksn) along channels within basin 14982 with landslide polygons (red); basemap tile layer from ESRI World Imagery.}
        \label{fig:ksn_basin_poly}
    \end{subfigure}
    
    \vspace{1em}
    
    \begin{subfigure}[t]{0.75\linewidth}
        \includegraphics[width=\linewidth]{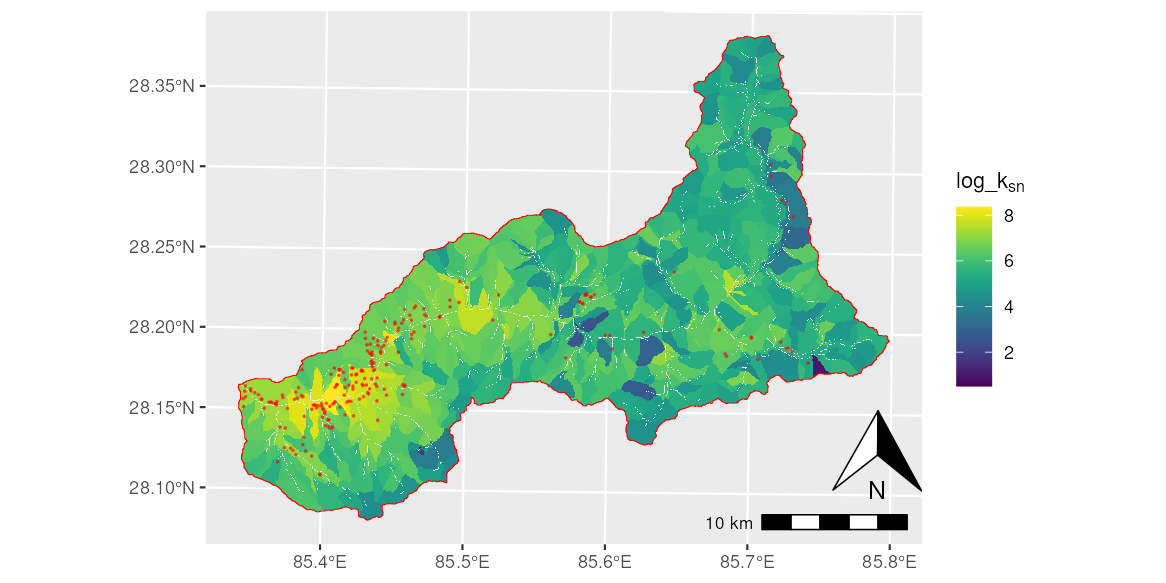}
        \caption{Channel steepness index ($\log$\ksn) tagged by hillslope polygons within basin 14982.}
        \label{fig:basin_tag}
    \end{subfigure}
    
    \vspace{1em}
    
    \begin{subfigure}[t]{0.75\linewidth}
        \includegraphics[width=\linewidth]{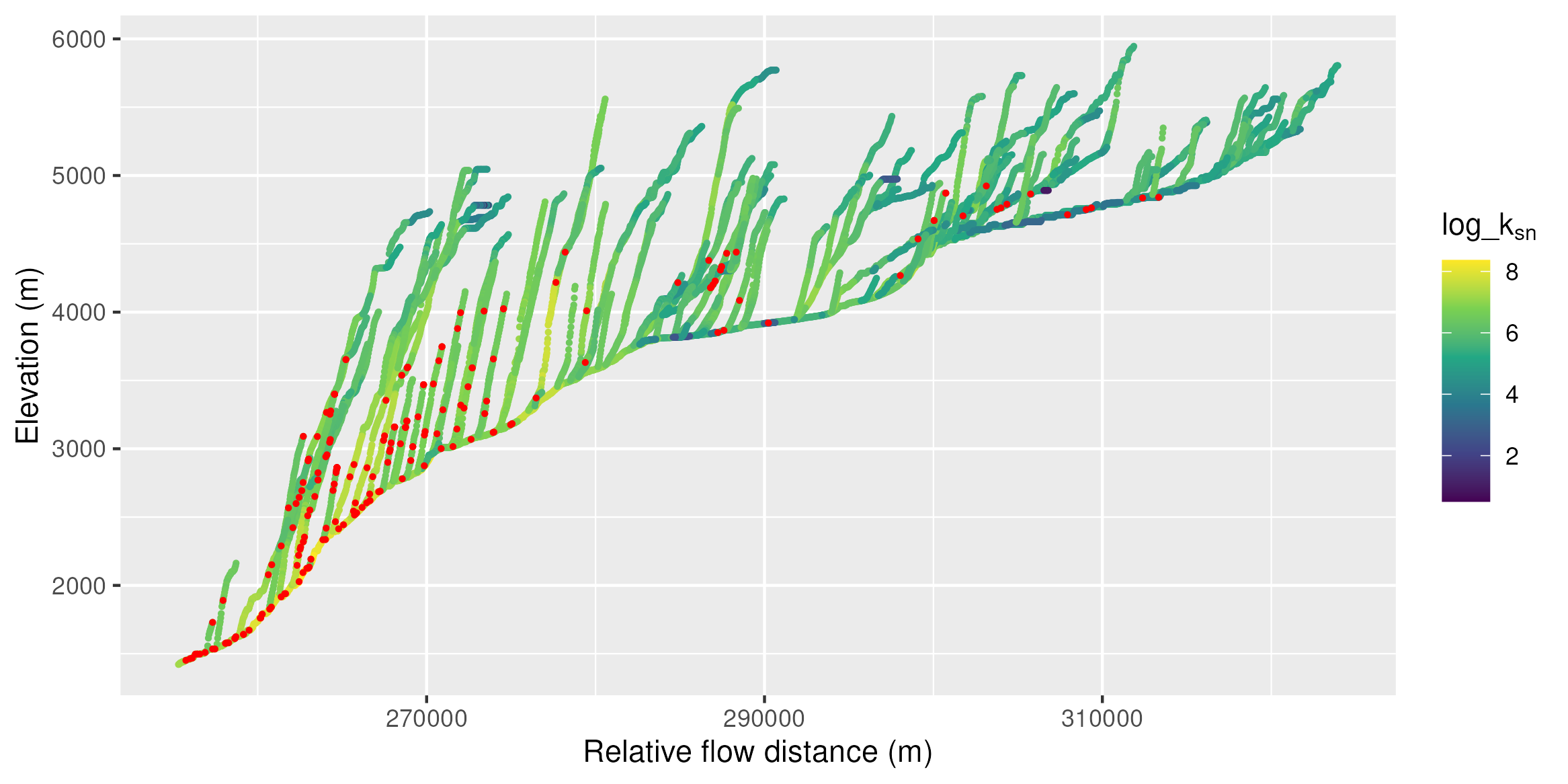}
        \caption{Channel profile plot of elevation against relative flow distance in basin 14982.}
        \label{fig:chi_analysis}
    \end{subfigure}
    
    \caption{Exploratory figures for basin 14982 showing channel steepness index (\ksn), landslide associations, and morphometric analysis.}
    \label{fig:glacial_basin_exploration}
\end{figure}

In Figure \ref{fig:glacial_basin_exploration}, the landslides in fluvial landscape slide toward the corresponding main channels, i.e.\ highest Strahler number, closest to them. In Figures \ref{fig:basin_tag} and \ref{fig:chi_analysis}, a landslide centroid was created for each polygon and was mapped to the nearest channel based on Euclidean distance. We can see that the landslide centroids are mostly lying at the channel nodes with higher \ksn{} and Strahler number. The former tells how steep the channel is and the latter provides clues on relative channel size and depth. The landslides in the glacial landscape further upstream of the channel (the red dots to the right in Figure \ref{fig:chi_analysis}) are affected by the influence of glacial processes on catchment morphology. This is because glacier flatten channel profiles and their impact on the landscape is more complex than simple flattening. Generally speaking, glacial erosion significantly alters valley morphology in several ways: 
\begin{enumerate*}[label=\roman*),itemjoin={\quad}]
    \item U-shaped valleys,
    \item Overdeepening \citep{anderson2006features},
    \item Longitudinal profile modification,
    \item Hanging valleys, and
    \item Undulating profiles \citep{grau2018subglacial}.
\end{enumerate*}  
Figure~\ref{fig:ksn_basin_poly} illustrates how the \ksn{} algorithm is affected by hanging valleys, where glacial processes disrupt standard geomorphic relationships and erosion patterns. Glacial activity significantly influences \ksn{} calculations, as well as related metrics such as the knickpoint extraction, all of which are sensitive to the concavity index \citep{gailleton2021impact}. Although the \ksn{} algorithm in \lsd{} is not explicitly designed for glaciated terrain, the concavity index, typically ranging from 0.4 to 0.6, can be adjusted to reflect variations in landscape processes. In this study, we use a fixed value of 0.5. However, glacial incision, which can vary seasonally, introduces additional complexity that falls outside the scope of this work and would require field-based validation.

In Figure~\ref{fig:basin_tag}, we assume a constant \ksn{} value along each hillslope polygon by tracing back to the corresponding donor node. The resulting map shows a clear clustering of landslide centroids on hillslope polygons with higher \ksn{} values.

\begin{figure}[htbp]
    \centering
    \begin{subfigure}[t]{0.48\linewidth}
        \includegraphics[width=\linewidth]{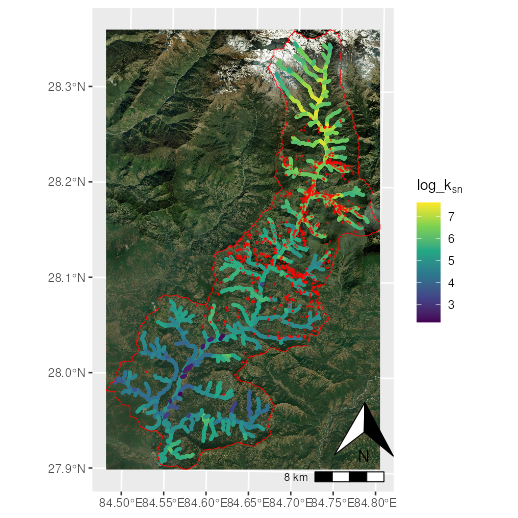}
        \caption{Channel steepness index ($\log$\ksn) along channels within basin 15329 with landslide polygons (red); basemap from ESRI World Imagery.}
        \label{fig:ksn_15329_basin_poly}
    \end{subfigure}
    \hfill
    \begin{subfigure}[t]{0.48\linewidth}
        \includegraphics[width=\linewidth]{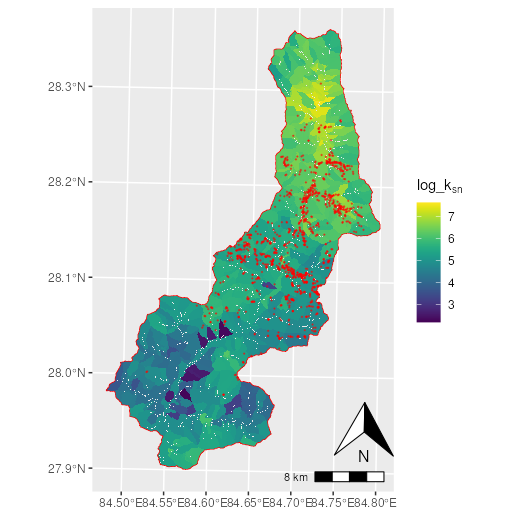}
        \caption{Channel steepness index ($\log$\ksn) tagged by hillslope polygons within basin 15329.}
        \label{fig:15329_basin_tag}
    \end{subfigure}

    \vspace{1em}
    \begin{subfigure}[t]{0.75\linewidth}
        \includegraphics[width=\linewidth]{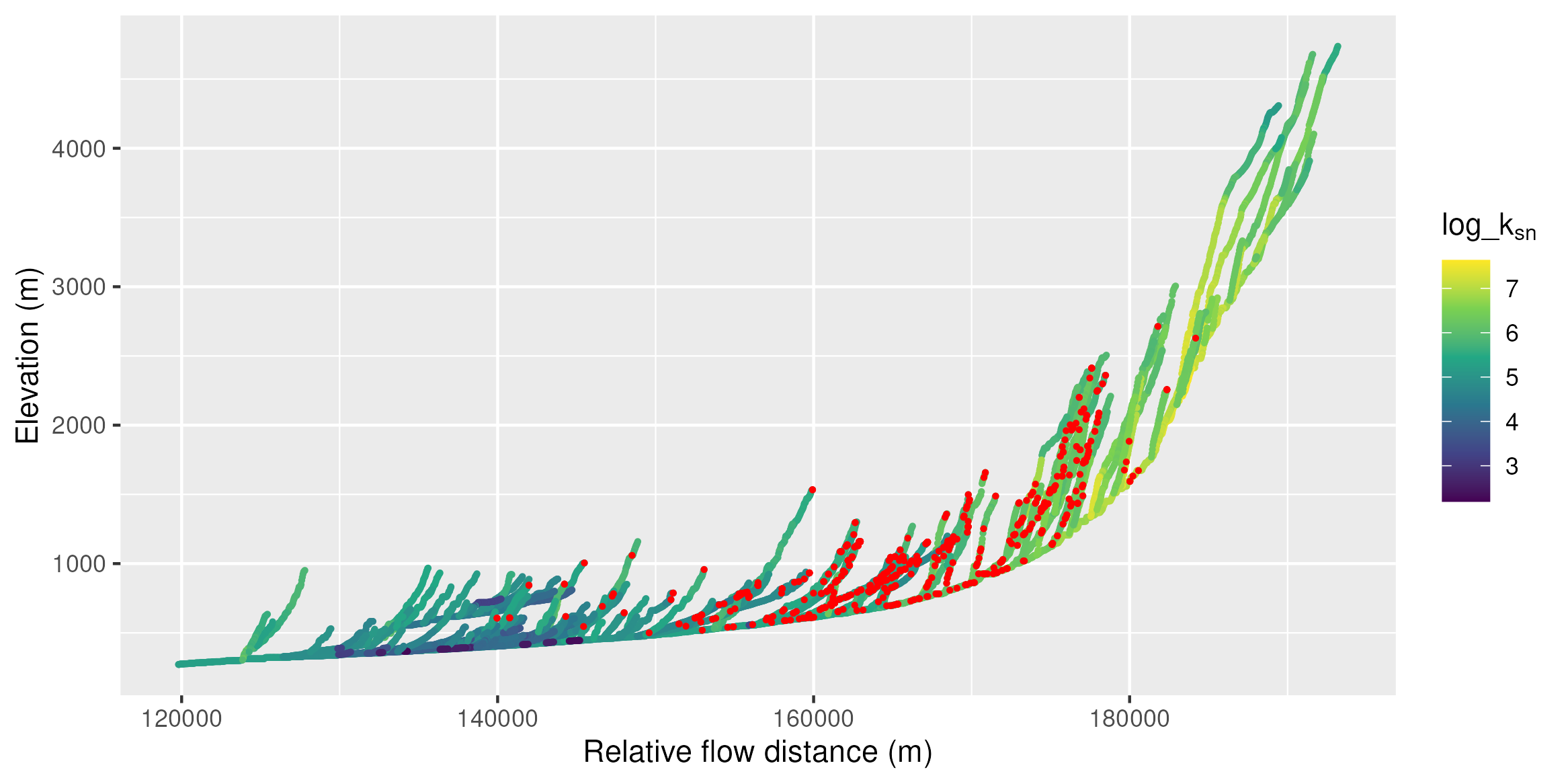}
        \caption{Channel profile plot of elevation against flow distance in basin 15329.}
        \label{fig:15329_chi_analysis}
    \end{subfigure}

    \caption{Exploratory visualizations for basin 15329 showing channel steepness (\ksn), landslide distribution, and corresponding morphometric analysis.}
    \label{fig:basin_15329_exploration}
\end{figure}

Figure~\ref{fig:basin_15329_exploration}, in contrast to the more glaciated basin shown in Figure~\ref{fig:glacial_basin_exploration}, represents a predominantly fluvial landscape. In this setting, the \ksn{} signal appears more consistent, and landslides are primarily concentrated in areas with high channel steepness. Notably, smaller but denser clusters of landslides are aligned along specific channel segments with elevated \ksn{} values, while such clustering is less evident in areas with lower \ksn{}.

In summary, the \ksn{} algorithm performs well in fluvial landscapes but is less reliable in glaciated regions, depending on the extent of glacial influence. However, by incorporating land cover covariates into the modelling framework, glacial landscapes can still be effectively captured, allowing the models to adjust according to the relevant land cover categories. Additional channel profile analyses are provided in Appendix~\ref{sec:further_ksn}. In both cases, landslides appear to correlate more strongly with \ksn{} values than with elevation.

\subsubsection{Concavity}
Ideally, one would fit an optimal concavity parameter that adapts spatially at fine resolution; however, this approach is computationally prohibitive. Concerns may be raised regarding the assumption of a fixed concavity value \(\theta = 0.5\) in this study region. For example, \citet{leonard2023isolating} highlight that deviations from this value may indicate spatial variations in erosional processes or reflect distinct geomorphological regimes. To evaluate the robustness of this assumption, we systematically varied the concavity parameter \(\theta\) across values of \(0.4\), \(0.5\) (default), and \(0.6\), along with adjusting the threshold for contributing pixels to \(500\), \(1{,}000\) (default), and \(2{,}000\).

Across these settings, we observed no significant changes in the spatial structure of the channel steepness index or in the delineation of drainage basins. To illustrate this, we selected representative landscapes encompassing both fluvial and glacial morphologies across appropriate spatial scales (see Figure~\ref{fig:concavity_wrap14982}). Additional results from the concavity sensitivity analysis are provided in Appendix~\ref{fig:concavity_wrap_all}.

\begin{figure}[htbp]
    \centering
    \includegraphics[width=\linewidth, , trim = 0 3.5cm 0 4cm, clip]{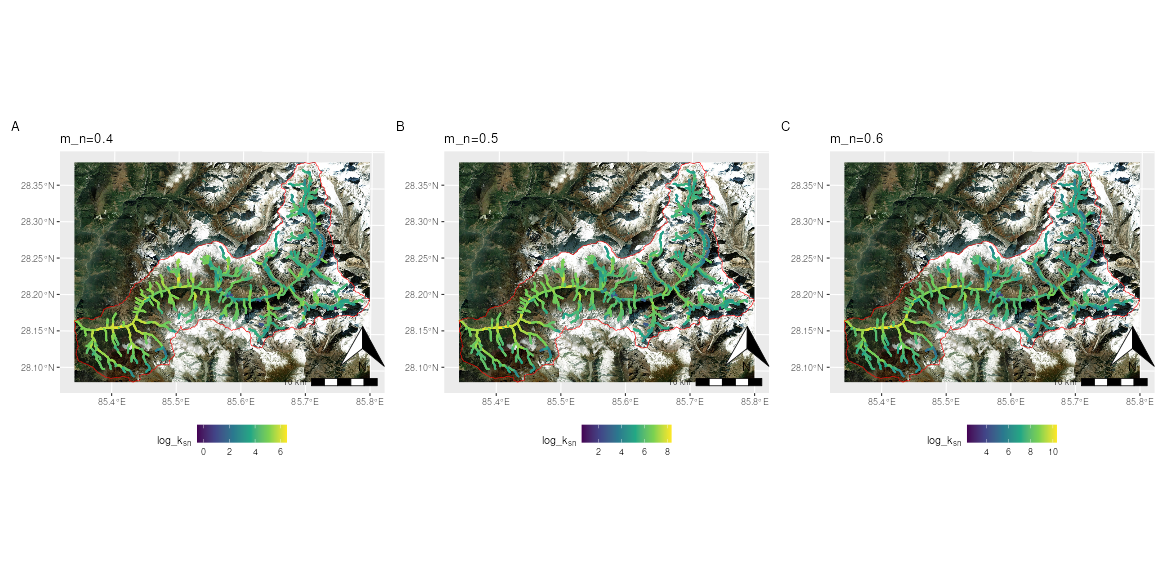}
    \caption{Example of populations of $\log k_{sn}$ calculated for three different concavity values (m\_n) $m/n= 0.4, 0.5, 0.6$  from left to right as in basin 14982.}
    \label{fig:concavity_wrap14982}
\end{figure}

\subsection{Cross Validation for Model Selection}\label{sec:p_score}

We focus on score differences for model comparison, using \texttt{fit1a}, the basic model, as the reference. Spatial score maps highlight regions where alternative models outperform \texttt{fit1a}, while empirical cumulative distribution functions (ECDFs) provide a global comparison of score distributions across models. For brevity, we present only the more powerful strictly proper scores: the Logarithmic Score (LS) and the Continuous Ranked Probability Score (CRPS). Although mean scores are often reported, they can obscure important spatial variation in model performance. Tables~\ref{tab:score_metrics_centroids} and~\ref{tab:score_metrics_logsize} summarise the mean scores for RMSE, Dawid–Sebastiani Score (DS), LS, and CRPS. We verified that 1,000 posterior samples yield sufficiently small Monte Carlo error. Approximate confidence intervals can be obtained using Monte Carlo standard errors. We begin with a detailed evaluation of the point pattern (centroid-based) models, followed by the Gaussian (log-size) models. The computation time and session information are listed in Appendix \ref{sec:sysinfo_ctime}. 

\subsubsection{Prediction Scores for Landslide Centroid Models}

 Examining the mean scores in Table~\ref{tab:score_metrics_centroids}, it is clear that Model \texttt{fit4a}, which models the DEM linearly, performs significantly worse than those incorporating the \ksn{}, particularly in terms of RMSE, DS, and CRPS. Although the LS scores are similar across all models, \texttt{fit6a} consistently delivers the best performance in RMSE, DS, and CRPS, underscoring its robustness and predictive reliability.

\begin{figure}[htbp]
    \centering

    \begin{minipage}{0.03\linewidth}
        \centering
        \rotatebox{90}{\parbox{2.5cm}{\centering \small Thinning CV (Set A)}}
    \end{minipage}%
    \hspace{0.5em}
    \begin{minipage}{0.85\linewidth}
        \includegraphics[width=\linewidth]{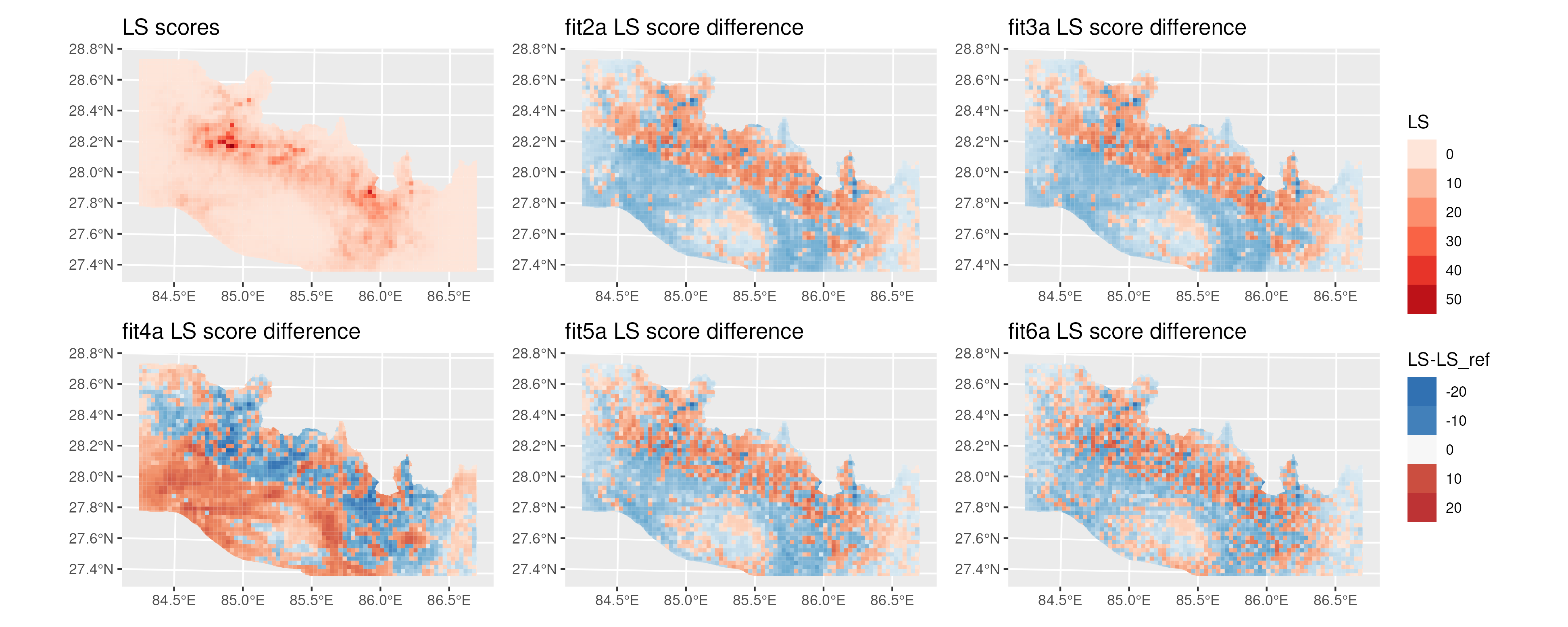}
    \end{minipage}

    \begin{minipage}{0.03\linewidth}
        \centering
        \rotatebox{90}{\parbox{2.5cm}{\centering \small Thinning CV (Set B)}}
    \end{minipage}%
    \hspace{0.5em}
    \begin{minipage}{0.85\linewidth}
        \includegraphics[width=\linewidth]{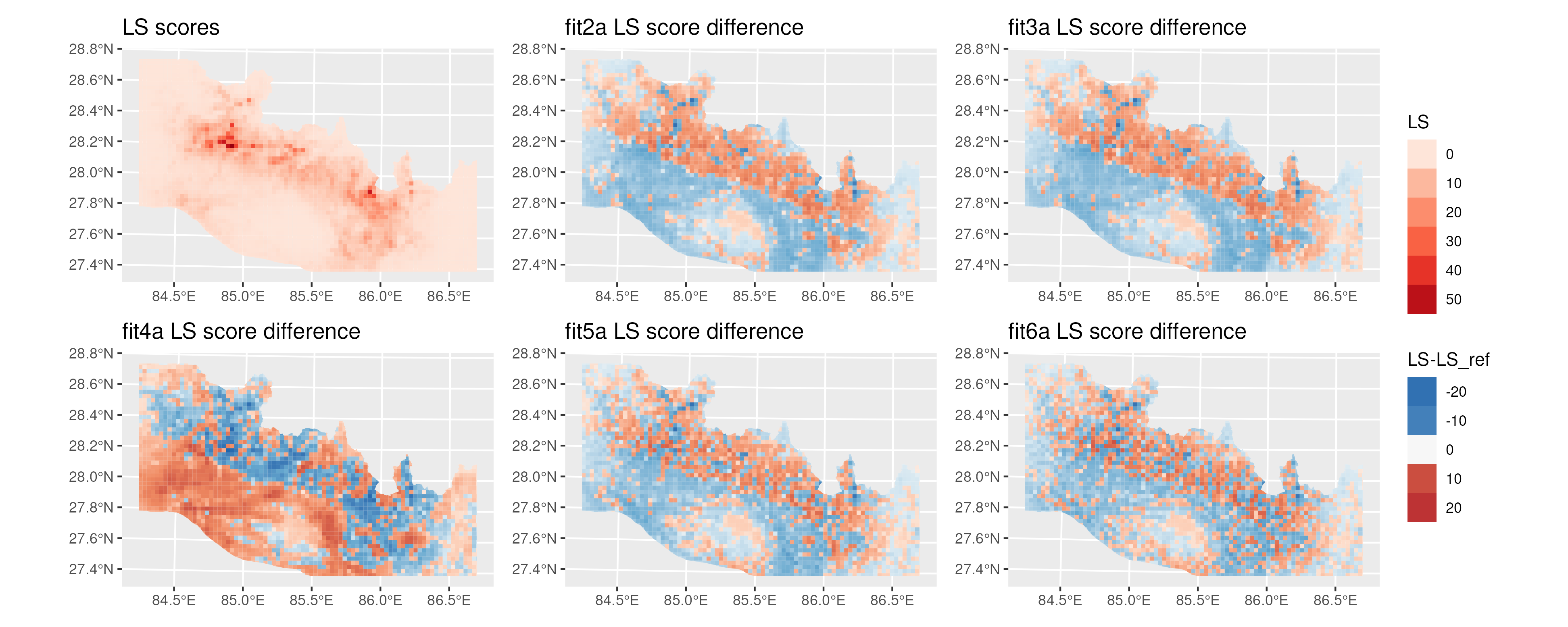}
    \end{minipage}

    \begin{minipage}{0.03\linewidth}
        \centering
        \rotatebox{90}{\parbox{2.5cm}{\centering \small Grid CV (White)}}
    \end{minipage}%
    \hspace{0.5em}
    \begin{minipage}{0.85\linewidth}
        \includegraphics[width=\linewidth]{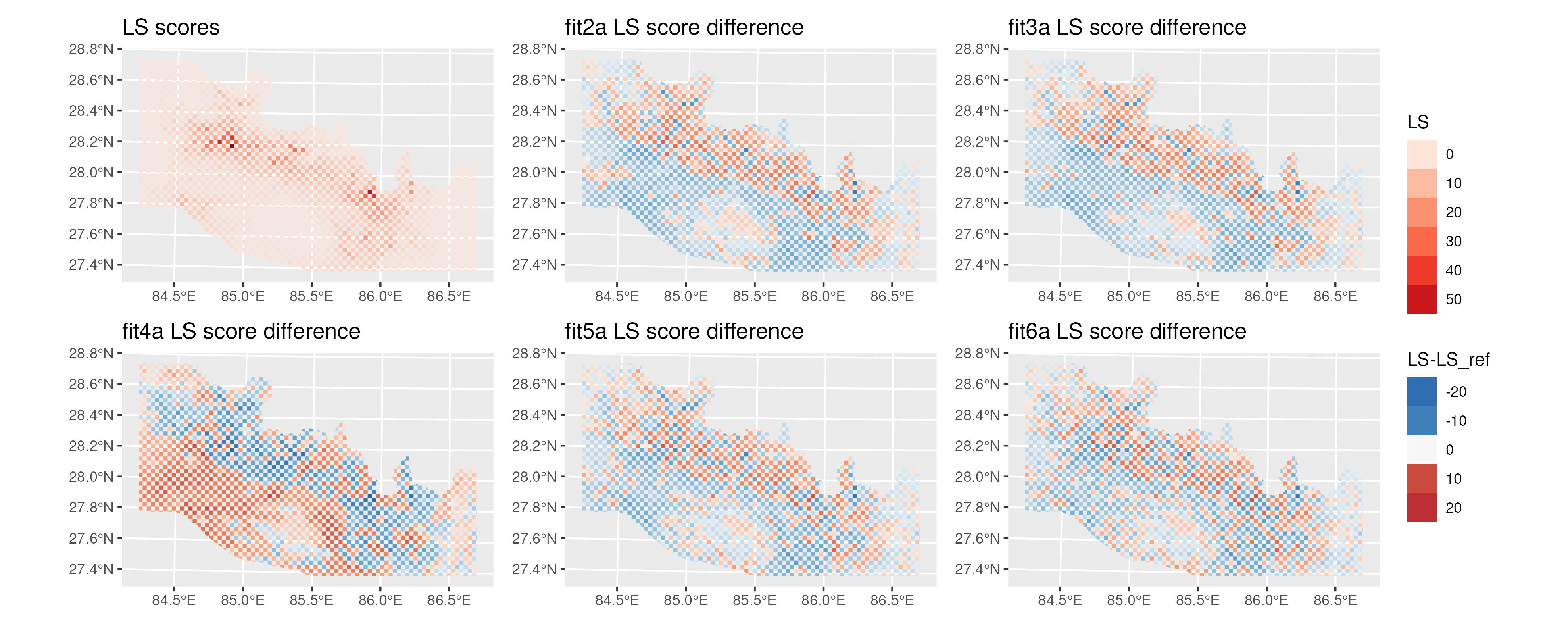}
    \end{minipage}

    \begin{minipage}{0.03\linewidth}
        \centering
        \rotatebox{90}{\parbox{2.5cm}{\centering \small Grid CV (Black)}}
    \end{minipage}%
    \hspace{0.5em}
    \begin{minipage}{0.85\linewidth}
        \includegraphics[width=\linewidth]{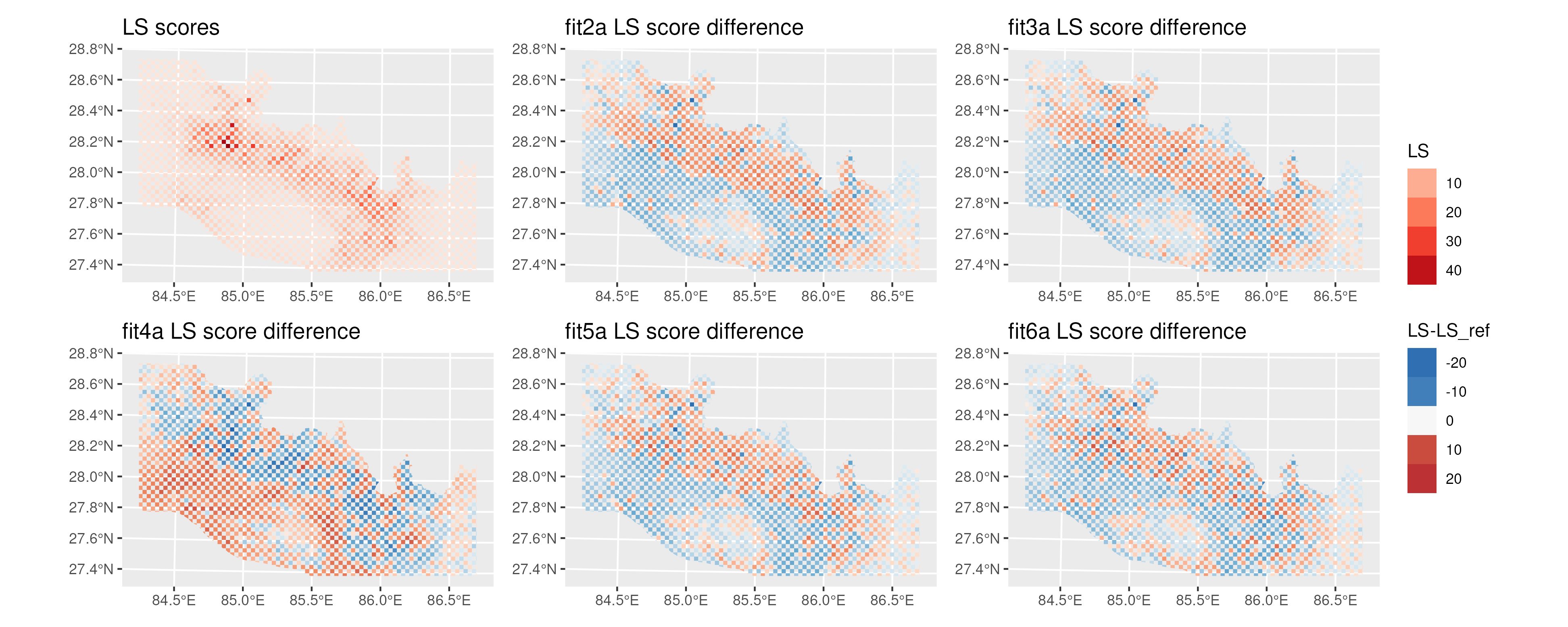}
    \end{minipage}

    \caption{Logarithmic Score (LS) differences for landslide centroid models across cross-validation settings, relative to reference model \texttt{fit1a}. Lower values indicate better predictive performance.}

    \label{fig:LS}
\end{figure}

\begin{figure}[htbp]
    \centering

    \begin{minipage}{0.03\linewidth}
        \centering
        \rotatebox{90}{\parbox{2.5cm}{\centering \small Thinning CV (Set A)}}
    \end{minipage}%
    \hspace{0.5em}
    \begin{minipage}{0.85\linewidth}
        \includegraphics[width=\linewidth]{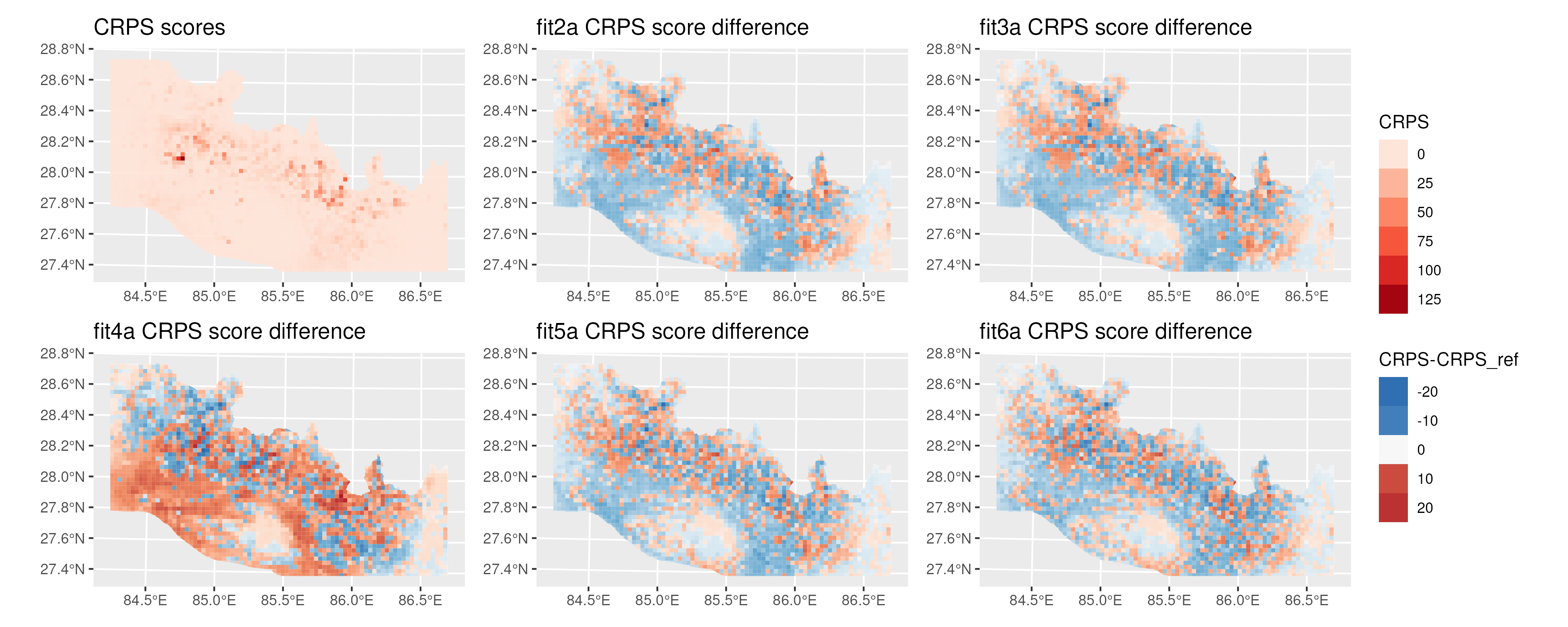}
    \end{minipage}

    \begin{minipage}{0.03\linewidth}
        \centering
        \rotatebox{90}{\parbox{2.5cm}{\centering \small Thinning CV (Set B)}}
    \end{minipage}%
    \hspace{0.5em}
    \begin{minipage}{0.85\linewidth}
        \includegraphics[width=\linewidth]{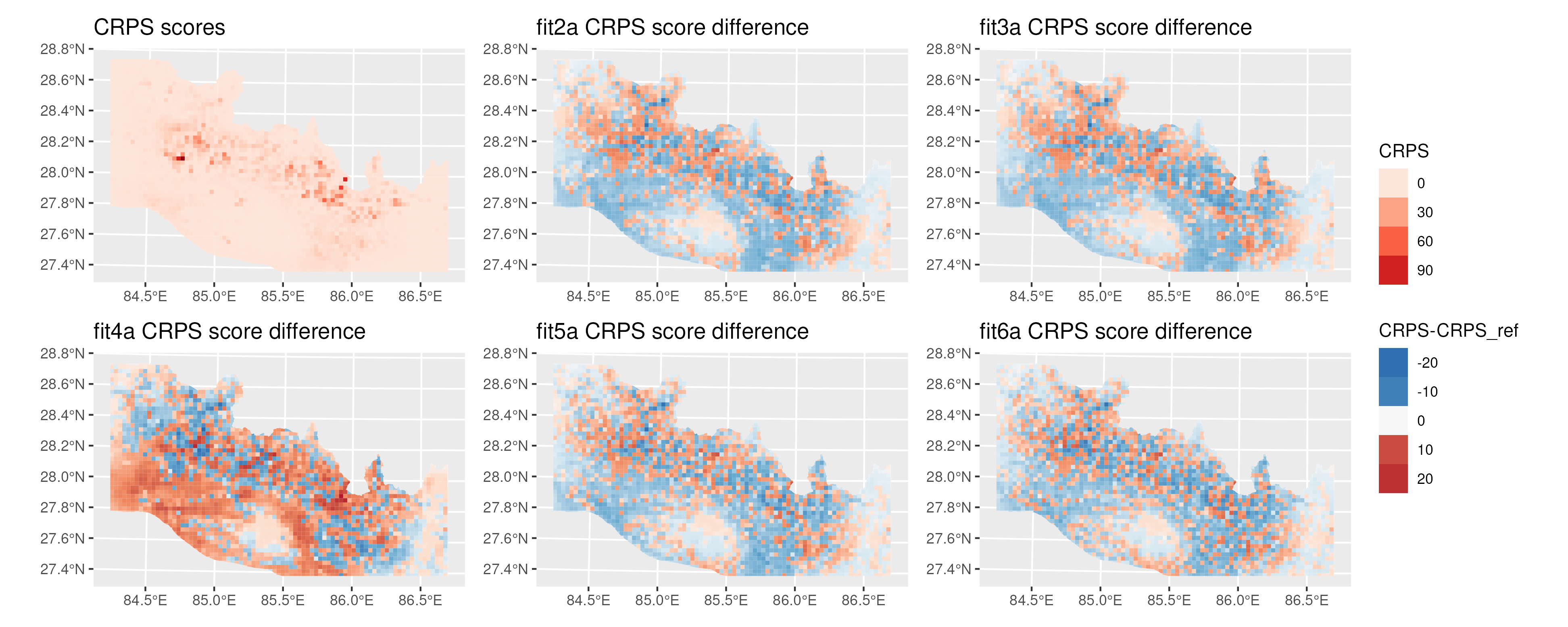}
    \end{minipage}

    \begin{minipage}{0.03\linewidth}
        \centering
        \rotatebox{90}{\parbox{2.5cm}{\centering \small Grid CV (White)}}
    \end{minipage}%
    \hspace{0.5em}
    \begin{minipage}{0.85\linewidth}
        \includegraphics[width=\linewidth]{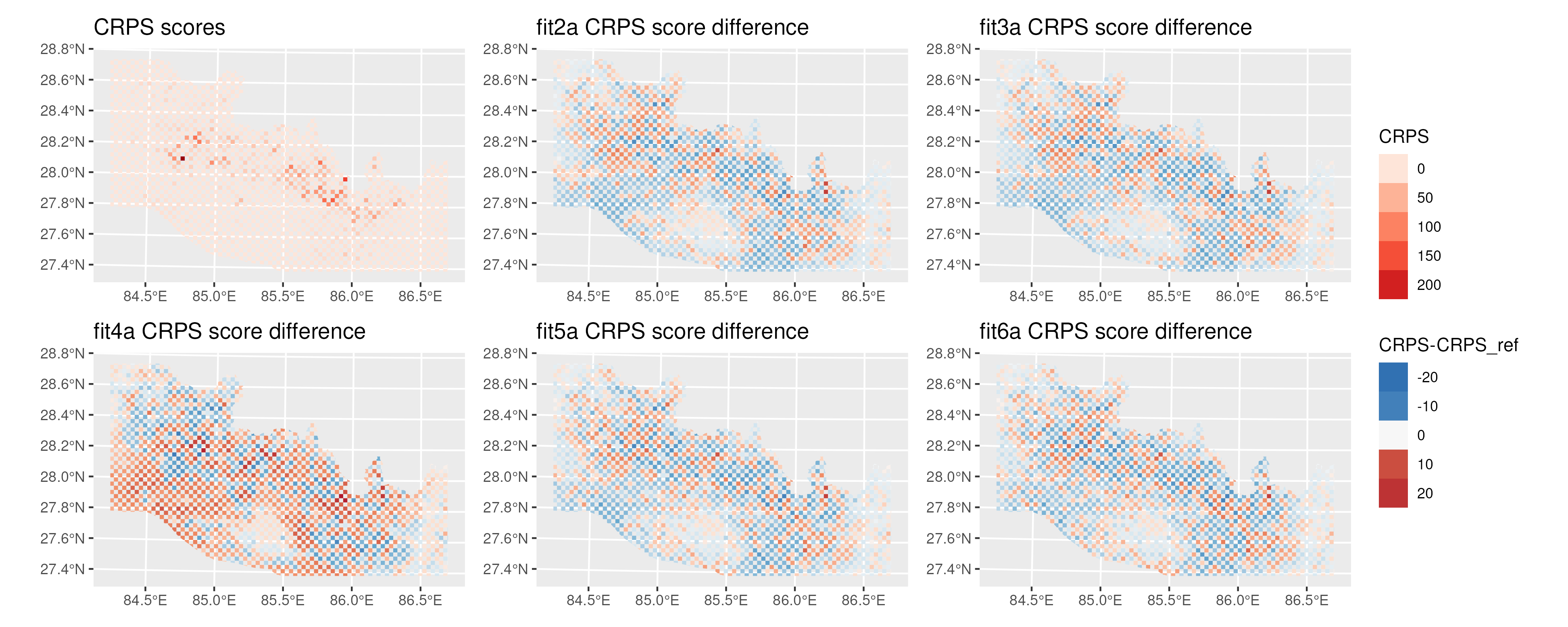}
    \end{minipage}

    \begin{minipage}{0.03\linewidth}
        \centering
        \rotatebox{90}{\parbox{2.5cm}{\centering \small Grid CV (Black)}}
    \end{minipage}%
    \hspace{0.5em}
    \begin{minipage}{0.85\linewidth}
        \includegraphics[width=\linewidth]{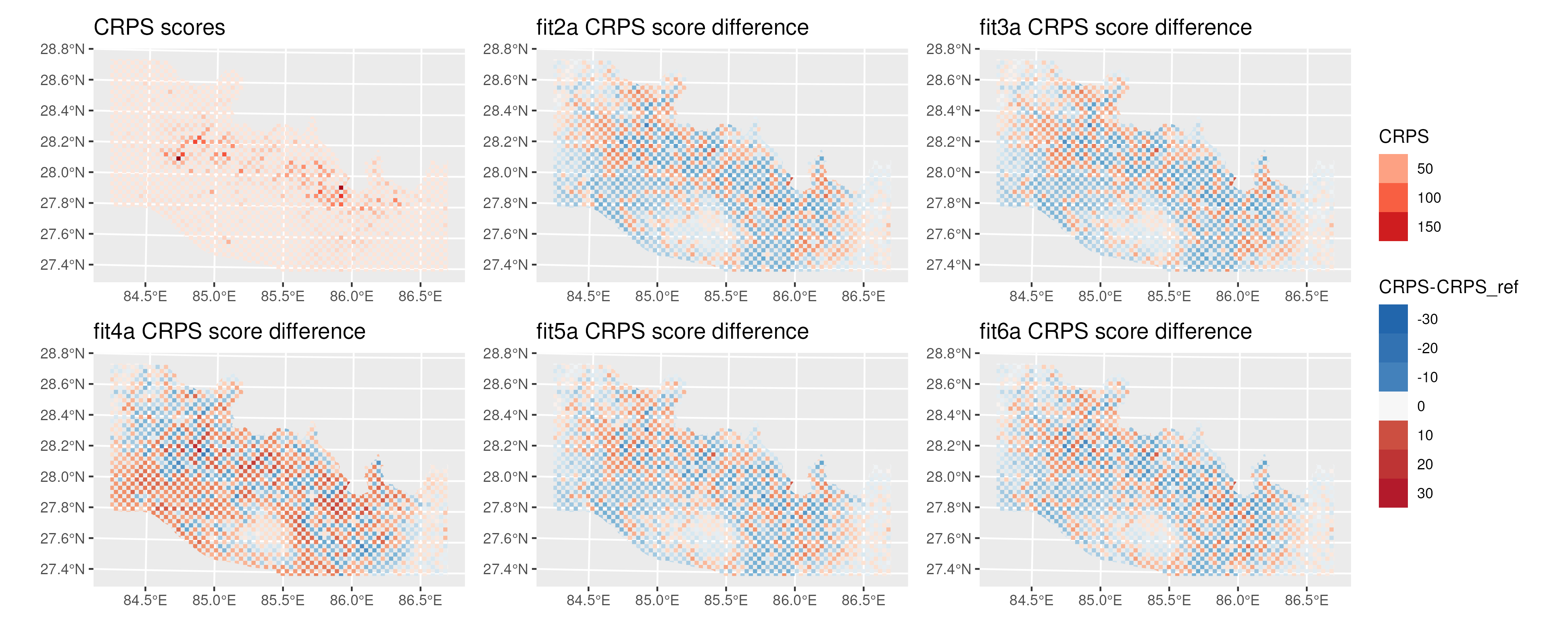}
    \end{minipage}

    \caption{Continuous Ranked Probability Score (CRPS) differences for landslide centroid models across cross-validation settings, relative to reference model \texttt{fit1a}. Lower values indicate better predictive performance.}

    \label{fig:CRPS}
\end{figure}

Since mean scores can be sensitive to outliers or skewed distributions, Figures~\ref{fig:LS} and~\ref{fig:CRPS} provide a more nuanced perspective by presenting spatial heat maps of the strictly proper scoring rules across all cross-validation folds. These visualizations confirm that model performance is generally stable across validation subsets, suggesting that the testing datasets are statistically comparable.

The heat maps show that \texttt{fit1a} generally performs well in distinguishing landslide-prone areas from stable regions but reveals underperformance in specific zones. Notably, clusters of underpredicted susceptibility emerge near Machha Khola (in LS) and Swara/Amale (in CRPS), both known for heightened landslide activity during the June–August monsoon season. These red-highlighted areas suggest that \texttt{fit1a} may fail to capture elevated susceptibility, likely due to missing covariates or unmodelled seasonal effects.

From the same figures, it is apparent that \texttt{fit3a} (which models the logarithm of \ksn{} linearly) and \texttt{fit4a} perform relatively well in regions where landslides occur. However, both tend to overestimate susceptibility in areas where no landslides are observed in the training data, as reflected by higher LS values. This pattern suggests that these models struggle to distinguish between true and false positives, leading to overprediction in stable areas. In contrast, the CRPS, which is more robust to outliers, provides a broader measure of model accuracy. The CRPS heat maps in Figure~\ref{fig:CRPS} show that \texttt{fit3a} and \texttt{fit4a} consistently underperform relative to other models across all cross-validation settings. This underscores the importance of using multiple scoring rules to comprehensively evaluate model performance.

\begin{figure}[htbp]
    \centering

    \begin{minipage}{0.03\linewidth}
        \centering
        \rotatebox{90}{\parbox{2.5cm}{\centering \small Thinning CV (Set A)}}
    \end{minipage}%
    \hspace{0.5em}
    \begin{minipage}{0.85\linewidth}
        \includegraphics[width=\linewidth]{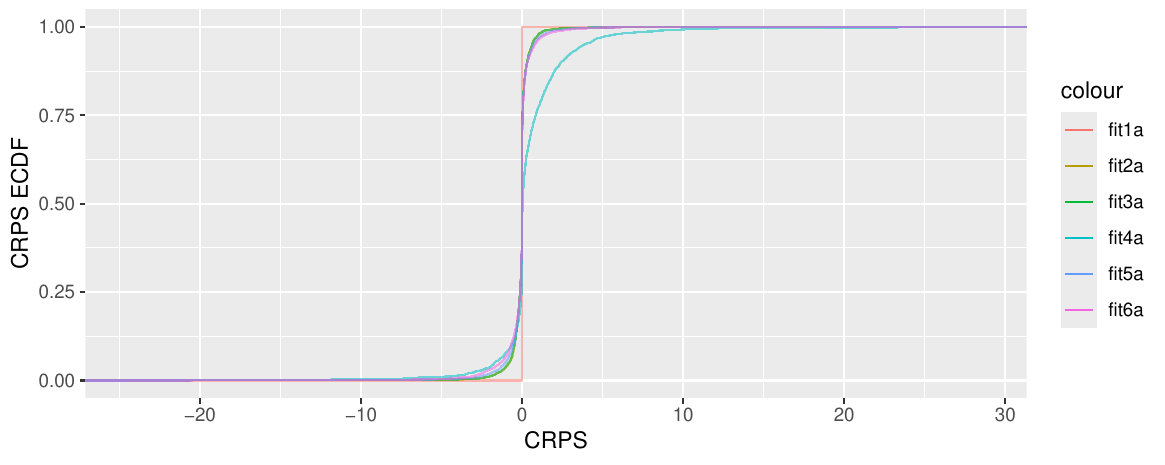}
    \end{minipage}

    \begin{minipage}{0.03\linewidth}
        \centering
        \rotatebox{90}{\parbox{2.5cm}{\centering \small Thinning CV (Set B)}}
    \end{minipage}%
    \hspace{0.5em}
    \begin{minipage}{0.85\linewidth}
        \includegraphics[width=\linewidth]{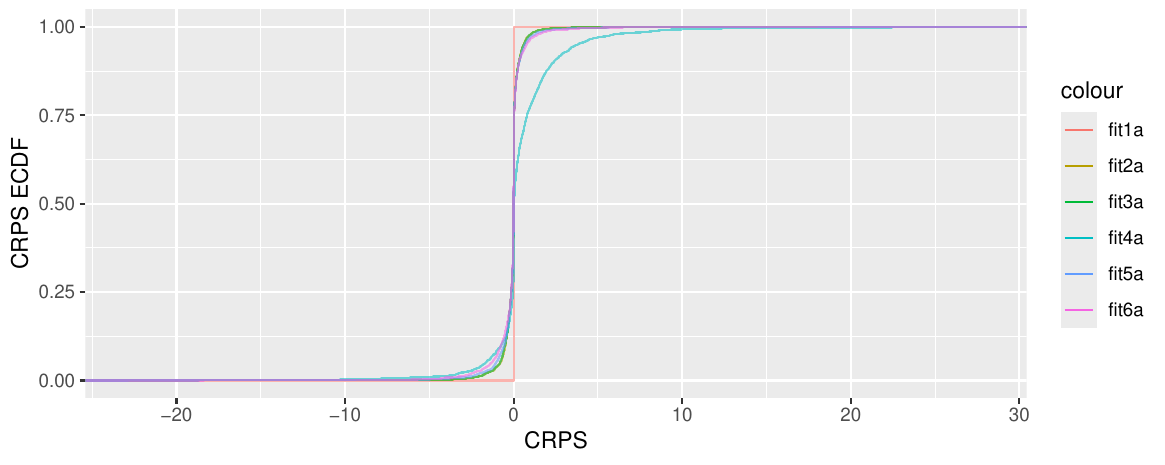}
    \end{minipage}

    \begin{minipage}{0.03\linewidth}
        \centering
        \rotatebox{90}{\parbox{2.5cm}{\centering \small Grid CV (White)}}
    \end{minipage}%
    \hspace{0.5em}
    \begin{minipage}{0.85\linewidth}
        \includegraphics[width=\linewidth]{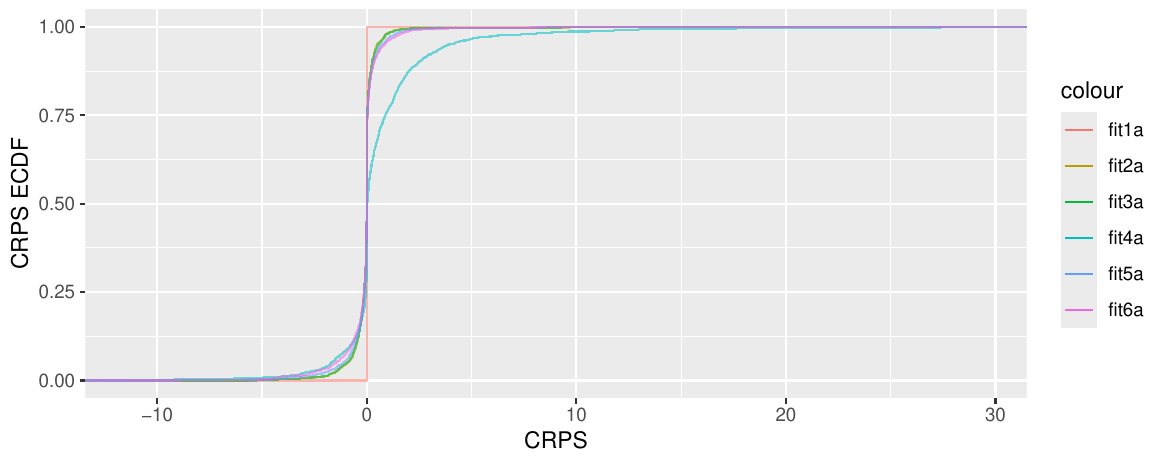}
    \end{minipage}

    \begin{minipage}{0.03\linewidth}
        \centering
        \rotatebox{90}{\parbox{2.5cm}{\centering \small Grid CV (Black)}}
    \end{minipage}%
    \hspace{0.5em}
    \begin{minipage}{0.85\linewidth}
        \includegraphics[width=\linewidth, draft = FALSE]{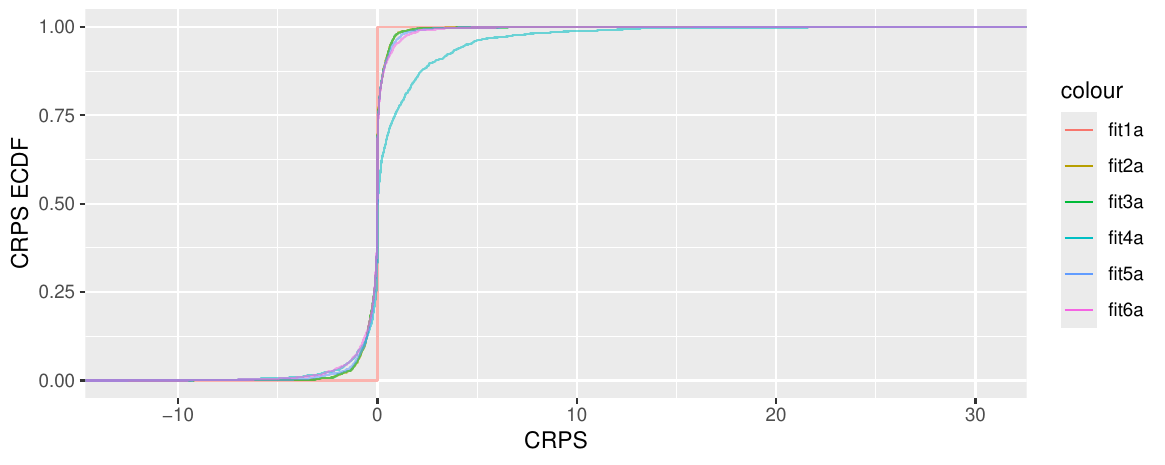}
    \end{minipage}
\caption{Empirical cumulative distribution function (ECDF) of Continuous Ranked Probability Score (CRPS) differences for landslide centroid models relative to \texttt{fit1a}. Negative values indicate better performance.}
    \label{fig:ecdf_crps}
\end{figure}

Turning to model refinements, the use of a second-order random walk (rw2) smoothing prior on $\log$(\ksn{}) and $\sqrt{\text{\ksn}}$ in \texttt{fit1a} and \texttt{fit2a} leads to more balanced and stable performance across both LS and CRPS, making them preferable to \texttt{fit3a} and \texttt{fit4a}. While the spatial heat maps may not fully reflect these improvements, the ECDF plots in Figures~\ref{fig:ecdf_ls} and~\ref{fig:ecdf_crps} clearly show that \texttt{fit5a} and \texttt{fit6a} outperform the other models, with \texttt{fit3a} and \texttt{fit4a} lagging behind. The mean score rankings in Table~\ref{tab:score_metrics_centroids} are consistent with the ECDF-based comparisons.

\subsubsection{Prediction Scores for Landslide Size Models}

\begin{figure}[htbp]
    \centering

    \begin{minipage}{0.03\linewidth}
        \centering
        \rotatebox{90}{\parbox{2.5cm}{\centering \small Thinning CV (Set A)}}
    \end{minipage}%
    \hspace{0.5em}
    \begin{minipage}{0.85\linewidth}
        \includegraphics[width=\linewidth]{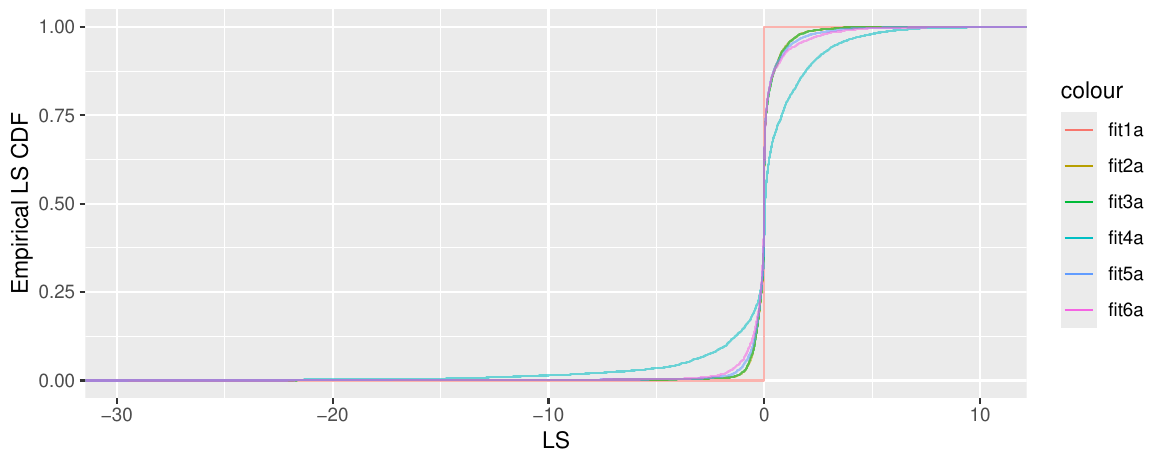}
    \end{minipage}

    \begin{minipage}{0.03\linewidth}
        \centering
        \rotatebox{90}{\parbox{2.5cm}{\centering \small Thinning CV (Set B)}}
    \end{minipage}%
    \hspace{0.5em}
    \begin{minipage}{0.85\linewidth}
        \includegraphics[width=\linewidth]{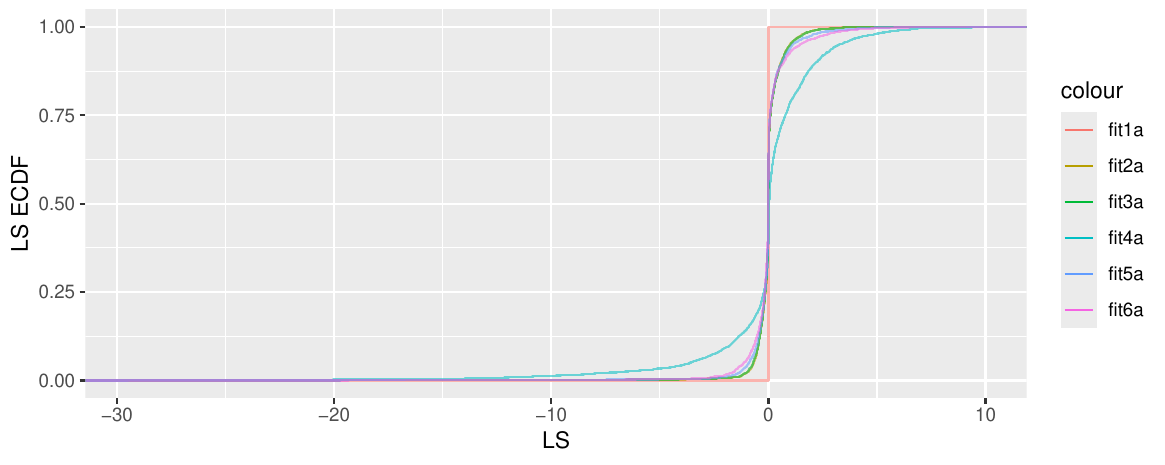}
    \end{minipage}

    \begin{minipage}{0.03\linewidth}
        \centering
        \rotatebox{90}{\parbox{2.5cm}{\centering \small Grid CV (White)}}
    \end{minipage}%
    \hspace{0.5em}
    \begin{minipage}{0.85\linewidth}
        \includegraphics[width=\linewidth]{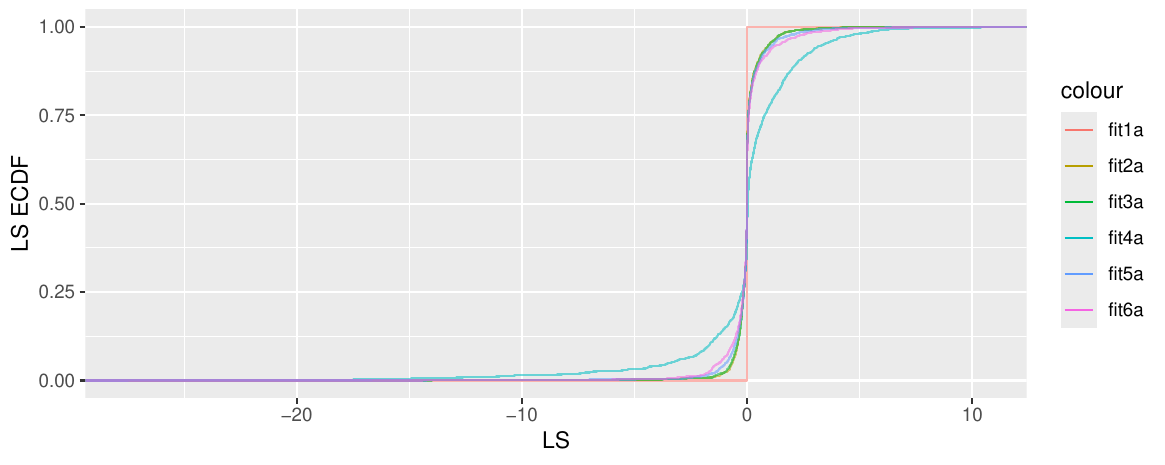}
    \end{minipage}

    \begin{minipage}{0.03\linewidth}
        \centering
        \rotatebox{90}{\parbox{2.5cm}{\centering \small Grid CV (Black)}}
    \end{minipage}%
    \hspace{0.5em}
    \begin{minipage}{0.85\linewidth}
        \includegraphics[width=\linewidth, draft = FALSE]{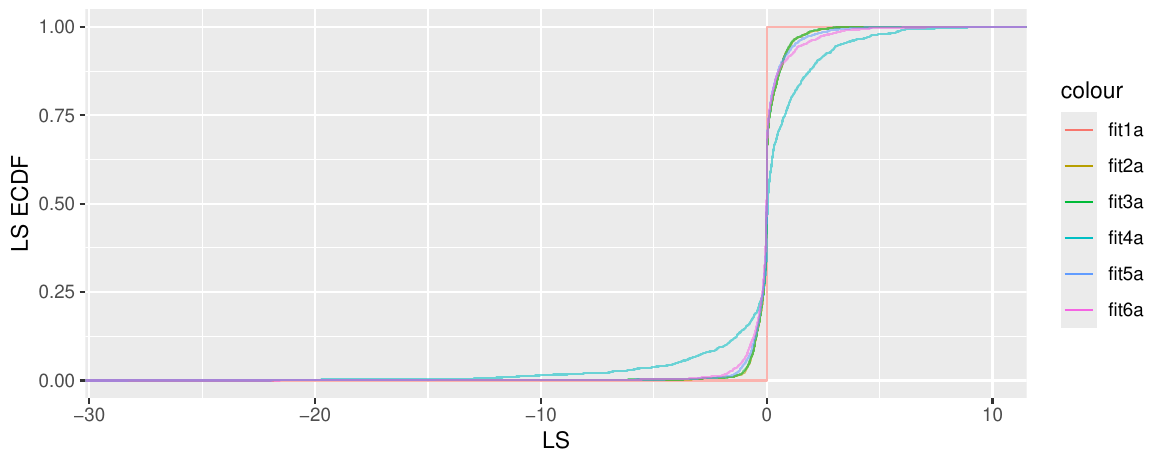}
    \end{minipage}
\caption{Empirical cumulative distribution function (ECDF) of Logarithmic Scores (LS) differences for landslide centroid models relative to \texttt{fit1a}. Negative values indicate better performance.
}
    \label{fig:ecdf_ls}
\end{figure}

For the log-size models, no discernible spatial pattern was observed in the prediction scores; therefore, heatmaps are not presented. Instead, Figures~\ref{fig:ecdf_ls} and Table~\ref{tab:score_metrics_logsize} consistently demonstrate that \texttt{fit6b} outperforms the alternative candidate models in terms of the logarithmic score (LS). This result is theoretically coherent, as the flow distance to the channel (Fd2Ch) is anticipated to exert a dominant influence on landslide size. Landslides generally propagate downslope towards channel networks, which were accurately delineated in this study using the algorithm. Consequently, the resulting landslide area can be considered proportional to the potential landslide volume along the flow path, effectively captured by the local flow distance. Thus, models incorporating Fd2Ch appropriately reflect a key geomorphological control on landslide magnitude that the longer the flow distance, the larger the potential size can travel further down to the channel.  

In this regard, the relief to the channel (Rf2Ch) exhibits a very similar predictive performance for log-size models. While Fd2Ch can correctly identify the primary flow path, relief appears to be less directly correlated with the landslide size compared to Fd2Ch, as relief does not fully account for the flow mass contributing to the landslide because direct vertical displacement towards the channel is not likely but rather along a flow path towards the channel. 

Interestingly, models \texttt{fit1b} to \texttt{fit3b}, which incorporate the steepness index (\ksn) as a covariate, perform less favourably in terms of LS but achieve reasonable results in terms of root mean squared error (RMSE) and Dawid–Sebastiani (DS) scores. This discrepancy may arise from the relatively homogeneous nature of \ksn{} across polygons, which limits its ability to capture fine-scale variations in landslide size. Thus, while \ksn{} effectively characterises broader landscape erosion processes, it appears less suited for modelling the detailed variability required for accurate landslide size prediction. 

\begin{figure}[H]
    \centering

    \begin{minipage}{0.03\linewidth}
        \centering
        \rotatebox{90}{\parbox{2.5cm}{\centering \small Thinning CV (Set A)}}
    \end{minipage}%
    \hspace{0.5em}
    \begin{minipage}{0.85\linewidth}
        \includegraphics[width=\linewidth]{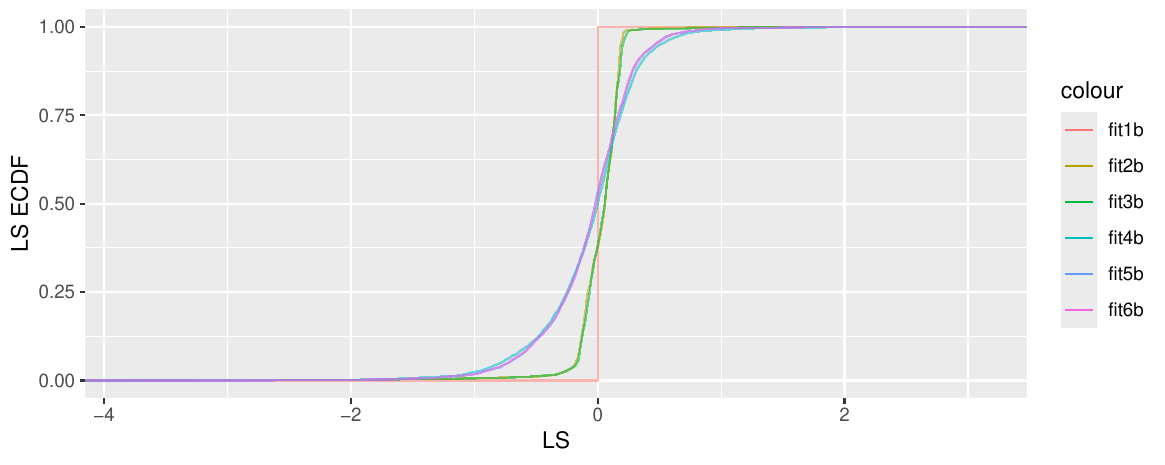}
    \end{minipage}

    \begin{minipage}{0.03\linewidth}
        \centering
        \rotatebox{90}{\parbox{2.5cm}{\centering \small Thinning CV (Set B)}}
    \end{minipage}%
    \hspace{0.5em}
    \begin{minipage}{0.85\linewidth}
        \includegraphics[width=\linewidth]{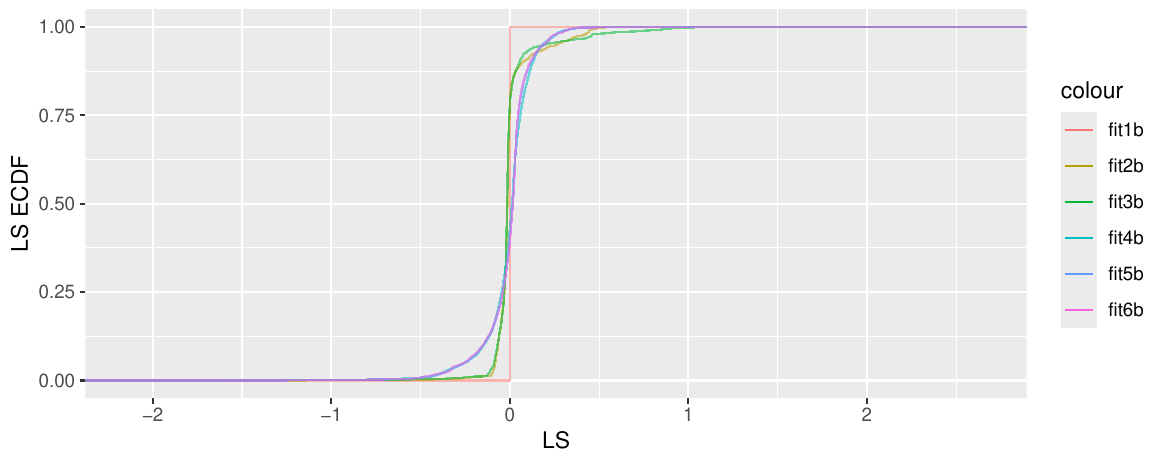}
    \end{minipage}

    \begin{minipage}{0.03\linewidth}
        \centering
        \rotatebox{90}{\parbox{2.5cm}{\centering \small Grid CV (White)}}
    \end{minipage}%
    \hspace{0.5em}
    \begin{minipage}{0.85\linewidth}
        \includegraphics[width=\linewidth]{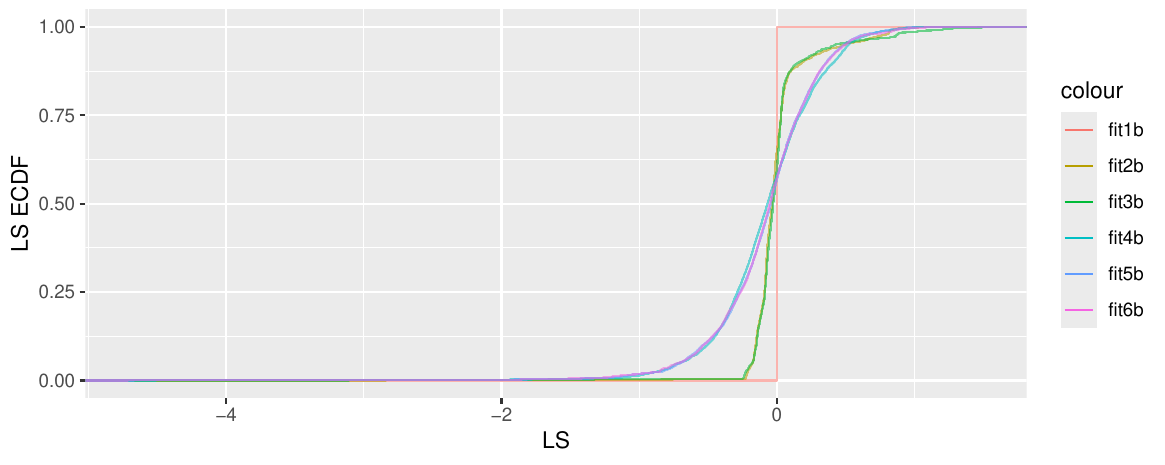}
    \end{minipage}

    \begin{minipage}{0.03\linewidth}
        \centering
        \rotatebox{90}{\parbox{2.5cm}{\centering \small Grid CV (Black)}}
    \end{minipage}%
    \hspace{0.5em}
    \begin{minipage}{0.85\linewidth}
        \includegraphics[width=\linewidth, draft = FALSE]{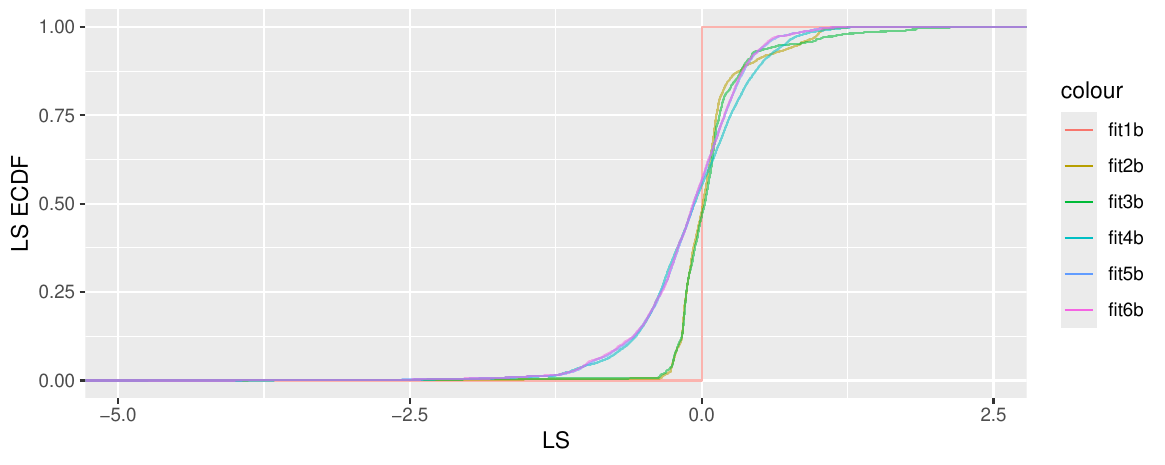}
    \end{minipage}
\caption{Empirical cumulative distribution function (ECDF) of Logarithmic Scores (LS) differences for landslide size models relative to \texttt{fit1a}. Negative values indicate better performance.}
    \label{fig:ecdf_ls_b}
\end{figure}





\begin{table}[htbp]
\centering
\caption{Model performance metrics (Centroids on $3 \text{km} \times 3 \text{km}$ grid).}
\begin{tabular}{ll|rrrr}
\toprule
\textbf{CV Type (Training Set)} & \textbf{Model} & \textbf{RMSE} & \textbf{DS} & \textbf{LS} & \textbf{CRPS} \\
\midrule
\multirow{6}{*}{Thinning (Set A)} 
& \texttt{fit1a} & 7.7924 & 9.9864 & 3.5801 & 2.8107 \\
& \texttt{fit2a} & 7.6679 & 10.2243 & 3.5866 & 2.7407 \\
& \texttt{fit3a} & 7.6638 & 10.1874 & 3.5899 & 2.7386 \\
& \texttt{fit4a} & 8.5267 & 12.3008 & \textbf{3.5799} & 3.3187 \\
& \texttt{fit5a} & 7.6290 & 10.0240 & 3.5905 & 2.7257 \\
& \texttt{fit6a} & \textbf{7.5856} & \textbf{9.9489} & 3.5892 & \textbf{2.6987} \\
\midrule
\multirow{6}{*}{Thinning (Set B)} 
& \texttt{fit1a} & 7.6555 & 9.2426 & 3.5735 & 2.8175 \\
& \texttt{fit2a} & 7.5419 & 9.4265 & 3.5783 & 2.7456 \\
& \texttt{fit3a} & 7.5336 & 9.3854 & 3.5791 & 2.7431 \\
& \texttt{fit4a} & 8.4260 & 11.7032 & \textbf{3.5710} & 3.3160 \\
& \texttt{fit5a} & 7.5103 & 9.2848 & 3.5788 & 2.7293 \\
& \texttt{fit6a} & \textbf{7.4566} & \textbf{9.1117} & 3.5807 & \textbf{2.7000} \\
\midrule
\multirow{6}{*}{Grid (White)} 
&\texttt{fit1a} & 17.7807 & 44.8746 & 3.5440 & 5.8747 \\
&\texttt{fit2a} & 17.6605 & 45.8413 & 3.5390 & 5.8099 \\
&\texttt{fit3a} & 17.6571 & 45.6890 & 3.5398 & 5.8071 \\
&\texttt{fit4a} & 18.7651 & 55.6164 & 3.5619 & 6.5040 \\
&\texttt{fit5a} & 17.6153 & 44.6942 & \textbf{3.5389} & 5.7994 \\
&\texttt{fit6a} & \textbf{17.5696} & \textbf{44.4830} & 3.5412 & \textbf{5.7709} \\
\midrule
\multirow{6}{*}{Grid (Black)} 
& \texttt{fit1a} & 16.5195 & 39.6084 & 3.6291 & 5.7203 \\
& \texttt{fit2a} & 16.3540 & 40.5994 & 3.6414 & 5.6400 \\
& \texttt{fit3a} & 16.3505 & 40.2834 & 3.6362 & 5.6381 \\
& \texttt{fit4a} & 17.5496 & 47.1120 & \textbf{3.6138} & 6.3775 \\
& \texttt{fit5a} & 16.3393 & 40.4048 & 3.6237 & 5.6390 \\
& \texttt{fit6a} & \textbf{16.2483} & \textbf{39.2304} & 3.6358 & \textbf{5.6063} \\
\bottomrule
\end{tabular}
\footnotesize

\vspace{0.5em}
\raggedright
\textbf{Note:} RMSE = Root Mean Squared Error, DS = Dawid–Sebastiani Score, LS = Logarithmic Score, CRPS = Continuous Ranked Probability Score.
\label{tab:score_metrics_centroids}
\end{table}

\begin{table}[htbp]
\centering
\caption{Model performance metrics (Log Size).}
\begin{tabular}{ll|rrrr}
\toprule
\textbf{CV Type(Training Set)} & \textbf{Model} & \textbf{RMSE} & \textbf{DS} & \textbf{LS} & \textbf{CRPS} \\
\midrule
\multirow{6}{*}{Thinning (Set A)} 
& \texttt{fit1b} & 1.0615 & 1.3371 & 2.5226 & -- \\
& \texttt{fit2b} & 1.0551 & 1.3260 & 2.5430 & -- \\
& \texttt{fit3b} & \textbf{1.0521} & \textbf{1.3229} & 2.5493 & -- \\
& \texttt{fit4b} & 1.0827 & 1.3640 & 2.4696 & -- \\
& \texttt{fit5b} & 1.0864 & 1.3723 & 2.4620 & -- \\
& \texttt{fit6b} & 1.0874 & 1.3740 & \textbf{2.4616} & -- \\
\midrule
\multirow{6}{*}{Thinning (Set B)} 
& \texttt{fit1b} & 1.3273 & 2.2579 & 1.3917 & -- \\
& \texttt{fit2b} & 1.3163 & 2.2548 & 1.3955 & -- \\
& \texttt{fit3b} & \textbf{1.3138} &\textbf{2.2536} & 1.3968 & -- \\
& \texttt{fit4b} & 1.3588 & 2.2690 & 1.3885 & -- \\
& \texttt{fit5b} & 1.3613 & 2.2703 & 1.3874 & -- \\
& \texttt{fit6b} & 1.3623 & 2.2705 & \textbf{1.3869} & -- \\
\midrule
\multirow{6}{*}{Grid (White)} 
& \texttt{fit1b} & 1.3445 & 2.2638 & 2.7517 & -- \\
& \texttt{fit2b} & 1.3344 & 2.2615 & 2.7584 & -- \\
& \texttt{fit3b} & \textbf{1.3303} & \textbf{2.2602} & 2.7620 & -- \\
& \texttt{fit4b} & 1.3696 & 2.2734 & 2.6828 & -- \\
& \texttt{fit5b} & 1.3784 & 2.2758 & 2.6815 & -- \\
& \texttt{fit6b} & 1.3802 & 2.2768 & \textbf{2.6798} & -- \\
\midrule
\multirow{6}{*}{Grid (Black)} 
& \texttt{fit1b} & 1.3351 & 2.2640 & 3.1305 & -- \\
& \texttt{fit2b} & 1.3315 & 2.2613 & 3.1897 & -- \\
& \texttt{fit3b} & \textbf{1.3257} & \textbf{2.2591} & 3.2012 & -- \\
& \texttt{fit4b} & 1.3683 & 2.2756 & 3.0570 & -- \\
& \texttt{fit5b} & 1.3714 & 2.2775 & 3.0226 & -- \\
& \texttt{fit6b} & 1.3727 & 2.2780 & \textbf{3.0217} & -- \\
\bottomrule
\end{tabular}
\footnotesize

\vspace{0.5em}
\raggedright
\textbf{Note:} RMSE = Root Mean Squared Error, DS = Dawid–Sebastiani Score, LS = Logarithmic score. CRPS not computed for log sizes.
\label{tab:score_metrics_logsize}
\end{table}

\subsection{Model Case Study}\label{sec:case_study}

In this case study, we focus on the best-performing models, \texttt{fit6a} and \texttt{fit6b}, and rerun them on the entire landslide inventory to explore their posterior distributions and Coefficient of Variation (\cv) in detail.

\subsubsection{Posterior Distribution for Centroids}

As illustrated in Figure~\ref{fig:pred}, the predicted susceptibility map, represented by the intensity function $\lambda$, effectively captures the spatial distribution of landslide centroids. The model successfully highlights high-susceptibility regions, aligning well with the observed landslide occurrences.

The spatial posterior mean in log-scale reveals that Peak Ground Acceleration (PGA) and the channel steepness index (\ksn{}) are the dominant contributors to landslide susceptibility. PGA reflects the seismic shaking intensity, showing relatively lower values around the Kathmandu Valley and attenuation towards both ends of the region. The \ksn{} covariate is particularly effective in identifying landslide-prone areas, capturing both the geomorphological structure of the Himalayas and the alignment of major drainage channels. In particular, \ksn{} helps delineate high-susceptibility zones along these channels, offering interpretable insights into terrain-driven risk.

Land cover also contributes meaningfully that areas classified as glacial landscapes (depicted in deep blue) are associated with a negative effect on landslide susceptibility, suggesting lower risk in regions dominated by ice and snow. Similarly, geological features such as recent alluvial deposits in Kathmandu, the Tibetan-Tethys Himalayan Zone, and Paleozoic granite formations are associated with reduced landslide likelihood.

Furthermore, the proximity to river channels, captured by the flow distance covariate (Fd2Ch), shows a strong inverse relationship with landslide risk: the closer to a channel, the higher the susceptibility.

Taken together, these covariate effects consistently indicate lower landslide susceptibility within the Kathmandu Valley. This result is in agreement with prior independent studies, reinforcing the reliability of our model's inferences.

Figure~\ref{fig:fp_a_cv} presents the \cv{} to illustrate how uncertainty is distributed across the study area. As expected, regions with few or no landslides tend to exhibit higher susceptibility uncertainty, reflecting limited data support. The $c_v$ for PGA is noticeably higher at lower values, likely due to a scarcity of landslides observed in those conditions. In terms of land cover, uncertainty is particularly elevated within the Kathmandu Valley and in glacial landscapes, where the terrain and surface characteristics are more heterogeneous.

This information is especially valuable when cross-referenced with the score plots in Figures~\ref{fig:LS} and~\ref{fig:CRPS}, which allow us to assess how the \cv{} relates to model performance. Notably, even though the \ksn{} metric was not designed for glacial environments, the model still performs reasonably well in these regions, which is an encouraging result that underscores the potential for broader applicability.

\begin{figure}[!b]
    \begin{subfigure}[t]{\linewidth}
    \centering
    \includegraphics[width=\linewidth, draft=FALSE, trim = 0 2cm 0 2.2cm, clip]{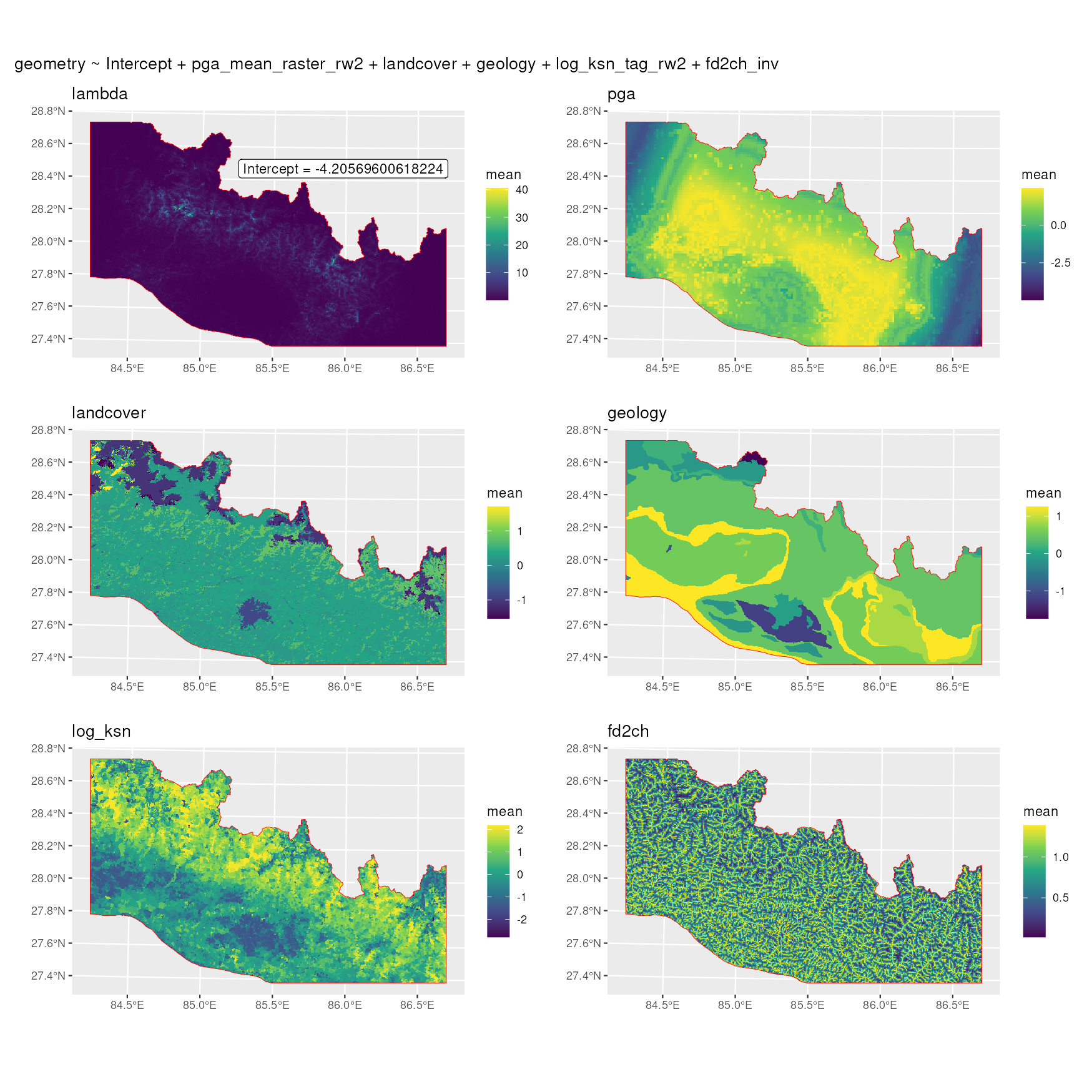}
    \caption{Posterior mean of modelled intensity (top left) and covariate effects.}
    \label{fig:pred}
    \end{subfigure}
\end{figure}

\begin{figure}[H]\ContinuedFloat
    \begin{subfigure}[t]{\linewidth}
    \centering
    \includegraphics[width=\linewidth, draft=FALSE, trim = 0 2cm 0 2.2cm, clip]{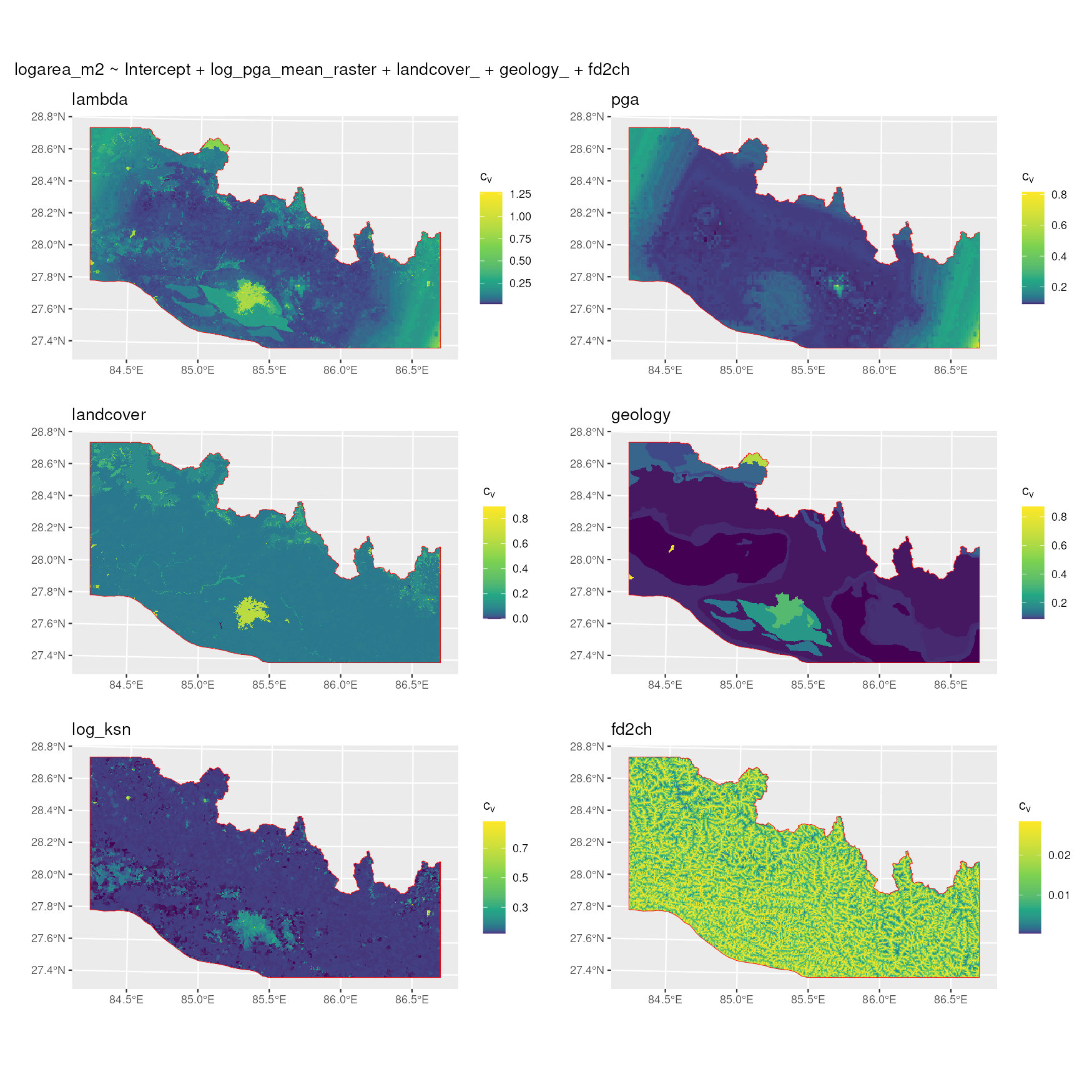}
    \caption{Coefficient of variation ($c_v$) of intensity and covariate effects on the exponential scale.}
    \label{fig:fp_a_cv}
    \end{subfigure}
    \caption{Model \texttt{fit6a} for the full set of landslide centroid observations, based on $100$ posterior samples. See Table \ref{tab:model_formulas_cov} for the description of covariate effects.}
\end{figure}

\subsubsection{Posterior Distribution for log Sizes}
Figures~\ref{fig:fp6b} and~\ref{fig:fp_b_cv} present the predicted landslide size, conditional on landslide occurrence, along with associated uncertainty measured by \cv. The posterior mean field exhibits clear spatial structure, with relatively low uncertainty in regions where landslides are frequently observed. These spatial patterns are largely influenced by covariates such as land cover and geology, which contribute meaningfully to the posterior estimates.

The model yields high confidence in regions with abundant landslide data, while uncertainty associated with PGA is elevated at the left and right ends of the study area due to sparse observations in those zones. Additionally, flow distance to channel appears positively associated with landslide size, suggesting that landslides further from channels tend to be larger. However, this relationship is characterised by higher uncertainty, as reflected by the \cv{} values.
 


\begin{figure}[!b]
    \begin{subfigure}[t]{\linewidth}
    \centering
    \includegraphics[width=\linewidth, draft=FALSE, trim = 0 2cm 0 2.25cm, clip]{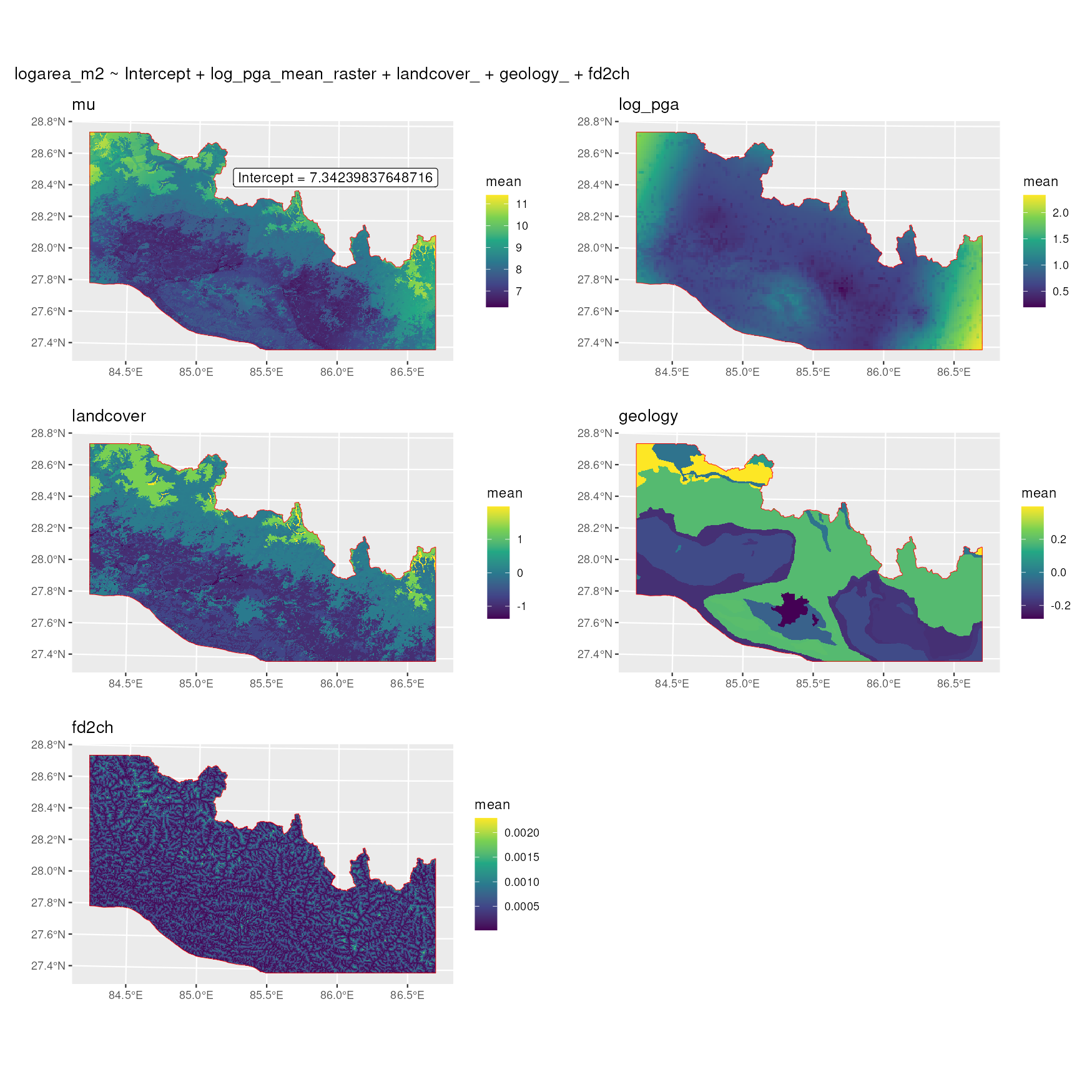}
    \caption{Posterior mean (top left) and covariate effects.}
    \label{fig:fp6b}
    \end{subfigure}
\end{figure}

\begin{figure}[ht]\ContinuedFloat
    \begin{subfigure}[t]{\linewidth}
    \centering
    \includegraphics[width=\linewidth, draft=FALSE, trim = 0 2cm 0 2.2cm, clip]{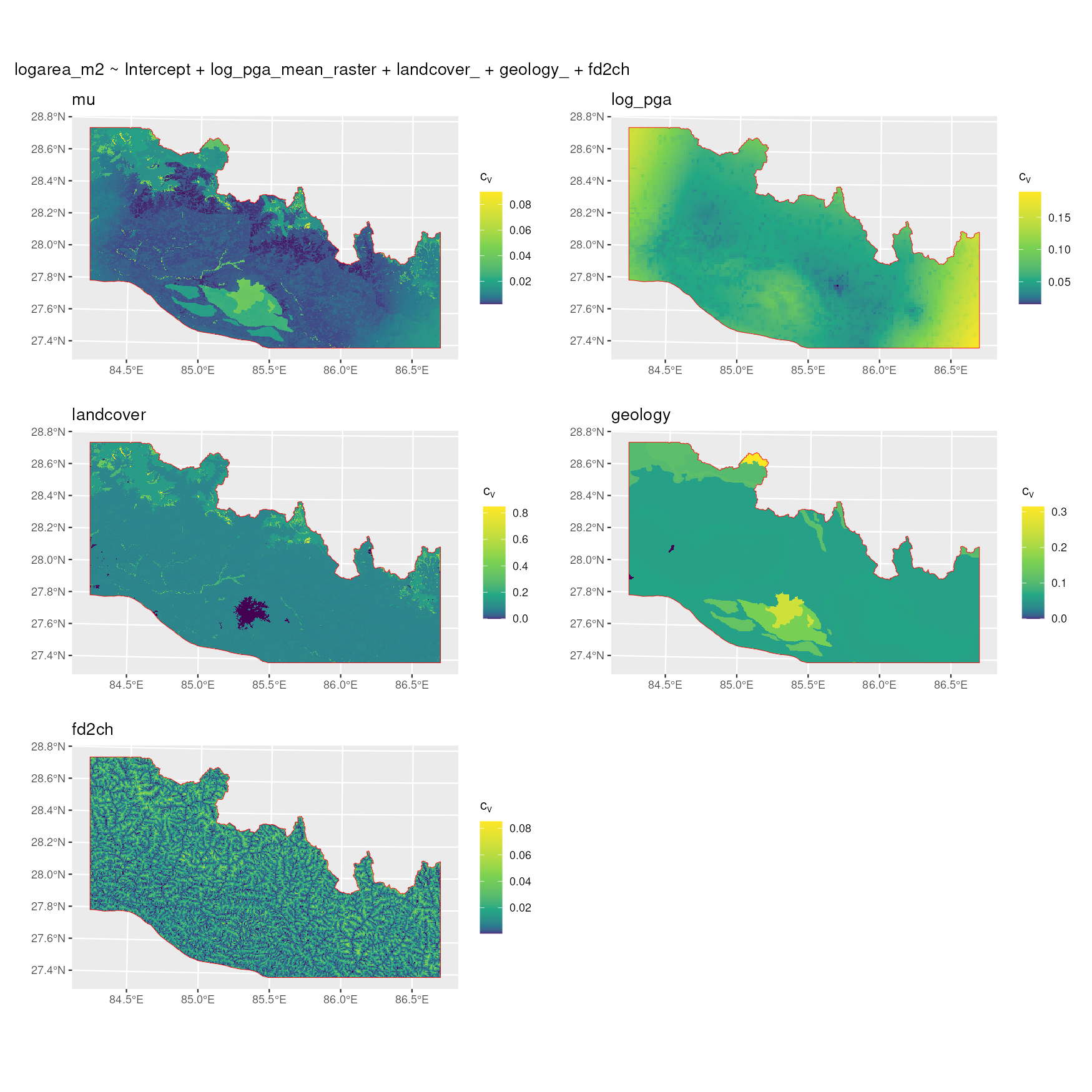}
    \caption{Coefficient of variation ($c_v$) of intensity and covariate effects on the exponential scale.}
    \label{fig:fp_b_cv}
    \end{subfigure}
    \caption{Model \texttt{fit6b} for the full set of landslide size observations, based on $100$ posterior samples. See Table \ref{tab:model_formulas_cov} for the description of covariate effects.}
\end{figure}

\subsubsection{Land Cover}

Tables \ref{tab:centroid_coefficients} and \ref{tab:logsize_coefficients} provide the statistical summary of ice and glacial related land cover coefficients derived from model \texttt{fit6a} for landslide presence and model \texttt{fit6b} for the log transformed landslide size.

Starting with landslide presence from \texttt{fit6a}, several land cover categories show clear and interpretable effects. Areas classified as 2SCO (Closed to Open Medium to High Shrubland, Thicket), 2HCO (Herbaceous Closed to Open Medium Tall Vegetation), and 2SSd (Sparse Dwarf Shrubs and Sparse Herbaceous) are significantly more prone to landslides. These environments often represent sparsely vegetated or transitional terrain on steeper slopes, making them more susceptible to surface failures. In contrast, categories such as 8SN (Perennial Snow) and 8SNs (Seasonal Snow) are associated with a much lower likelihood of landslides. This may reflect both stabilising conditions provided by snow cover and the fact that snow can obscure the observation of smaller failures.

Regarding landslide size based on \texttt{fit6b}, conditional on a landslide occurring, some glacial and ice related areas tend to be associated with larger landslides. Specifically, 8ICEr (Perennial Ice, Moving, Revised) and 8SN (Perennial Snow) show positive effects on landslide size, suggesting that failures in these environments are likely to be deeper or more extensive. These may reflect the influence of ice melt, long term slope weakening, or underlying structural instabilities. However, the uncertainty in these coefficients is relatively large. On the other hand, categories such as 1HSs (Small Sized Fields of Herbaceous Crops on Sloping Land) and 2SCO tend to produce smaller landslides, even though they are more likely to experience landslide events.

Together, these results provide insight into not only where landslides are likely to occur but also their potential size. This highlights the importance of considering both land cover type and glacial influence when assessing landslide hazard.

\subsubsection{Second-order Random Walk (RW2) Smoothing Covariate Effect}

\begin{figure}[htbp]
    \centering
    \begin{subfigure}[b]{0.49\linewidth}
        \centering
        \includegraphics[width=\linewidth, trim = 0 0 0 .7cm, clip]{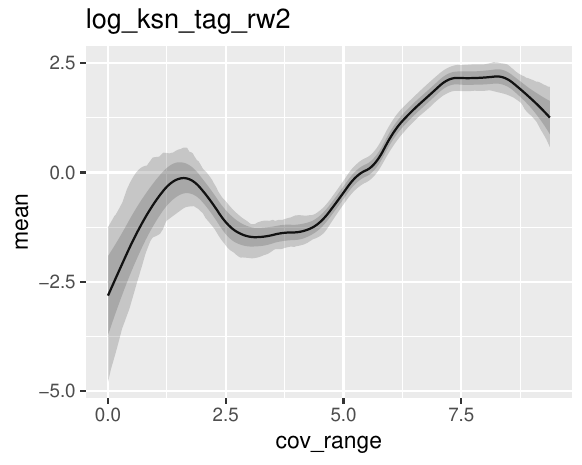}
        \caption{Smoothed $\log$ \ksn{} effect}
    \end{subfigure}
    \hfill
    \begin{subfigure}[b]{0.49\linewidth}
        \centering
        \includegraphics[width=\linewidth, trim = 0 0 0 .7cm, clip]{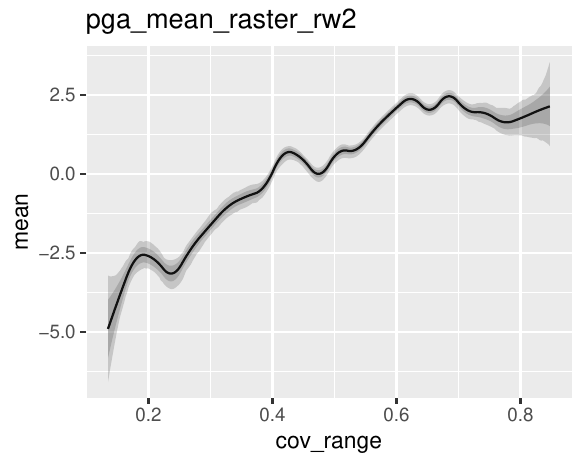}
        \caption{Smoothed PGA effect}
    \end{subfigure}
    \caption{Posterior estimates of the smoothed $\log$ \ksn{} (left) and PGA (right) covariate effects using an RW2 prior, illustrating the non-linear relationships in \texttt{fit6a} using the full dataset.}
    \label{fig:rw2}
\end{figure}

Figure~\ref{fig:rw2} shows that, although the relationship between landslide occurrence and \ksn{} is not strictly linear, there is a clear trend indicating increased susceptibility with higher $\log$ \ksn{} values above 2.5. Notably, only $0.09\%$ of the landslides fall into where $\log$ \ksn{} $< 2.5$. At the upper end of the scale, the curve begins to flatten, potentially indicating a geomorphic equilibrium. Since \ksn{} quantifies relative channel steepness normalized by drainage area, extremely steep channels may reach a threshold where additional steepening does not proportionally increase landslide risk. This plateau likely reflects stabilizing geomorphic processes, such as bedrock incision or sediment transport adjustments, that limit further morphological change.

In contrast, the second-order random walk (RW2) smoothing curve for PGA reveals a more complex pattern, with several local fluctuations and a pronounced bump around a PGA value of approximately 
0.42. This feature may indicate that the model is implicitly capturing residual spatial structure through the covariate. In the absence of an explicit spatial random field, spatially structured variability may be absorbed by the covariates themselves, potentially leading to spurious or overfitted patterns. To investigate this, we attempted to spatially trace the source of the bump by mapping areas with PGA values near 
0.42 and comparing them to the distribution of observed landslide centroids. However, the diffuse spatial spread and density of the covariate made it difficult to attribute this feature to specific regions or events. This highlights a broader modelling challenge that interpreting smoothed covariate effects becomes especially complex when spatial random effect is not explicitly accounted for. Additionally, the posterior credible interval for PGA broadens substantially at lower values (below 
0.2), reflecting the lack of landslide observations in those areas.

\subsubsection{Selected Basins Case Study}
It is important to note that the susceptibility map represents a potential intensity distribution of landslide occurrences under a hypothetical recurrence of the earthquake. In other words, it reflects the model’s probabilistic expectation of where landslides are likely to occur, given the current covariate configuration.

To explore this in greater detail, we examine a set of representative basins discussed earlier, focusing on the landslide susceptibility patterns generated by model \texttt{fit6a}. Although this model estimates the spatial intensity of landslide centroids based solely on covariate information, the resulting maps in Figure~\ref{fig:fit6a_thinA_zoom} clearly reflect underlying drainage structures and show strong agreement with the observed distribution of landslide centroids.

Figure~\ref{fig:fit6a_zm_2228} corresponds to one of the larger basins with a substantial number of landslide events and is the same basin where we manually filled gaps in the channel steepness index (\ksn; see Figure~\ref{fig:glacier} in Appendix \ref{sec:ksn_fix}). Figures~\ref{fig:fit6a_zm_15329} and \ref{fig:fit6a_zm_14982} show smaller basins: basin 15329 lies farther from glacial influence, while basin 14982 includes areas with glacial features where some landslide activity is less well explained by \ksn{} alone. Nonetheless, the model effectively captures the clustering of landslides across all basins, including those in glaciated terrain, where flow distance to channel appears to partially account for the observed distribution (see Figure~\ref{fig:fit6a_zm_14982}).

While a few landslide centroids fall outside high-susceptibility regions, the majority are well captured by the covariate-only model. This supports the model’s capacity to identify key spatial drivers of landslide occurrence. At the same time, the presence of unexplained centroids highlights the likely influence of unmeasured factors or spatial dependence, suggesting future opportunities for model refinement and extension.

\begin{figure}[!b]
    \begin{subfigure}[t]{0.48\linewidth}
        \centering
        \includegraphics[width=\linewidth, trim = 0.21cm 2cm 0.25cm 2.5cm, clip]{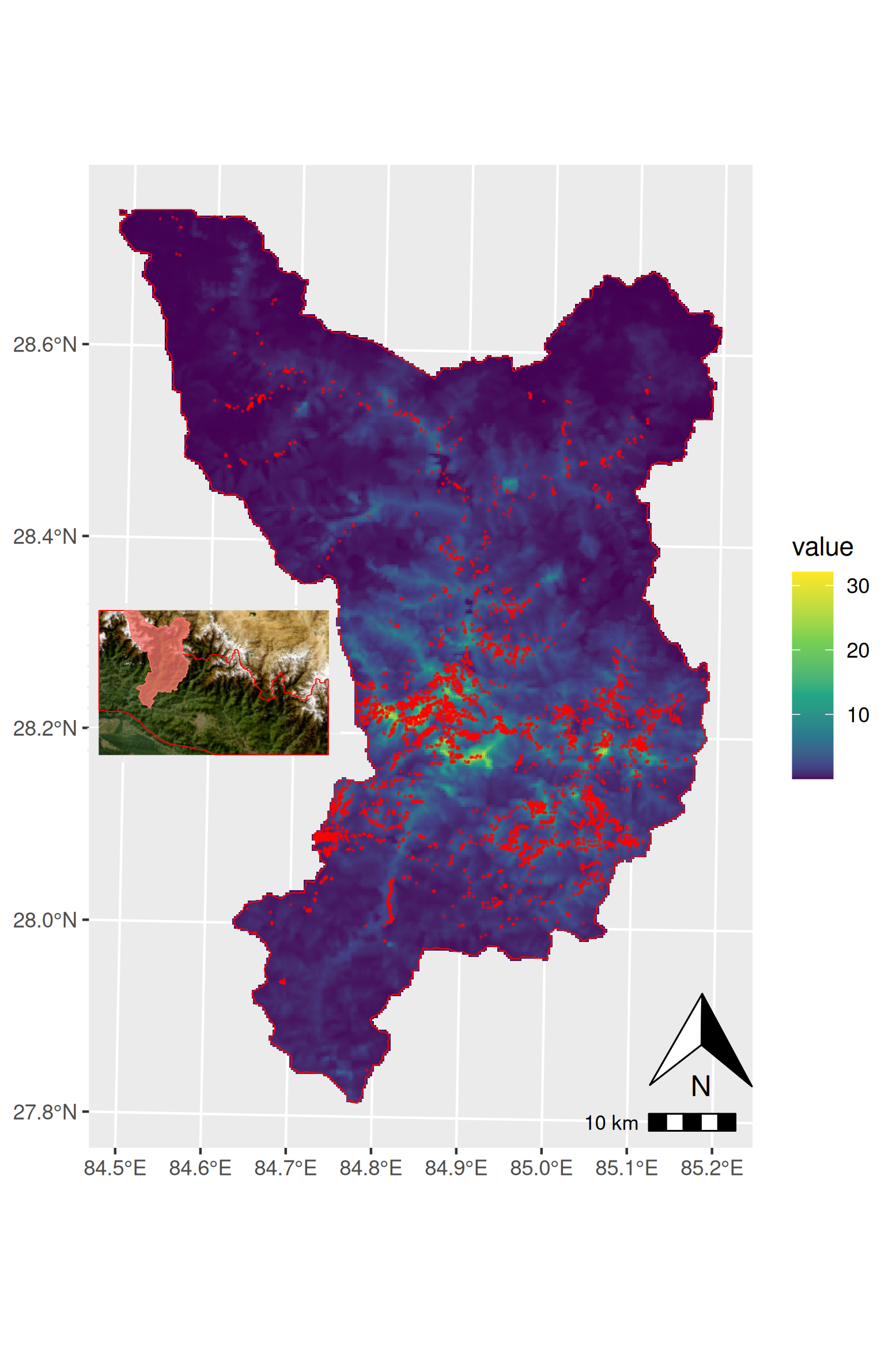}
        \caption{Posterior mean field (based on 100 samples) for basin 2228, zoomed in to highlight local structure.}
        \label{fig:fit6a_zm_2228}
    \end{subfigure}
    \hfill
    \begin{subfigure}[t]{0.48\linewidth}
        \centering
        \includegraphics[width=\linewidth, trim = 0 1.5cm 0 2cm, clip]{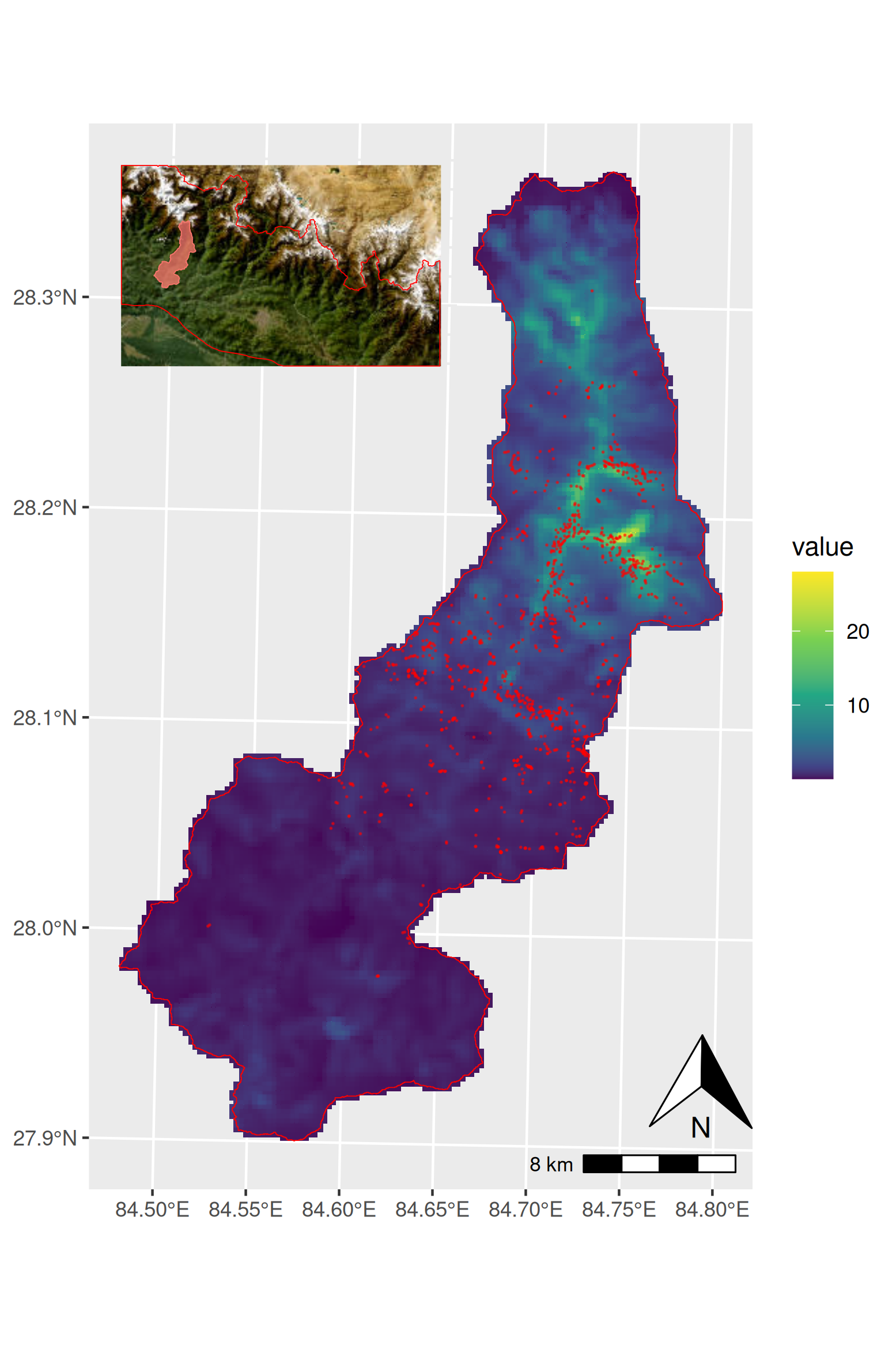}
        \caption{Posterior mean field (based on 100 samples) for basin 15329, zoomed in to highlight local structure.}
        \label{fig:fit6a_zm_15329}
    \end{subfigure}
\end{figure}

\begin{figure}[ht]\ContinuedFloat
    \begin{subfigure}[t]{\linewidth}
        \centering
        \includegraphics[width=.65\linewidth, trim = 0 .5cm 0 0, clip]{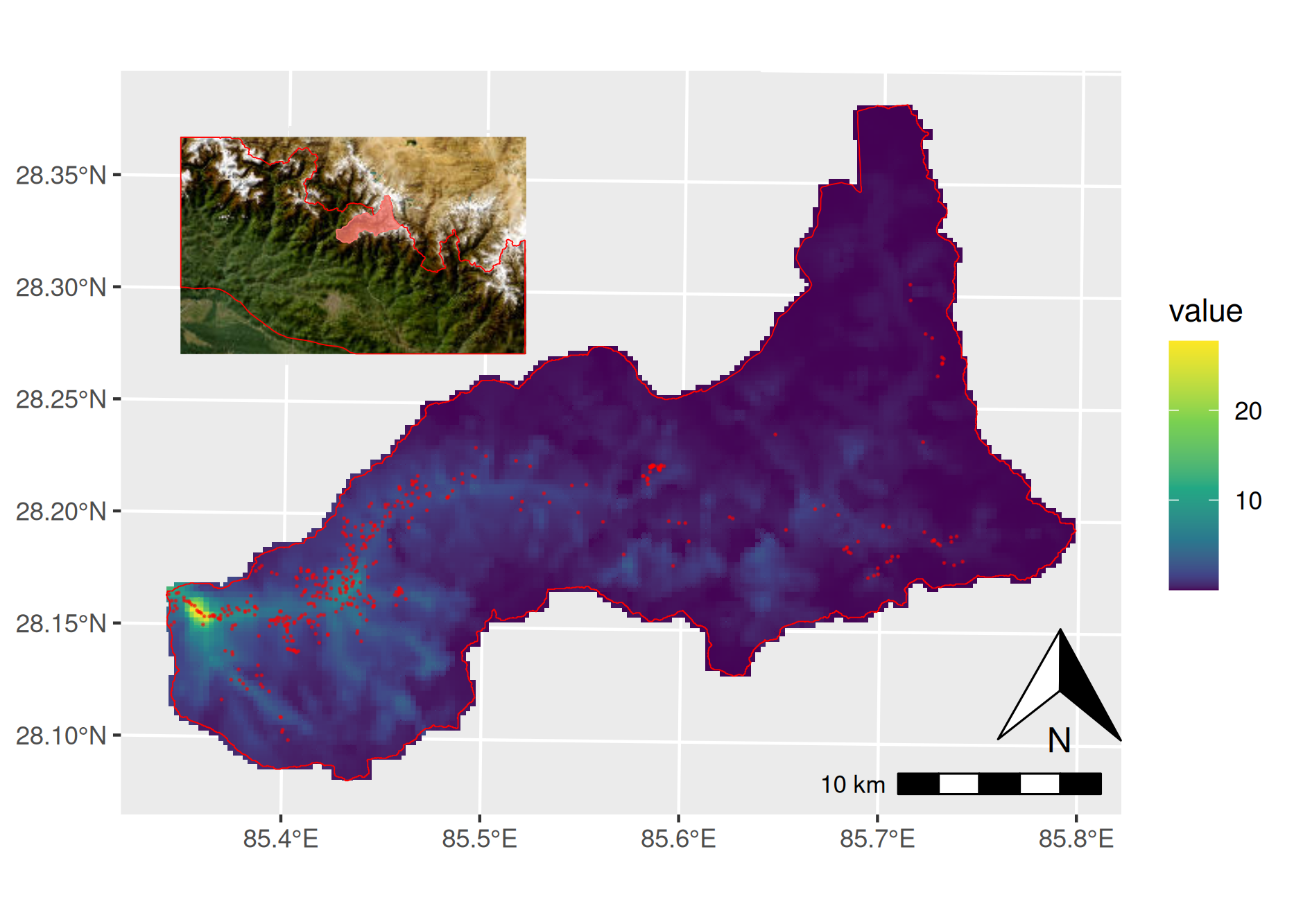}
        \caption{Posterior mean field (based on 100 samples) for basin 14982, zoomed in to highlight local structure.}
        \label{fig:fit6a_zm_14982}
    \end{subfigure}
    
    \caption{Zoomed-in posterior susceptibility mean fields from model \texttt{fit6a} using full dataset for basins 15329, 2228, and 14982. Red points indicate all observed landslide centroids; inset maps show basin locations.}
    \label{fig:fit6a_thinA_zoom}
\end{figure}

\section{Discussion}
\subsection{Limitations}
The landslide inventory in this study was collected over a sequence of seismic events, but it does not record which specific shock triggered each landslide. We tested two approaches to incorporate the $M_w$ 7.8 mainshock and $M_w$ 7.3 aftershock into the model: (i) stacking both ground shaking rasters as a single covariate, and (ii) including them as two separate covariates. However, incorporating the aftershock did not improve model scoring results. An alternative would be to fit separate models for the main shock and aftershock. However, since the landslide inventory lacks time stamps for individual events, we did not feel comfortable (i) assigning landslides to shocks based solely on proximity to epicentres or (ii) restricting the study regions accordingly.

Additionally, the spatial score patterns indicate that regions exhibiting systematic underprediction prior to the earthquake and the monsoon season (typically June to August) may have undergone preconditioning processes, such as progressive soil weakening or slope destabilisation, before seismic triggering. These areas are already known to be highly susceptible to monsoon-induced landslides, suggesting that antecedent geomorphic conditions (omitted from the current model) may partially account for the spatially coherent discrepancies between observed and predicted landslide susceptibility. The absence of relevant meteorological or hydrological covariates likely contributes to this mismatch, highlighting the need for incorporating such information in future spatio-temporal model extensions.

A likely contributing factor is the absence of meteorological covariates such as antecedent rainfall or cumulative soil moisture, which may interact with seismic shaking to elevate landslide risk. Without incorporating such hydrometeorological variables, the model may underestimate susceptibility in areas where precipitation had already primed the terrain for failure.

\subsection{Future works}
\subsubsection{Spatio-temporal Extension}
The study region exhibits substantial heterogeneity in landslide behaviour, including areas with sparse or no activity, zones dominated by numerous small events, and regions characterised by fewer but significantly larger landslides. It remains an open question which of these spatial patterns contributes most prominently to the overall area affected by landslides. A potential extension involves identifying whether certain landslides belong to the same “event family,” analogous to methods used in ecology or animal tracking, where multiple detections are attributed to a single group or origin. This can be conceptualised as a self-exciting process, such as a Hawkes process, where smaller initial landslides may act as parent events that reduce slope stability locally, thereby triggering additional failures in a cascading fashion until geomorphic equilibrium is reached. Conversely, large-scale failures may act as stabilising events that inhibit subsequent landsliding, which could explain why larger landslides tend to be more spatially isolated.

Such grouping would allow for a more physically grounded registration of multiple landslide occurrences as components of a single geomorphic response. However, inferring initiation points or reconstructing temporal sequences remains inherently difficult, particularly in the absence of high-resolution spatio-temporal data. This ambiguity underscores the broader need to improve our understanding of the spatio-temporal dynamics of landslide generation and their interdependencies, especially in the context of modelling co-occurring or cascading hazard processes.

\subsubsection{Landslides Classification}
In particular, landscapes shaped by fluvial incision often display extended hillslope-to-channel coupling. When individual hillslope units respond relatively independently to channel downcutting, this configuration can give rise to a high frequency of small-scale landslides. Conversely, in other geomorphic settings, large but less frequent failures may dominate.

Each landslide event in the dataset is associated with a set of covariate attributes, including size and external triggers such as ground shaking. This information presents an opportunity to explore whether distinct types of landslide processes can be identified. For instance, it may be possible to distinguish between landslides driven by river incision and those influenced by glacial dynamics.

Such differentiation suggests the potential utility of unsupervised classification methods to uncover latent groupings within the data. While these techniques are sometimes approached with caution in the geosciences, owing to interpretability challenges, they can nonetheless help delineate clusters that correspond to geomorphic processes. These insights could guide and strengthen subsequent analyses using structured statistical frameworks. This approach aligns with the long-standing tradition in geomorphology and geography, where classification is fundamental to understanding landscape-forming mechanisms.

\section{Conclusion}
Accurately capturing both the occurrence and magnitude of earthquake-induced landslides (EQILs) requires a statistical framework that reflects the complex interplay of geomorphic and seismic processes. In this study, we proposed a spatial modelling framework that simultaneously accounts for landslide centroids and their associated sizes. Our approach incorporates relevant covariates, Peak Ground Acceleration (PGA), channel steepness index (\ksn), land cover, geology, and flow distance to the channel, while addressing spatial misalignment between the response and explanatory variables through a mesh-based disaggregation strategy. The model is evaluated using spatial and thinning cross-validation, ensuring robustness and generalisability to unobserved geographic regions.

The case study presented in Section~\ref{sec:case_study} provides strong empirical support for the hypothesis that elevated \ksn{} values, which reflect enhanced fluvial incision, are associated with higher landslide susceptibility. Notably, this association pertains to the likelihood of landslide initiation rather than to landslide size, indicating that \ksn{} modulates triggering probability more than failure volume. Our results also show that glacially derived landslides are well captured under the proposed model framework.

The channel steepness index (\ksn{}) emerges as a key covariate in our approach, not only for its geomorphic significance, but also for its objectivity and reproducibility, being derived from robust topographic algorithms. This contrasts with slope unit-based methods, which are often sensitive to parameter tuning and lack transferability. By leveraging \ksn{} as an off-the-shelf, data-pragmatic alternative, our framework avoids the need for region-specific slope unit delineation, enhancing reproducibility and facilitating broader applicability.

In summary, this work contributes a generalisable and interpretable model that integrates both landslide occurrence and size within a unified spatial framework, offering a valuable tool for hazard assessment in seismically active mountainous terrains.

\appendix
\section{Table}
\begin{table}[H]
\centering
\caption{Some geological zones and their corresponding lithological characteristics.}
\begin{tabular}{lp{11cm}}
\toprule
\textbf{Geological Zone} & \textbf{Lithological Characteristics} \\
\midrule
Terai & Quaternary loose sediment (clay, silt, gravel) \\
Siwalik & Middle Miocene-Pleistocene sedimentary rocks (mudstone, sandstone, conglomerate) \\
Lesser Himalayan & Precambrian-Palaeozoic low grade metamorphic/metasedimentary rocks (slate, phyllite, limestone, metasandstone, schist, quartzite, gneiss) \\
Higher Himalayan & Precambrian crystalline rocks (quartzite, marble, gneiss) \\
Tibetan-Tethys Himalayan & Mesozoic-Cenozoic sedimentary rocks (siltstone, sandstone, limestone) \\
\bottomrule
\end{tabular}
\label{tab:geological_zones}
\end{table}

\begin{table}[H]
\centering
\caption{Filtered Land Cover Classification and Modelled Coefficients.}
\label{tab:merged_lcc_and_models}

\begin{subtable}[t]{\textwidth}
\centering
\caption{Filtered Land Cover Classification (LCC) Labels and Map Codes.}
\label{tab:lcc_labels_filtered}
\begin{tabular}{l p{12cm}}
\toprule
\textbf{LCC Code} & \textbf{LCC Label} \\
\midrule
1H & Herbaceous Crop(s) \\
1HI & Permanently Cropped Area With Irrigated Herbaceous Crop(s) \\
1HSs & Small Sized Field(s) Of Herbaceous Crop(s), Major Landclass: Sloping Land \\
1HSv & Small Sized Field(s) Of Herbaceous Crop(s), Major Landclass: Level Land, Valley Floor \\
2HCO & Herbaceous Closed to Open Medium Tall Vegetation \\
2HS & Sparse Short Herbaceous Vegetation \\
2HS//6BR & Sparse Short Herbaceous Vegetation // Bare Rock And/Or Coarse Fragments \\
2SCO & Closed to Open Medium To High Shrubland (Thicket) \\
2SOd & Dwarf Shrubland with Herbaceous \\
2SSd & Sparse Dwarf Shrubs and Sparse Herbaceous \\
2TCObe & Broadleaved Evergreen Closed to Open Trees \\
2TCOne & Needleleaved Evergreen Closed to Open Trees \\
2TCOne//2TCObe & Needleleaved Evergreen Closed to Open Trees // Broadleaved Evergreen Closed to Open Trees \\
5UI & Non-Linear Built Up Area(s) \\
6BR & Bare Rock(s) \\
6GR & Gravels, Stones And/Or Boulders \\
8ICE & Perennial Ice (Moving) \\
8ICEr & Perennial Ice (Moving) (Revised) \\
8SN & Perennial Snow \\
8SNs & Seasonal Snow \\
8WF & Natural Waterbodies (Flowing) \\
8WNP & Non-Perennial Natural Waterbodies (Standing) (Surface Aspect: Bare Rock) \\
\bottomrule
\end{tabular}
\footnotesize\vspace{0.5em}
\raggedright\textbf{Note:} See \citet{garcia2022general}.
\end{subtable}
\end{table}

\pagebreak
\begin{table}[H]
\ContinuedFloat
\begin{subtable}[t]{\textwidth}
\centering
\caption{Posterior means, standard deviations, and 95\% credible intervals for centroid-based modelled coefficients in \texttt{fit6a}. Significant effects (credible intervals excluding zero) are shown in bold.}
\label{tab:centroid_coefficients}
\begin{tabular}{lrrl}
\toprule
Land Cover & Mean & SD & 95\% CI \\
\midrule
1H              & -0.816 & 0.606 & $(-2.077,\ 0.314)$ \\
1HI             & -0.372 & 0.677 & $(-1.755,\ 0.921)$ \\
\textbf{1HSs}   & \textbf{0.352} & \textbf{0.060} & $\mathbf{(0.236,\ 0.470)}$ \\
1HSv            &  0.084 & 0.167 & $(-0.244,\ 0.413)$ \\
\textbf{2HCO}   & \textbf{0.805} & \textbf{0.060} & $\mathbf{(0.688,\ 0.924)}$ \\
2HS             & -0.378 & 0.675 & $(-1.758,\ 0.911)$ \\
\textbf{2HS//6BR} & \textbf{-0.443} & \textbf{0.117} & $\mathbf{(-0.673,\ -0.212)}$ \\
\textbf{2SCO}   & \textbf{0.697} & \textbf{0.063} & $\mathbf{(0.574,\ 0.822)}$ \\
\textbf{2SOd}   & \textbf{0.214} & \textbf{0.063} & $\mathbf{(0.092,\ 0.339)}$ \\
\textbf{2SSd}   & \textbf{1.716} & \textbf{0.214} & $\mathbf{(1.295,\ 2.135)}$ \\
2TCObe          & -0.753 & 0.612 & $(-2.024,\ 0.393)$ \\
\textbf{2TCOne} & \textbf{0.291} & \textbf{0.059} & $\mathbf{(0.177,\ 0.408)}$ \\
\textbf{2TCOne//2TCObe} & \textbf{0.176} & \textbf{0.061} & $\mathbf{(0.057,\ 0.297)}$ \\
5UI             & -0.843 & 0.619 & $(-2.132,\ 0.311)$ \\
6BR             & -0.674 & 0.389 & $(-1.450,\ 0.077)$ \\
6GR             & -0.198 & 0.147 & $(-0.486,\ 0.092)$ \\
8ICE            & -0.258 & 0.525 & $(-1.303,\ 0.763)$ \\
8ICEr           &  0.091 & 0.213 & $(-0.325,\ 0.508)$ \\
\textbf{8SN}    & \textbf{-1.086} & \textbf{0.095} & $\mathbf{(-1.271,\ -0.900)}$ \\
\textbf{8SNs}   & \textbf{-1.534} & \textbf{0.256} & $\mathbf{(-2.043,\ -1.039)}$ \\
8WF             & -0.234 & 0.237 & $(-0.699,\ 0.231)$ \\
8WNP            & -0.119 & 0.737 & $(-1.601,\ 1.321)$ \\
\bottomrule
\end{tabular}
\end{subtable}

\vspace{1em}

\begin{subtable}[t]{\textwidth}
\centering
\caption{Posterior means, standard deviations, and 95\% credible intervals for log-size modelled coefficients in \texttt{fit6b}. Significant effects (credible intervals excluding zero) are shown in bold.}
\label{tab:logsize_coefficients}
\begin{tabular}{lrrl}
\toprule
Land Cover & Mean & SD & 95\% CI \\
\midrule
\textbf{1HSs}       & \textbf{-0.981} & \textbf{0.081} & $\mathbf{(-1.142,\ -0.823)}$ \\
\textbf{1HSv}       & \textbf{-1.302} & \textbf{0.228} & $\mathbf{(-1.750,\ -0.858)}$ \\
2HCO               & -0.087 & 0.081 & $(-0.249,\ 0.071)$ \\
2HS//6BR           &  0.253 & 0.162 & $(-0.064,\ 0.569)$ \\
\textbf{2SCO}       & \textbf{-0.389} & \textbf{0.086} & $\mathbf{(-0.559,\ -0.223)}$ \\
2SOd               &  0.039 & 0.085 & $(-0.128,\ 0.204)$ \\
2SSd               &  0.454 & 0.254 & $(-0.043,\ 0.953)$ \\
2TCOne             &  0.014 & 0.080 & $(-0.145,\ 0.169)$ \\
\textbf{2TCOne//2TCObe} & \textbf{-0.692} & \textbf{0.084} & $\mathbf{(-0.858,\ -0.530)}$ \\
6BR                &  0.124 & 0.536 & $(-0.926,\ 1.182)$ \\
6GR                & -0.013 & 0.199 & $(-0.404,\ 0.378)$ \\
8ICE               &  0.368 & 0.675 & $(-0.934,\ 1.727)$ \\
\textbf{8ICEr}      & \textbf{1.965} & \textbf{0.294} & $\mathbf{(1.394,\ 2.545)}$ \\
\textbf{8SN}        & \textbf{1.279} & \textbf{0.130} & $\mathbf{(1.024,\ 1.534)}$ \\
8SNs               &  0.323 & 0.364 & $(-0.389,\ 1.038)$ \\
\textbf{8WF}        & \textbf{-1.355} & \textbf{0.331} & $\mathbf{(-2.010,\ -0.713)}$ \\
\bottomrule
\end{tabular}
\end{subtable}

\end{table}

\section{Channel Steepness Index (\ksn)}

\subsection{Hanging Valley}\label{sec:ksn_fix}
\begin{figure}[H]
    \centering
    \includegraphics[width=\linewidth, draft=FALSE]{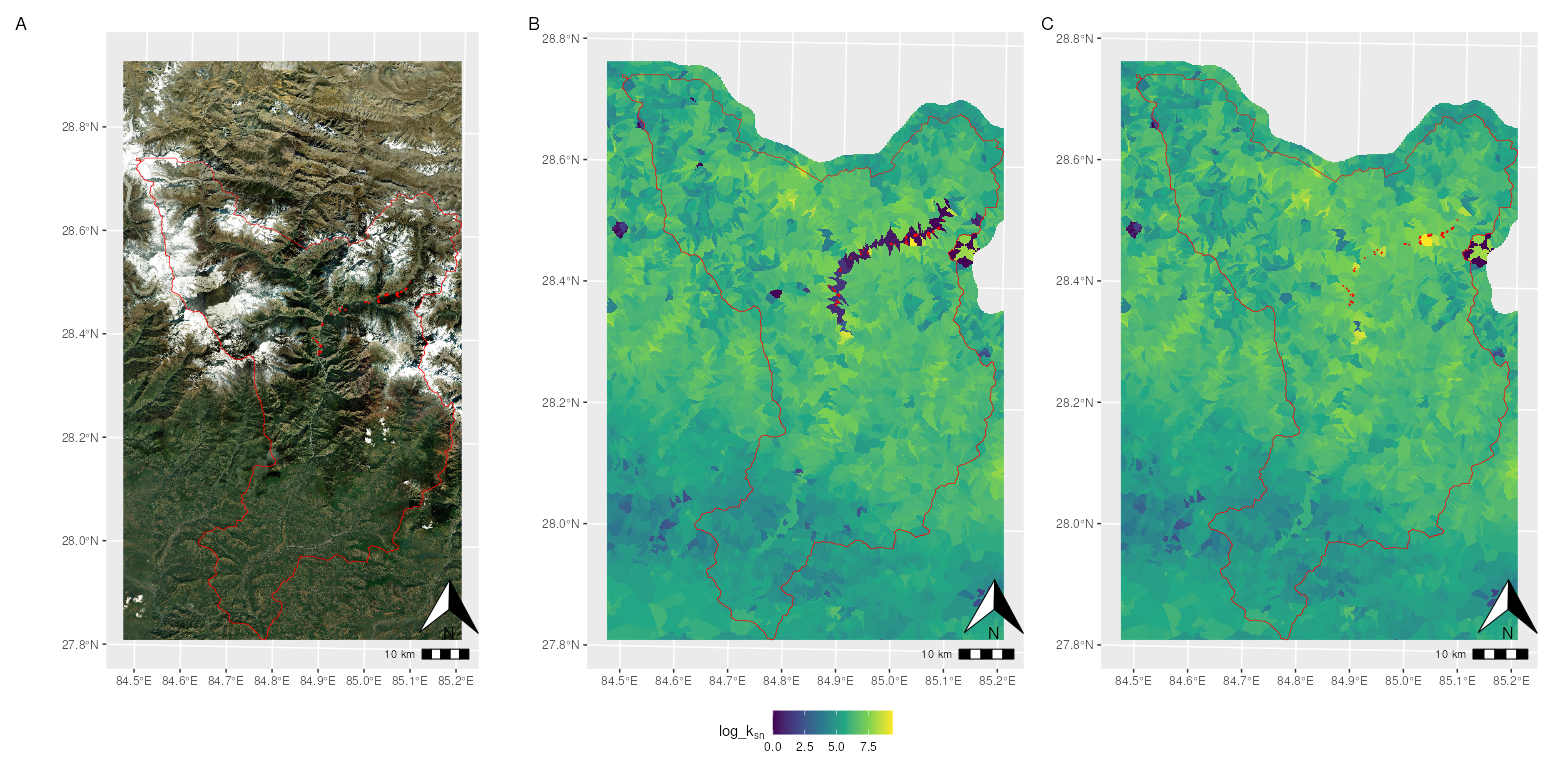}
    \caption{\textbf{A}: Landslide centroids falling into glacier landscapes for basin 2228 with basemap tile layer from ESRI World Imagery. \textbf{B}: \ksn \ basin map with failure of algorithm due to glacier landscapes. \textbf{C}: \ksn \ basin map after filling underperforming area with nearest neighbour values. Basin boundary in red and landslide centroids with  $\log k_{sn} < 1.5 $ in m in red.}

    \label{fig:glacier}
\end{figure}

Here we illustrate how we replace the channel steepness index (\ksn) algorithm due to hanging valley with nearby values from Section \ref{sec:lcl}, see Figure \ref{fig:glacier}.

\subsection{Algorithm Technicality}\label{sec:ksn_tech}
As mentioned in Section \ref{sec:ksn}, we spell out some details of the algorithm in \lsd. To compute the DEM derivatives, we use channel extraction methods based on the FASTSCAPE algorithm \citep{braun2013very}, which implements the D8 method for determining drainage directions. The D8 method calculates the slope between a pixel and its steepest downslope neighbour. Using a DEM and a specified threshold of stream order, where a minimum stream order defines a channel, we can compute the \textbf{Relief to Channel} as a raster following the relief to the nearest channel node. We also experimented with both the \textbf{Relief/ Flow Distance to Far Ridge}, i.e.\ ridgetops, but it did not perform well in modelling, and thus, it is not included in this analysis.

\section{Cross-Validation Grid}\label{sec:cv_chess}
\begin{figure}[H]
    \centering
    \includegraphics[width=.75\linewidth, trim = {0 3cm 0 4cm}, clip]{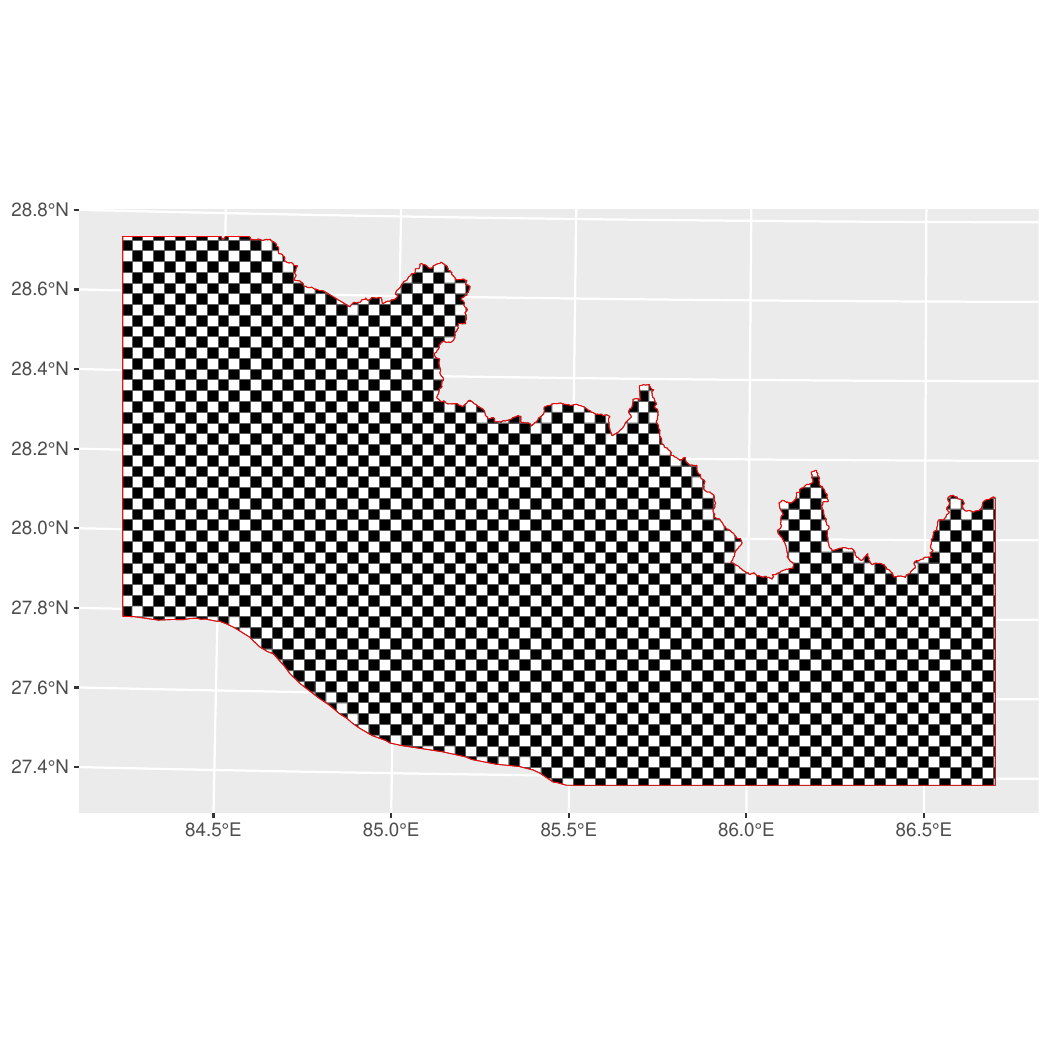}
    \caption{Train Test Data Split for Cross Validation with grid size $3 \text{km} \times 3 \text{km}$}
    \label{fig:cv_chess}
\end{figure}

Figure \ref{fig:cv_chess} shows the Cross-Validation $3 \text{km} \times 3 \text{km}$ grid mentioned in Section \ref{sec:cv}.

\section{Landslide Sizes Exploratory Analysis}\label{sec:lds_explore}
\begin{figure}[H]
    \centering
    \includegraphics[width=\linewidth, draft=FALSE]{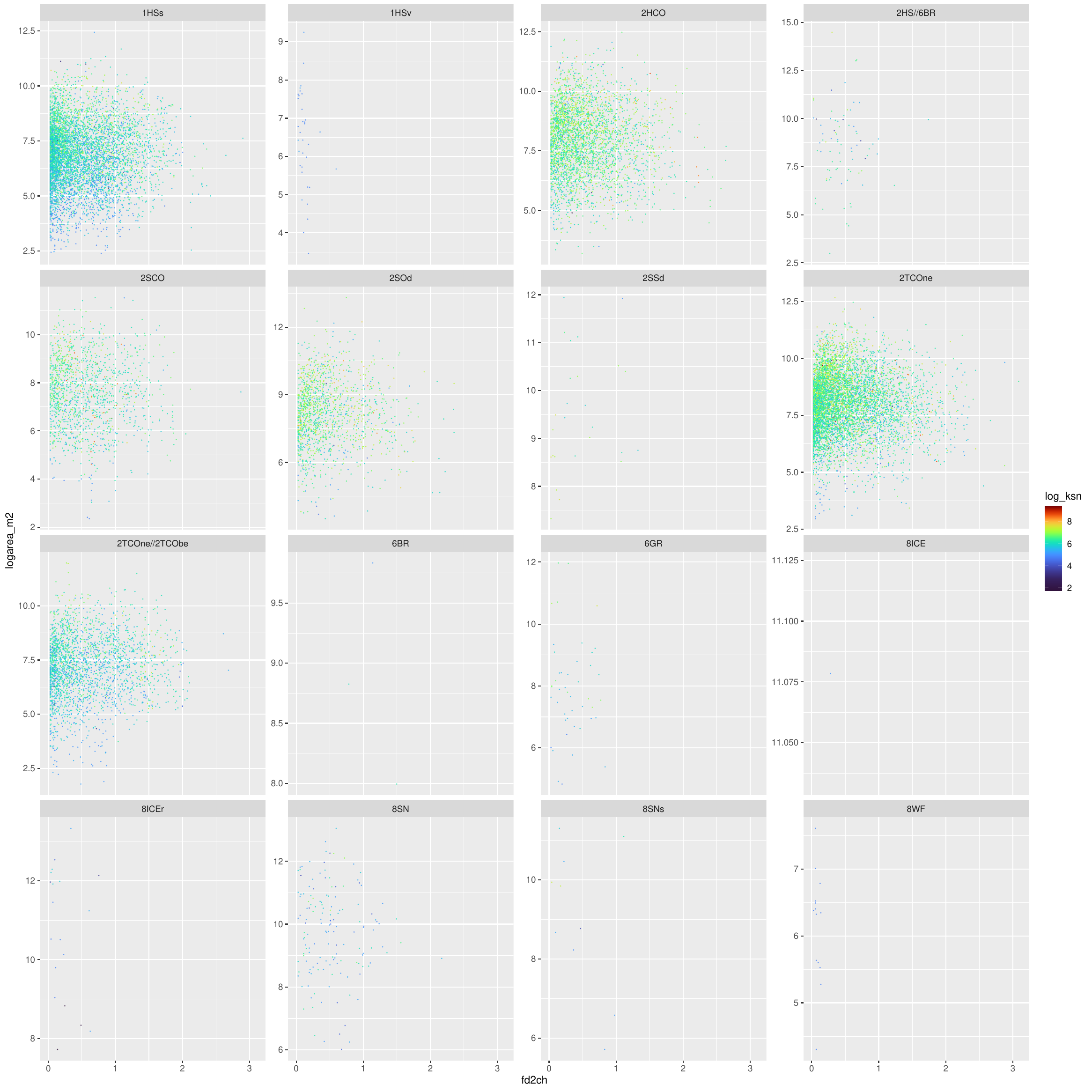}
    \caption{Scatter Plot of log landslide size against flow distance to channel (fd2ch) coloured by $\log$ \ksn{} stratified by Land Cover. See Table \ref{tab:lcc_labels_filtered} for Land Cover Code.}
    \label{fig:lds_fd}
\end{figure}

\begin{figure}[H]
    \centering
    \includegraphics[width=\linewidth, draft=FALSE]{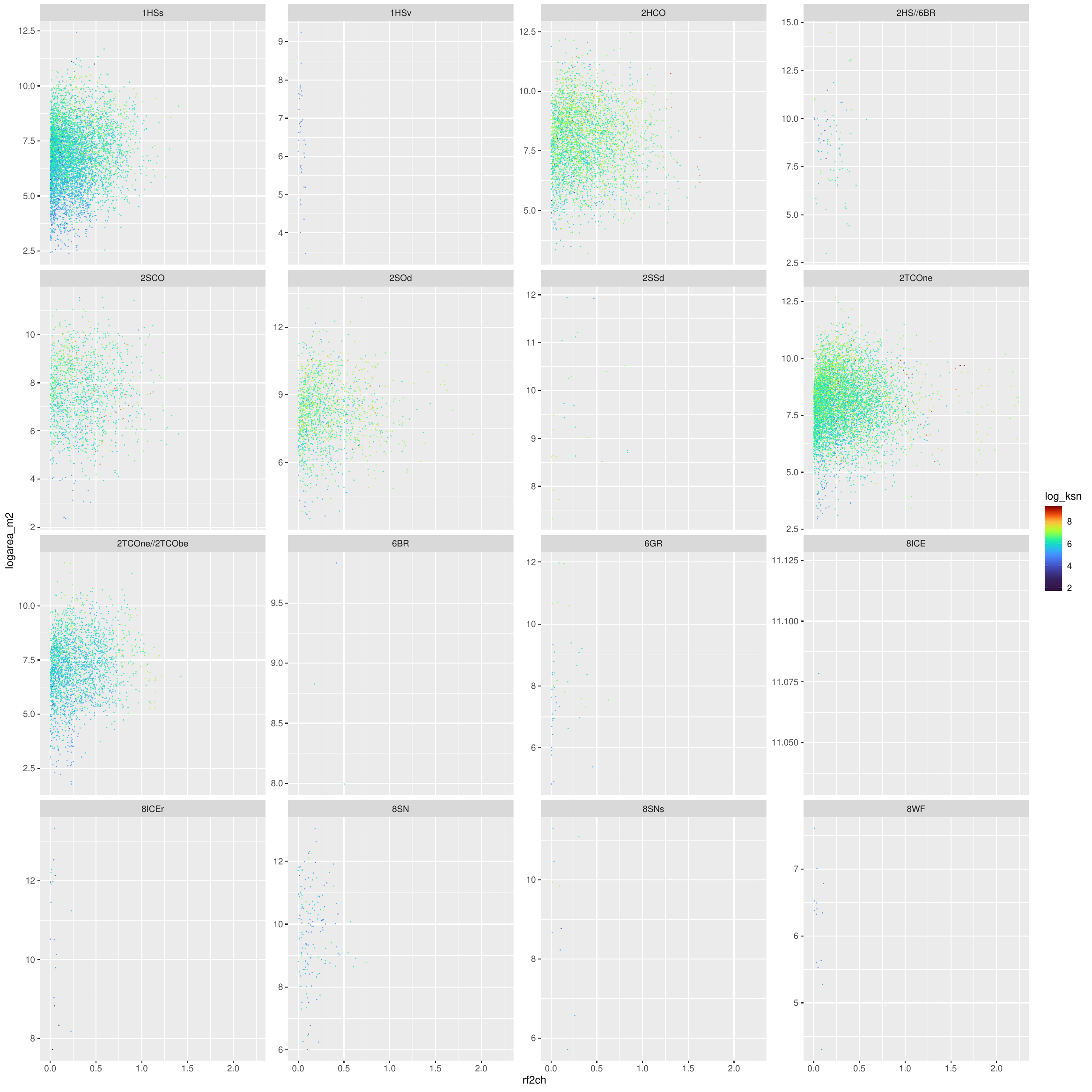}
    \caption{Scatter Plot of log landslide size against relief to channel (rf2ch) coloured by $\log$ \ksn{}  stratified by Land Cover. See Table \ref{tab:lcc_labels_filtered} for Land Cover Code.}
    \label{fig:lds_rf}
\end{figure}

Figures~\ref{fig:lds_fd} and~\ref{fig:lds_rf} present scatter plots of $\log$ landslide size against two distance metrics to the channel, stratified by land cover categories and coloured by $\log$ \ksn{}. Both distance metrics exhibit similar scatter patterns. In particular, smaller landslide sizes are associated with lower $\log$ \ksn{} values, especially in land cover classes 1HSs, and 2TCOne//2TCObe, where landslide observations are more frequent.


\subsection{Further Channel Profile Analysis}\label{sec:further_ksn}
\begin{figure}[H]
    \centering

    \begin{subfigure}[t]{0.7\linewidth}
        \centering
        \includegraphics[width=\linewidth, draft=FALSE]{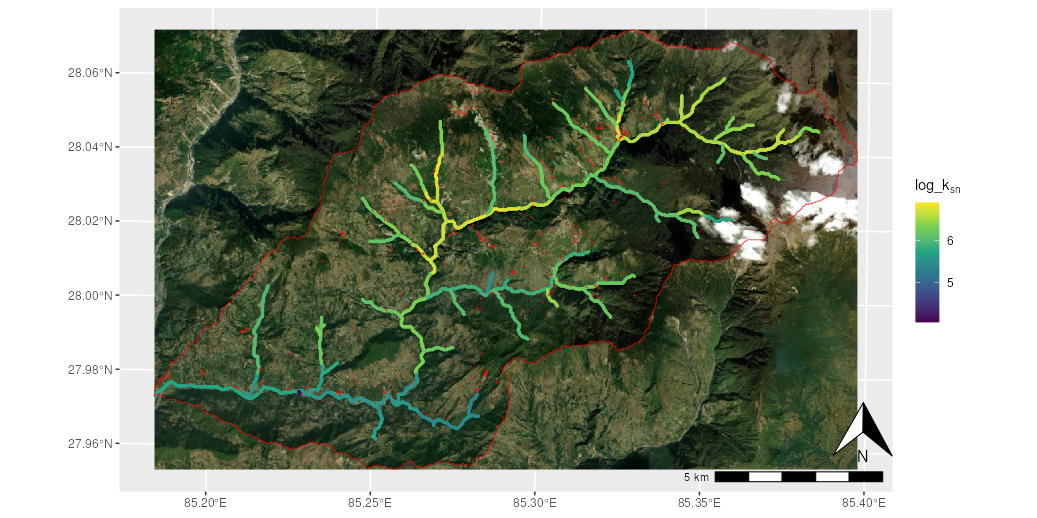}
        \caption{Channel steepness index ($\log$\ksn) along channels with landslide polygons (red); basemap from ESRI World Imagery.}
        \label{fig:ksn_22048_basin_poly}
    \end{subfigure}

    \vspace{1em}

    \begin{subfigure}[t]{0.7\linewidth}
        \centering
        \includegraphics[width=\linewidth, draft=FALSE]{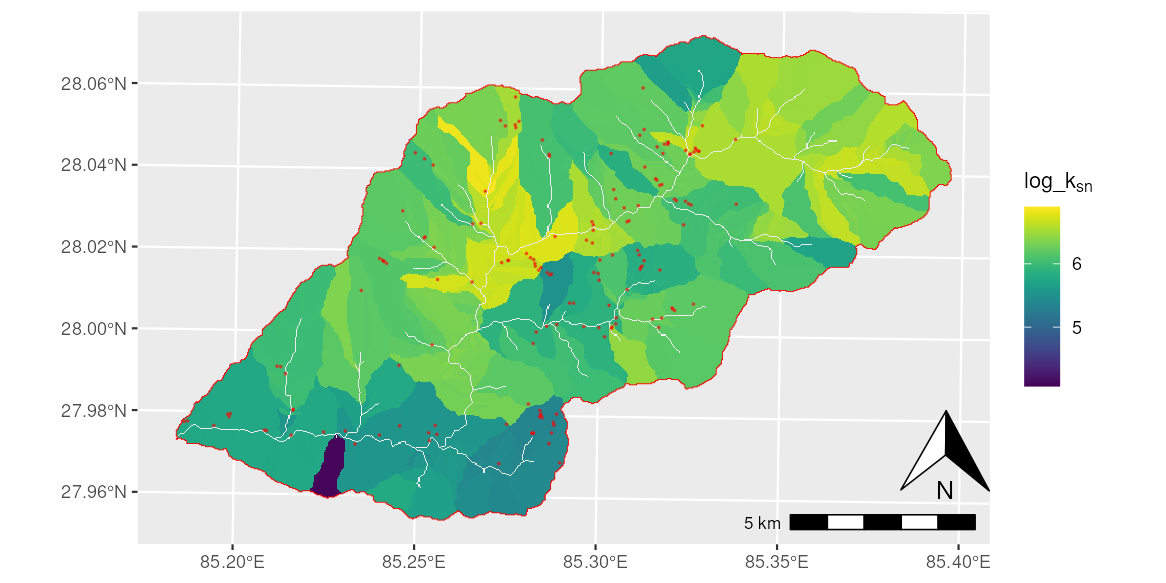}
        \caption{Channel steepness index ($\log$\ksn) tagged by hillslope polygons.}
        \label{fig:22048_basin_tag}
    \end{subfigure}

    \vspace{1em}

    \begin{subfigure}[t]{0.8\linewidth}
        \centering
        \includegraphics[width=\linewidth, draft=FALSE]{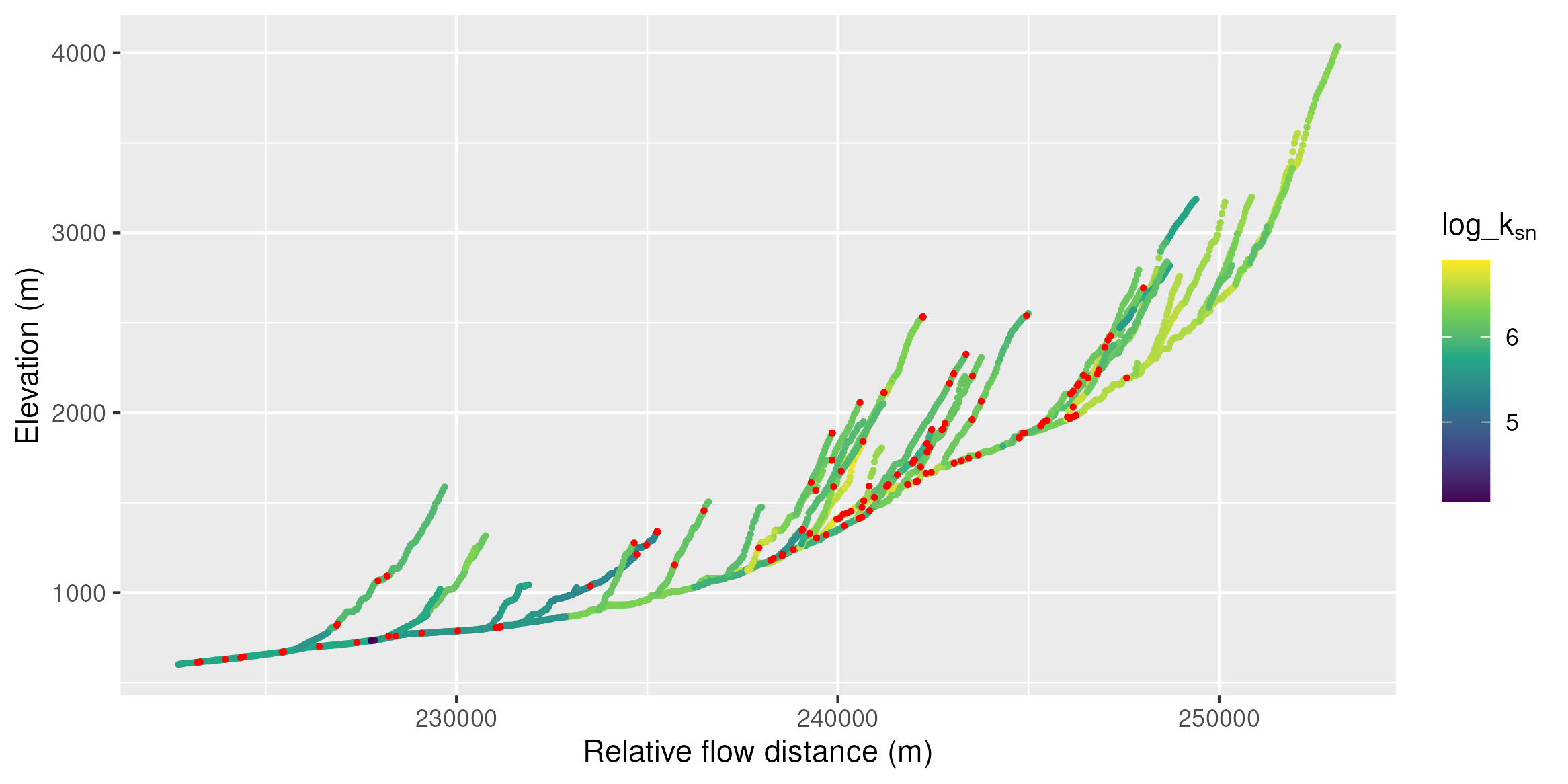}
        \caption{Channel profile plot of elevation against flow distance.}
        \label{fig:22048_chi_analysis}
    \end{subfigure}

    \caption{Channel characteristics and analysis for basin 22048.}
    \label{fig:22048_all}
\end{figure}

Further channel profile analysis from Section \ref{sec:channel_profile} is presented in Figure \ref{fig:22048_all}.






\subsection{Further Concavity Analysis}
\begin{figure}[H]
    \centering

    \begin{subfigure}[t]{\linewidth}
        \centering
        \includegraphics[width=\linewidth, draft=FALSE]{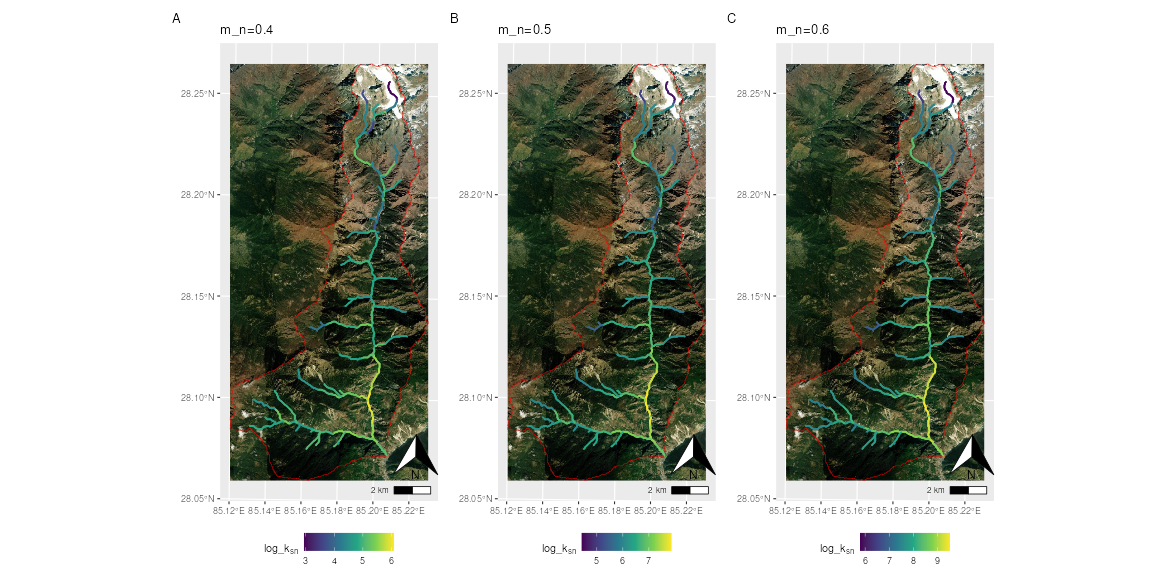}
        \caption{Basin 17757: $\log k_{sn}$ vs. drainage area for three concavity values $m/n$ (m\_n).}
        \label{fig:concavity_wrap17757}
    \end{subfigure}
    
    \vspace{1em}
    
    \begin{subfigure}[t]{\linewidth}
        \centering
        \includegraphics[width=\linewidth, draft=FALSE]{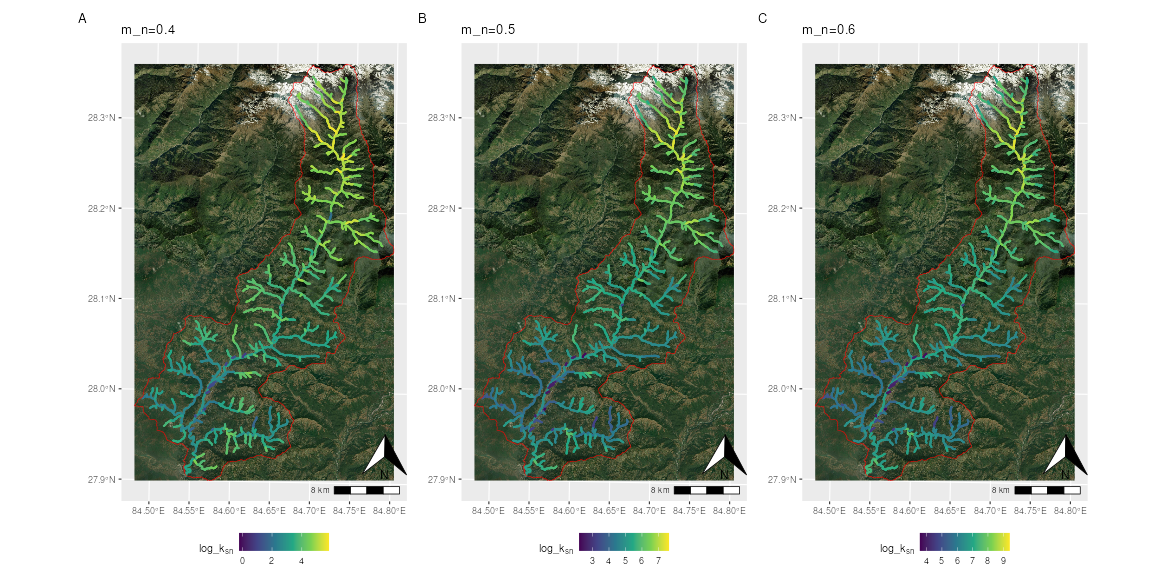}
        \caption{Basin 15329: $\log k_{sn}$ vs. drainage area for three concavity values $m/n$ (m\_n).}
        \label{fig:concavity_wrap15329}
    \end{subfigure}
    
    

    \caption{Examples of $\log k_{sn}$ distributions for selected basins under varying channel concavity values $m/n$ ($m_n$).}
    \label{fig:concavity_wrap_all}
\end{figure}

Here we show more concavity plots (see Figures \ref{fig:concavity_wrap_all}) to illustrate the consistent results from different concavity. We see the colour pattern of the $\log$ \ksn{} remains the same, hence it would not affect the spatial modelling results. 




\section{Computation Time and Session Information}\label{sec:sysinfo_ctime}
The Channel steepness index (\ksn) computation was run on the latest development version of \lsd. All analyses were conducted using R version 4.5.0, with \texttt{INLA} package version 25.05.07,  \inlabru{} version 2.12.0.9015, and \texttt{fmesher} version 0.3.0.

All spatial models were performed limited to 10 threads on a Four Intel Xeon E5-2680 v3 2.5GHz, 30M Cache, 9.6 GT/s QPI 192 GB RAM machine (i.e.\ 48 cores in total and 2 threads each core). 

\begin{figure}[H]
    \centering
    \begin{subfigure}[t]{0.48\linewidth}
            \includegraphics[width=\linewidth]{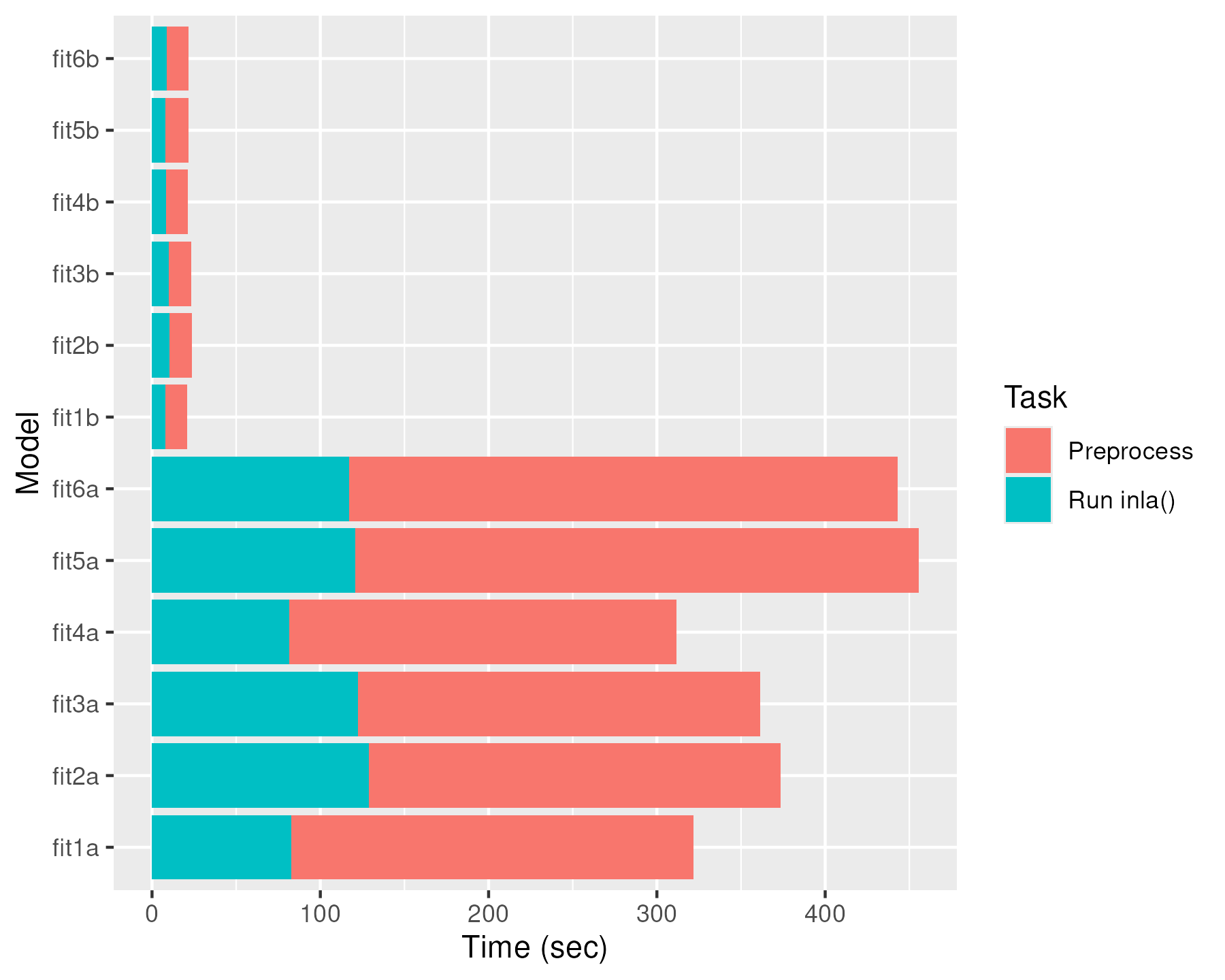}
    \caption{Computation time for Thinning CV (Set A)}
    \label{fig:ctime_thinA}
    \end{subfigure}
    \hfill
    \begin{subfigure}[t]{0.48\linewidth}
            \includegraphics[width=\linewidth]{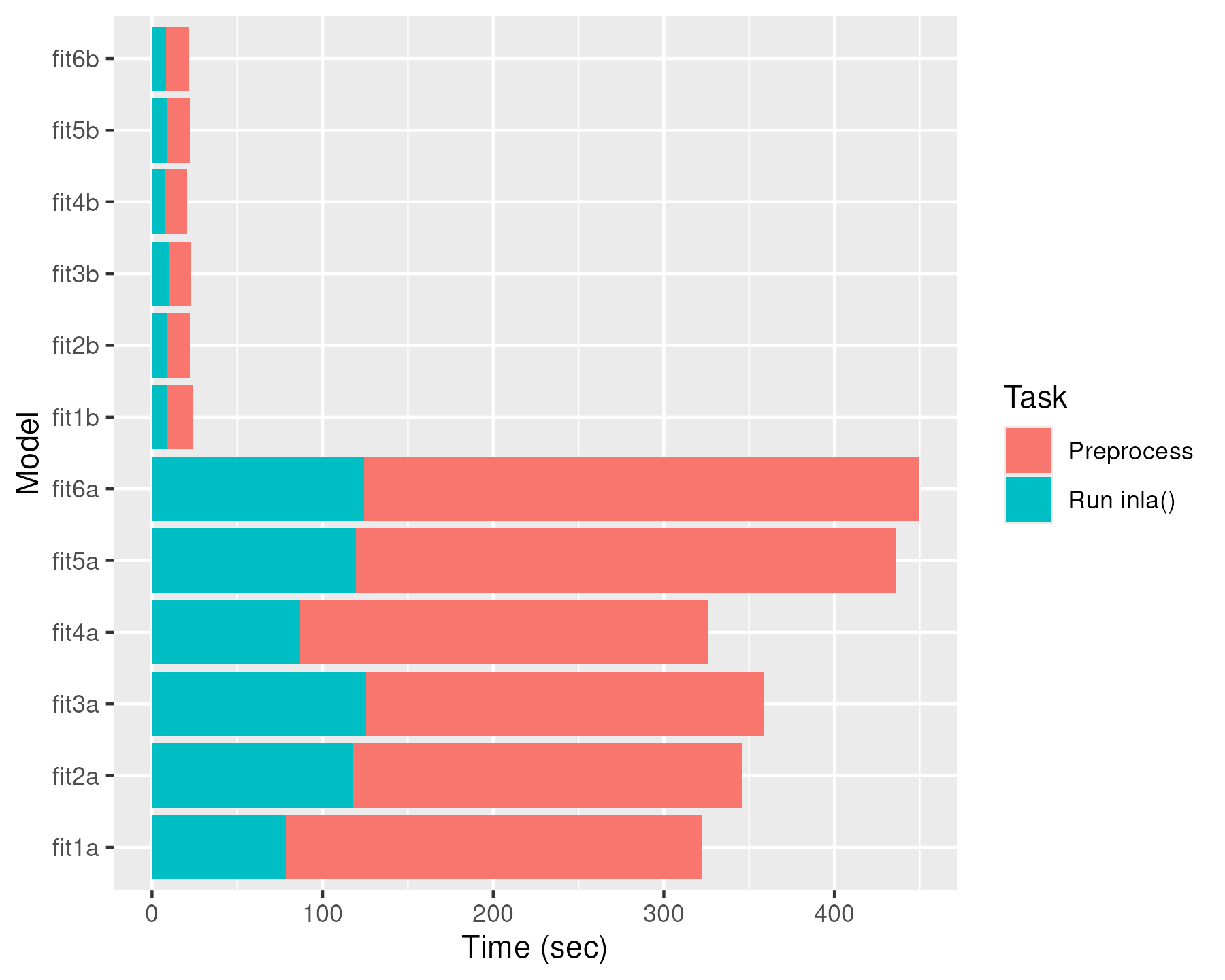}
    \caption{Computation time for Thinning CV (Set B)}
    \label{fig:ctime_thinB}
    \end{subfigure}
    
    \begin{subfigure}[t]{0.48\linewidth}
            \includegraphics[width=\linewidth]{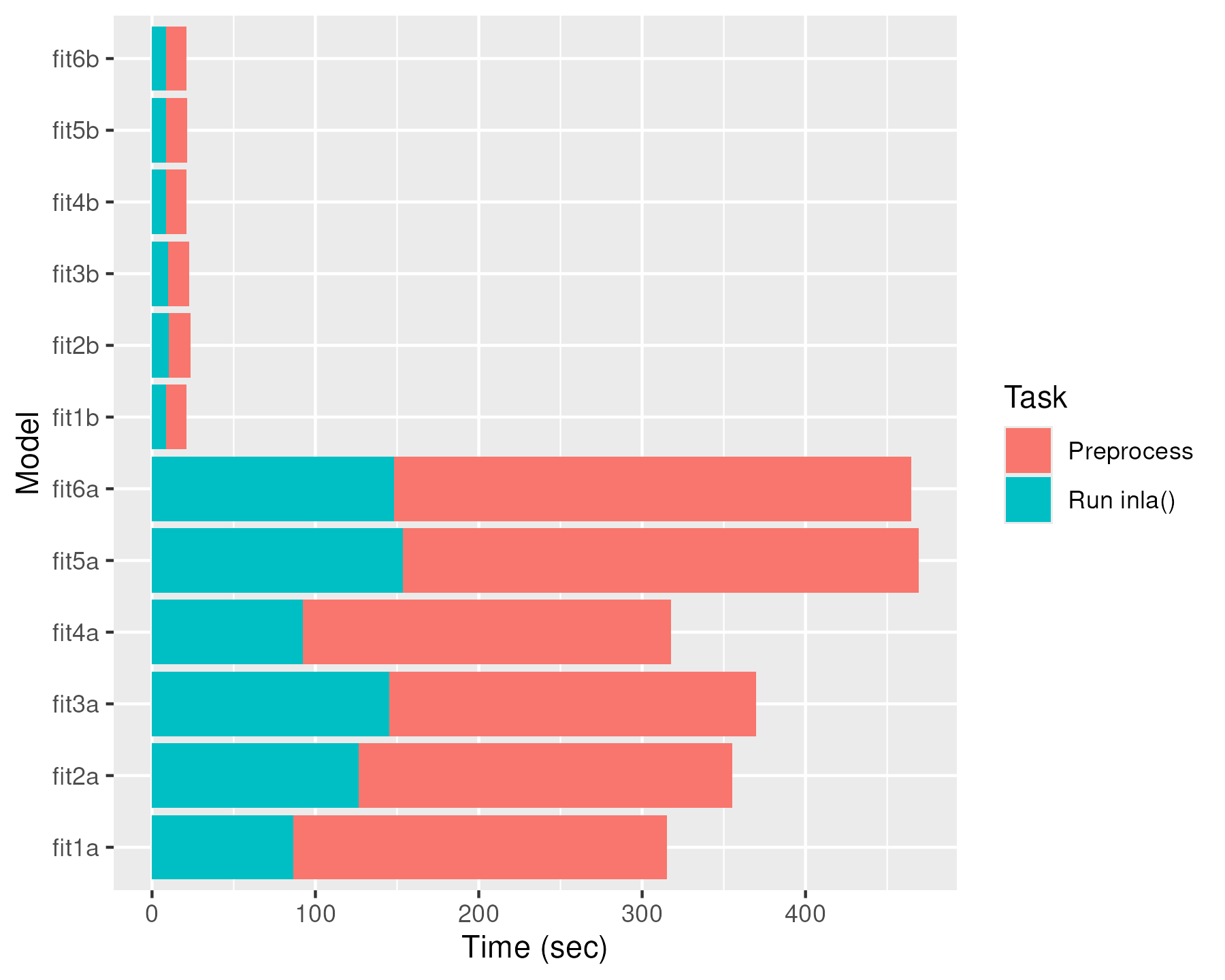}
    \caption{Computation time for Grid CV (Black)}
    \label{fig:ctime_black}
    \end{subfigure}
    \hfill
    \begin{subfigure}[t]{0.48\linewidth}
            \includegraphics[width=\linewidth]{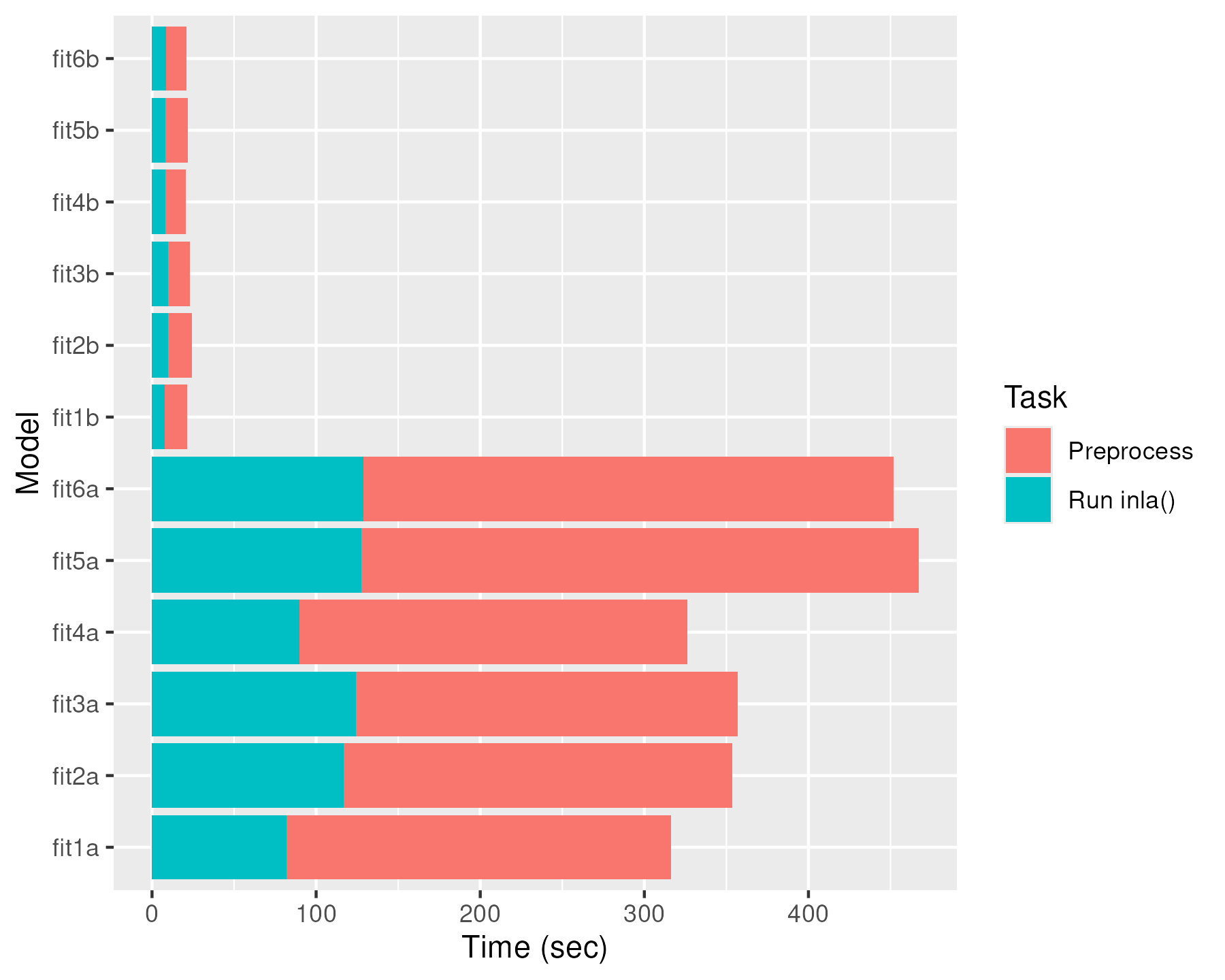}
    \caption{Computation time for Grid CV (White)}
    \label{fig:ctime_white}
    \end{subfigure}
    \caption{Computation time for all models in all cross-validation (CV) settings}
    \label{fig:ctime}
\end{figure}

\bibliography{gorkha}

\end{document}